\documentclass[11pt,a4paper]{article}
\pdfoutput=1
\usepackage{jheppub}
\usepackage[utf8x]{inputenc}
\usepackage{color}
\usepackage{amsmath}
\usepackage{amssymb}
\allowdisplaybreaks 
\usepackage{xargs}
\usepackage{dcolumn}
\usepackage{array}
\usepackage{caption}
\usepackage{pbox}
\usepackage{float}
\usepackage{graphicx}
\usepackage[all]{xy}

\numberwithin{equation}{section}

\begin{document}

\thispagestyle{empty}

\rightline{\small}

\vskip 3cm
\noindent
{\LARGE \bf 
On the  ${\cal N}=1^\ast$ Gauge Theory on a Circle
and
}
\vskip .19cm
\noindent
{\LARGE \bf  Elliptic Integrable Systems
}
\begin{center}
\linethickness{.06cm}
\line(1,0){447}
\end{center}

\vskip .8cm

\noindent
{\large \bf Antoine Bourget and Jan Troost}
\vskip 0.2cm
{\em \hskip -.05cm Laboratoire de Physique Th\'eorique\footnote{Unit\'e Mixte du
CNRS et
    de l'Ecole Normale Sup\'erieure associ\'ee \`a l'universit\'e Pierre et
    Marie Curie 6, UMR
    8549.}}
    \vskip -.05cm
{\em \hskip -.05cm Ecole Normale Sup\'erieure}
 \vskip -.05cm
{\em \hskip -.05cm 24 rue Lhomond, 75005 Paris, France}

\vskip 1cm

\vskip 0.6 cm

\noindent {\sc Abstract :}
We continue our study of the ${\cal N}=1^\ast$ supersymmetric gauge theory on $\mathbb{R}^{2,1} \times S^1$ 
and its relation to elliptic integrable systems. Upon compactification on a circle, we show that
the semi-classical analysis of the massless and massive vacua depends on the classification of nilpotent orbits,
as well as on the conjugacy classes of the component group of their centralizer. 
We demonstrate that semi-classically massless vacua can be lifted by Wilson lines in unbroken
discrete gauge groups.
The pseudo-Levi subalgebras that play a classifying role in the nilpotent orbit theory are also key in defining generalized Inozemtsev limits of (twisted) elliptic integrable systems. We illustrate our analysis
in the  ${\cal N}=1^\ast$ theories
with gauge algebras $su(3)$, $su(4)$, $so(5)$ and for the exceptional
gauge algebra $G_2 $. 
We map out modular duality diagrams of the massive and massless vacua. Moreover, we provide an analytic description of 
the branches of massless vacua in the case of the $su(3)$ and the $so(5)$ theory. 
The description of these branches in terms of the complexified Wilson lines on the circle
invokes the Eichler-Zagier technique for inverting the 
elliptic Weierstrass function.  After fine-tuning the coupling to elliptic points of order three, we identify the Argyres-Douglas singularities of the $su(3)$ ${\cal N}=1^\ast$ theory.

\vskip  0.15cm

\noindent

\newpage

{
\tableofcontents }


\section{Introduction}
The infrared dynamics of supersymmetric gauge theories is a rich and fruitful
subject. The classification of massless and massive
vacua, and the analysis of their symmetry and duality properties are basic features of the theory.
For pure ${\cal N}=1$ supersymmetric gauge theory in four dimensions, which is massive in
the infrared, we understand the supersymmetric index \cite{Witten:1982df,Witten:1997bs,Keurentjes:1999qf,Kac:1999gw,Witten:2000nv}, as well as the transformation properties
of the vacua under the broken non-anomalous R-symmetry. It is natural to extend the  study of the vacua
to other ${\cal N}=1$ supersymmetric gauge theories. We recently completed   the census
of massive vacua in the ${\cal N}=1^\ast$ theory that arises from mass deformation of the maximally
supersymmetric ${\cal N}=4$ theory in $\mathbb{R}^4$ \cite{Donagi:1995cf,Naculich:2001us,
Wyllard:2007qw,Bourget:2015lua}. This is an interesting playground due to 
duality symmetries inherited by the ${\cal N}=1^\ast$ theory from the celebrated duality properties
of ${\cal N}=4$ supersymmetric Yang-Mills theory. In the semi-classical count of vacua, nilpotent
orbit theory plays a central role, since $sl(2)$ representations solve the F-term equations of motion, and $sl(2)$ embeddings are intimately related to nilpotent orbits through the Jacobson-Morozov theorem \cite{
Bourget:2015lua}.

Upon further compactification of the gauge theory on a circle, there is a method to derive a low-energy effective superpotential for the ${\cal N}=1^\ast$ theory with any gauge group.
The method is based on a soft breaking of ${\cal N}=2^\ast$ supersymmetry by a
third mass deformation, as well as the identification of the integrable system 
for ${\cal N}=2^\ast$ theory and
its quadratic Hamiltonian \cite{Seiberg:1996nz,Dorey:1999sj,D'Hoker:1998yg,D'Hoker:1998yh,Kumar:2001iu}. The effective potential on $\mathbb{R}^4$ can then be recuperated from the radius independent potential on $\mathbb{R}^{2,1} \times S^1$. However, in this procedure it is clear that one should be mindful about 
the global distinctions between the gauge theory on $\mathbb{R}^4$ and the theory on $\mathbb{R}^{2,1} 
\times S^1$. An example of such subtlety is provided by the supersymmetric index of pure ${\cal N}=1$ theories which indeed depends on those global properties. In that case, the choice of the center of the gauge group and the spectrum of line operators is crucial in computing the vacuum structure after compactification on $S^1$ \cite{Aharony:2013hda,Aharony:2013kma}. 

The comparison of semi-classical calculations in ${\cal N}=1^\ast$ gauge theory
to the properties of the 
corresponding twisted elliptic Calogero-Moser integrable system  
allows to construct a beautiful bridge between ${\cal N}=1$ gauge theories and integrable systems 
\cite{Dorey:1999sj,Kumar:2001iu,Bourget:2015cza}. The further detailed comparison of the global features of the theory on 
$\mathbb{R}^{2,1} \times S^1$ will add ornaments to this bridge.
In this paper, we argue that upon circle
compactification, more intricate features of nilpotent orbit theory come into play. Indeed, the non-trivial
topology allows for turning on Wilson lines that can increase the number of massive vacua through 
various mechanisms \cite{Bourget:2015cza}. One such mechanism is the presence of a non-trivial component group in the
unbroken gauge group. The Wilson lines can then take values in the component group, thus enhancing
the number of semi-classical vacua. A second mechanism is the breaking of gauge groups with abelian factors through a Wilson line expectation value \cite{Bourget:2015cza}.
Thus, a classification of nilpotent orbits along with the
conjugacy classes of their component groups becomes pertinent. A crucial step in that classification is the listing of pseudo-Levi subalgebras \cite{BC1,BC2,Sommers}. We show that the latter also play a leading role in
listing the semi-classical limits of elliptic integrable systems that generalize the Inozemtsev limit of \cite{I}.

We have structured our presentation as follows.
Our paper contains advanced nilpotent orbit theory, complexified integrable system analysis, as well as
intricate aspects of ${\cal N}=1$ gauge theories in four dimensions upon circle compactification.
We have therefore decided to first illustrate many features of the generic analysis in the 
example of ${\cal N}=1^\ast$ theory with gauge algebra $so(5)$, where a lot of details can be worked through by hand.
We include a description of the consequences of the choice of global gauge group and the spectrum of line
operators, which neatly complements the analysis of \cite{Aharony:2013hda,Aharony:2013kma} in an example that
is intermediate between ${\cal N}=4$ and pure ${\cal N}=1$ supersymmetric gauge theory in four dimensions.
Section \ref{so5} serves to study a tree before exploring the forest. The finer
features of the $so(5)$ example will motivate the later sections.

In section \ref{semiclassical} we make a link between the 
classification of nilpotent orbits and the conjugacy classes of the component group of their centralizer on the one hand, and limits of elliptic integrable systems on the other. We illustrate features of this
analysis in section \ref{sun}
in the example of the gauge algebra $su(N)$, which will allow to demonstrate the  existence of branches 
of massless vacua, as well as the use of generalized Inozemtsev
limits. We will obtain an explicit analytic description of the  massless vacua 
of the ${\cal N}=1^\ast$ theory with $su(3)$ gauge algebra, and their duality properties. The description of the massless
branch
in terms of the elliptic system variables invokes intricate aspects of the theory of elliptic functions. 
The branch of massless vacua has a (Argyres-Douglas) singularity. 
The singularity also will show up as a point of monodromy for the 
position of the massless vacua described in terms of the complexified Wilson lines on the torus.
The singularity lies at the elliptic point of order three on the boundary of the fundamental domain of the modular group.
At the hand of the gauge algebra $su(4)$, we illustrate further aspects
that pop up at higher rank.

We also discuss the ${\cal N}=1^\ast$ theory with gauge group of exceptional type
$G_2$. 
A first reason to study this case is that $G_2$ is a gauge group of limited rank, allowing for an elaborate numerical analysis of the duality properties of the massive vacua. A second reason is that the group $G_2$  exhibits an orbit with an unbroken discrete gauge group. This will allow us to cleanly illustrate the role played by the discrete group in the identification of the extrema of the integrable system with massive gauge theory vacua on $\mathbb{R}^{2,1} \times S^1$. This aspect puts into focus the difference between the gauge theory on $\mathbb{R}^4$ and the gauge theory compactified on a circle. 

In section \ref{discrete}, we thus provide a large amount of detail of the semi-classical analysis of the vacua of ${\cal N}=1^\ast$ theory on $\mathbb{R}^{2,1} \times S^1$ with gauge group $G_2$, including a nilpotent orbit classification with their pertinent properties,
and the low-energy quantum dynamics in the corresponding phases. Moreover, we perform an 
in-depth analysis of the associated twisted elliptic Calogero-Moser integrable system, and we make a comparison with the semi-classically predicted vacua. We also provide the duality
diagram of the massive vacua and a first estimate of a point of monodromy.
In section \ref{so5massless}, we tie up a loose end, and analytically describe the branch of massless vacua for the $so(5)$ theory.
We conclude in section \ref{conclusions} with a summary, and a partial list of open problems on the intersection of 
supersymmetric gauge theory, nilpotent orbit theory, integrability, modularity and the theory of elliptic functions.

 \section{The ${\cal N}=1^\ast$ Theory with Gauge Algebra $so(5)$}
\label{so5}

To illustrate  finer points that crop up when analyzing ${\cal N}=1^\ast$ gauge theories
with generic gauge group upon circle compactification, we concentrate in this section
on the study of ${\cal N}=1^\ast$ theory with gauge algebra $so(5)$, and the associated
 twisted elliptic
integrable system with root system $B_2$ \cite{Kumar:2001iu}.
Our analysis in this and the following sections is a continuation of the work presented in \cite{Bourget:2015cza,Bourget:2015lua}. In particular, we refer to \cite{Bourget:2015cza} for the detailed discussion
of the correspondence between the gauge theory and the numerical results on the elliptic integrable system, and
we relegate to \cite{Bourget:2015lua} the full explanation of the relevance of nilpotent orbit theory to the 
semi-classical gauge theory on $\mathbb{R}^4$. We moreover refer to \cite{C,CM,LT,LS} for pedagogical introductions to
nilpotent orbit theory. We creatively combine these sources in the following.

\subsection{The Semi-Classical Analysis and Nilpotent Orbit Theory}

The ${\cal N}=4$ supersymmetric Yang-Mills theory on $\mathbb{R}^4$ has fields in one vector and three chiral multiplet representations of the ${\cal N}=1$ supersymmetry algebra. All fields transform in the adjoint representation of the gauge algebra. 
After triple mass deformation to ${\cal N}=1^\ast$ gauge theory, 
the F-term equations of motion (divided by the complexified gauge group) for the
three adjoint chiral scalars
have solutions classified by embeddings of $sl(2)$ commutation relations inside the adjoint
of the gauge algebra. By the Jacobson-Morozov theorem, these $sl(2)$ triples are in one-to-one
correspondence with nilpotent orbits, which have been classified for simple algebraic groups \cite{C,CM,LT,LS}.

Nilpotent orbits of the classical groups can be enumerated by partitions that correspond to the
dimensions of the $sl(2)$ representations that arise upon embedding in the gauge algebra.\footnote{For the case
of gauge algebra $so(2N)$ and the adjoint gauge group $SO(2N)$, the very even partitions (having only even parts with even multiplicity) give rise
to two distinct nilpotent orbits. For this gauge algebra, each orbit gives rise to its own vacua. When the outer
automorphism of $so(2N)$ is joined to the adjoint gauge group, we obtain the gauge group $O(2N)$ in which these
orbits and the corresponding vacua are identified. See \cite{Bourget:2015lua} for a detailed discussion.}
The Lie algebra of the centralizer has been computed, and non-abelian centralizers  give rise to
effective pure ${\cal N}=1$ gauge theories that have a number of quantum vacua equal to the dual Coxeter
number of the unbroken gauge group. The partition, the unbroken gauge algebra, and the number of massive quantum
vacua they give rise to on $\mathbb{R}^4$ for the gauge algebra $so(5)$ are enumerated in the first three
columns in table \ref{so5orbits}. 
For instance, the $2+2+1$ partition of $5$ corresponds to a configuration
for the adjoint scalar expectation values that represent a particular orbit (via the correspondence between
$sl(2)$ embeddings and nilpotent orbits), and these vacuum expectation values leave a $C_1=A_1$ gauge algebra unbroken.
The resulting pure ${\cal N}=1$ gauge theory at low energy gives rise to two massive vacua. See
\cite{Donagi:1995cf,Naculich:2001us,
Wyllard:2007qw,Bourget:2015cza,Bourget:2015lua}.
\begin{table}[H]
\centering
\begin{tabular}{|c|c|c|c|c|c|c|c|c|}
\hline
Orbit Partition & Unbroken  &  Massive Vacua on $\mathbb{R}^4$ & W-class  & Levi 
\\
\hline
$1+1+1+1+1$ & $B_2$ &  3 & $\emptyset$  &  0 
\\
$2+2+1$ & $C_1$ &  2 & $\{ \alpha_1 \}$ &   $C_1$ 
\\
 $3+1+1$ &  $u(1)$ & 0 & $\{ \alpha_2 \}$ &  $\tilde{A}_1$ 
\\
$5$ & 1 & 1 & $\{ \alpha_1,\alpha_2 \}$ & $B_2$ 
\\
\hline
\end{tabular}
\caption{Nilpotent orbit data for $so(5)$. }
\label{so5orbits}
\end{table}
\noindent
The last two columns in table \ref{so5orbits} are related to the Bala-Carter theory of nilpotent orbits 
\cite{BC1,BC2} that associates a Weyl group equivalence class of subsets of the set of simple roots to each Levi subalgebra
of the gauge algebra. The reader may revert to studying these columns after reading section \ref{semiclassical}. See also section \ref{g2section} for  Bala-Carter theory with an example worked out in detail.
When we compactify the gauge theory on $\mathbb{R}^{2,1} \times S^1$, properties of the centralizer
beyond its Lie type become crucial.  A refined classification of the nilpotent orbits, including
the conjugacy classes of the component group\footnote{The component group is the quotient of the group by its
identity component.} of the unbroken gauge group (by Bala, Carter and Sommers
\cite{BC1,BC2,Sommers})
gives rise to table \ref{so5orbitsrefined}.

\begin{table}[H]
\centering
\begin{tabular}{|c|c|c|c|c|c|c|}
\hline
  Orbit & Centr. &   C. C. & Massive Vac. & W-classes & PLS    \\
\hline
 $1+1+1+1+1$ & $B_2$ & 1 &  3 
 & $\emptyset$   &  $0$   \\
  $2+2+1$ & $A_1$ &   1 & 2 
  & $\{ \alpha_0 \}$ , $\{ \alpha_1 \}$  & $C_1$  \\
   $3+1+1$ &  $u(1)$ & 1 & 0  & $\{ \alpha_2 \}$  &  $\tilde{A}_1$  \\
   $3+1+1$ & $u(1)$ &  (12) &  1 
   & $\{ \alpha_0, \alpha_1 \}$  &  $D_2 $ \\ 
   $5$ & 0 &     1  & 1 
   & $\{ \alpha_0, \alpha_2 \}$ , $\{ \alpha_1, \alpha_2 \}$  & $B_2$  \\
\hline
\end{tabular}
\caption{The Bala-Carter-Sommers classification of nilpotent orbits with their centralizers,
including the conjugacy classes (C.C.) of the component group of the centralizer.
The first column gives the partition labeling the orbit, the second the Lie type of the centralizer (i.e. the unbroken gauge algebra
for given adjoint vacuum expectation values), the third the conjugacy class of the discrete part of the centralizer corresponding to the chosen pseudo-Levi subalgebra (PLS) in the last column, the fourth the number of massive vacua and the previous to last the Weyl conjugacy classes of subsystems
of simple roots of the affine root system. In each case, there is only one distinguished parabolic subalgebra, which is the principal one. This analysis is valid for the adjoint group and will be further refined when we take into account the choice of global properties of  the gauge group (see table \ref{centralizerswithcentre}).}
\label{so5orbitsrefined}
\end{table}

At this stage, we wish to take away the elementary fact that the partition $3+1+1$ appears twice in the
first column of table \ref{so5orbitsrefined}, because there is a discrete $\mathbb{Z}_2$ component subgroup 
of the centralizer. The $\mathbb{Z}_2$ component group has two conjugacy
classes, namely the trivial one, and the non-trivial one (labeled by the cyclic permutation
$(12)$). The importance of the second occurrence is the fact that we can turn on a Wilson line 
on the circle equal to this conjugacy class while still satisfying the equations of motion (as discussed in detail
in section \ref{discrete}). Because 
the $\mathbb{Z}_2$ forms a semi-direct product with the $SO(2)$ unbroken gauge group
for the $3+1+1$ partition, turning on the Wilson line breaks the abelian gauge group,
and generates a new massive vacuum on $\mathbb{R}^{2,1} \times S^1$  \cite{Bourget:2015cza}.
Finally, we note that we also have a massless branch of rank one.

\subsection{The Elliptic Integrable System}
We turn to how the physics of the ${\cal N}=1^\ast$ theory with gauge algebra $so(5)$ is coded
in the twisted elliptic integrable system of type $B_2$ that was proposed  to be the low-energy effective superpotential for the model \cite{Kumar:2001iu}. In as far as this constitutes a review of the results presented in \cite{Bourget:2015cza},
we will again be concise, while new features will be emphasized. 

 \begin{figure}[H]
 \centering
 \includegraphics[width=0.25\textwidth]{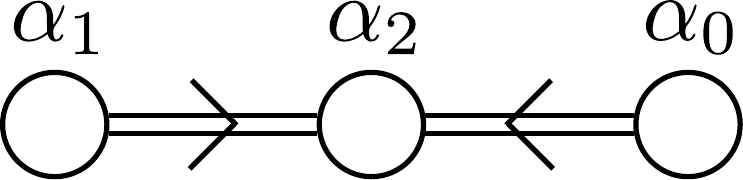}
 \caption{The Dynkin diagram of the affine algebra $\widehat{so(5)}=B_2^{(1)}$ with our convention for the numbering of long and short roots. }
\label{DynkinB2}
 \end{figure}
\begin{figure}
\begin{minipage}{\linewidth}
\begin{tabular}{p{4.6cm}p{4.6cm}p{4.6cm}}

{%
\setlength{\fboxsep}{8pt}%
\setlength{\fboxrule}{0pt}%
\fbox{\includegraphics[width=3.5cm]{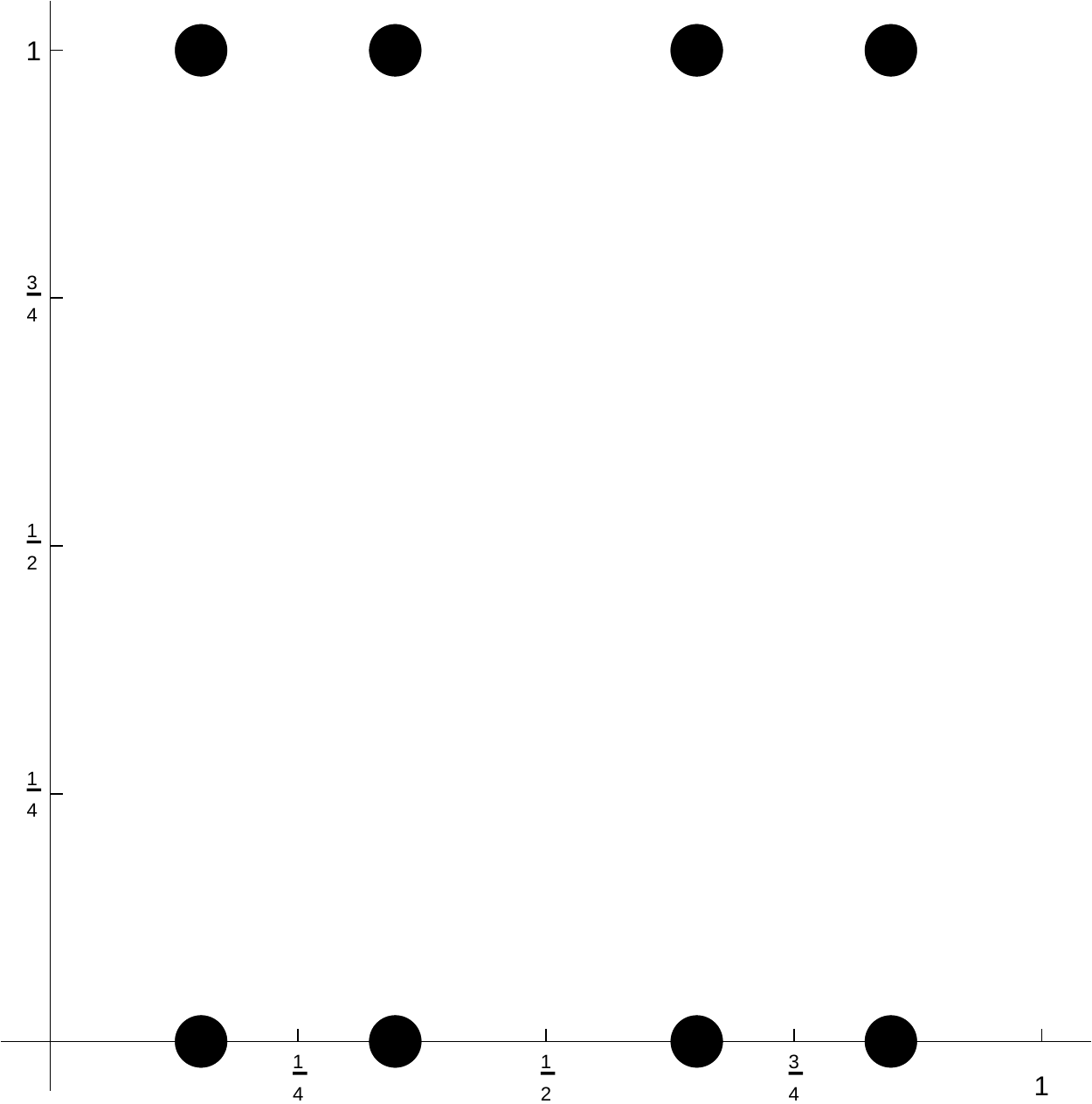}}%
}%

\begin{center}
{Extremum 1}
\end{center}

&
{%
\setlength{\fboxsep}{8pt}%
\setlength{\fboxrule}{0pt}%
\fbox{\includegraphics[width=3.5cm]{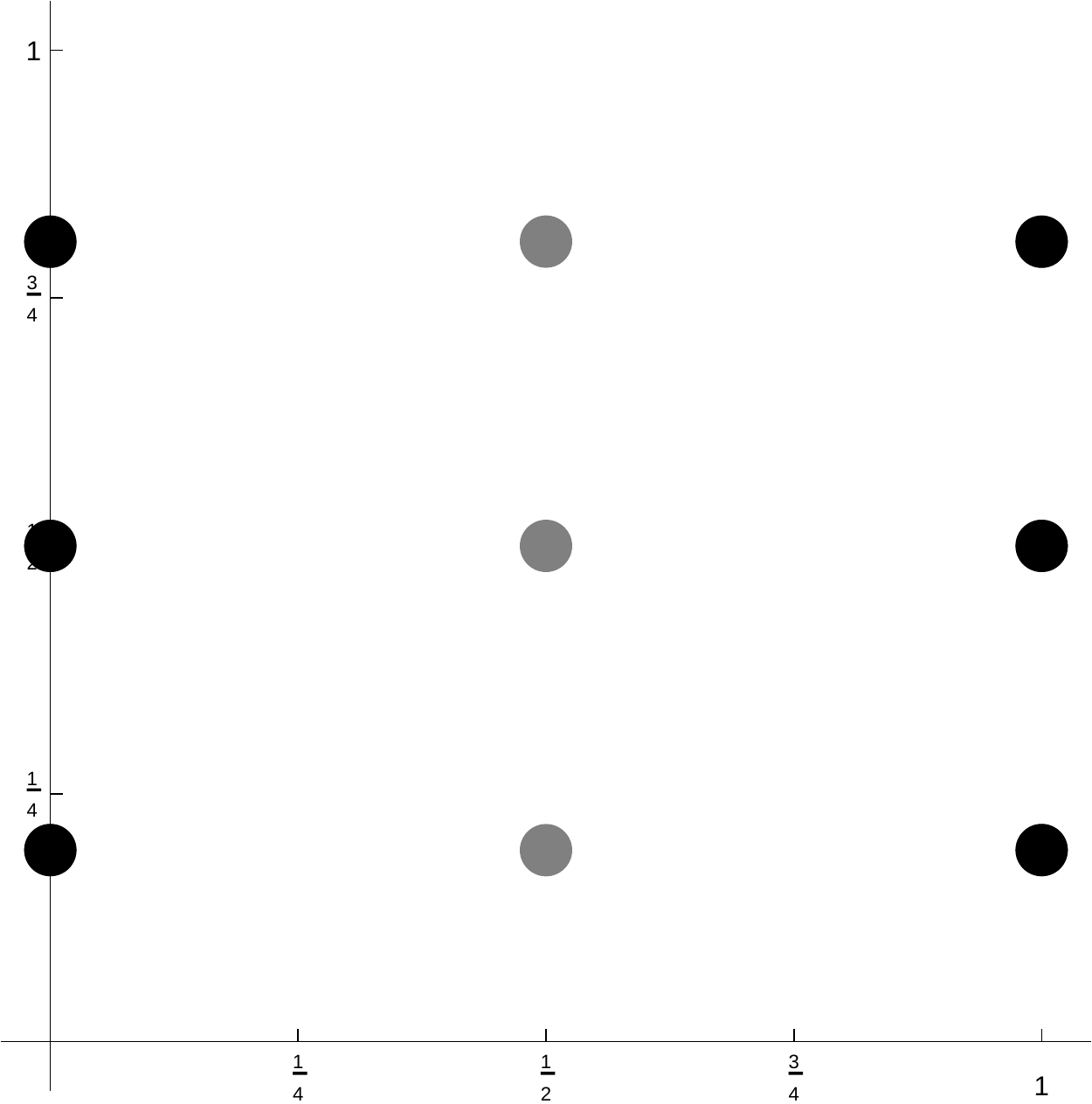}}%
}%

\begin{center}
Extremum 2
\end{center}

 &
 {%
\setlength{\fboxsep}{8pt}%
\setlength{\fboxrule}{0pt}%
\fbox{\includegraphics[width=3.5cm]{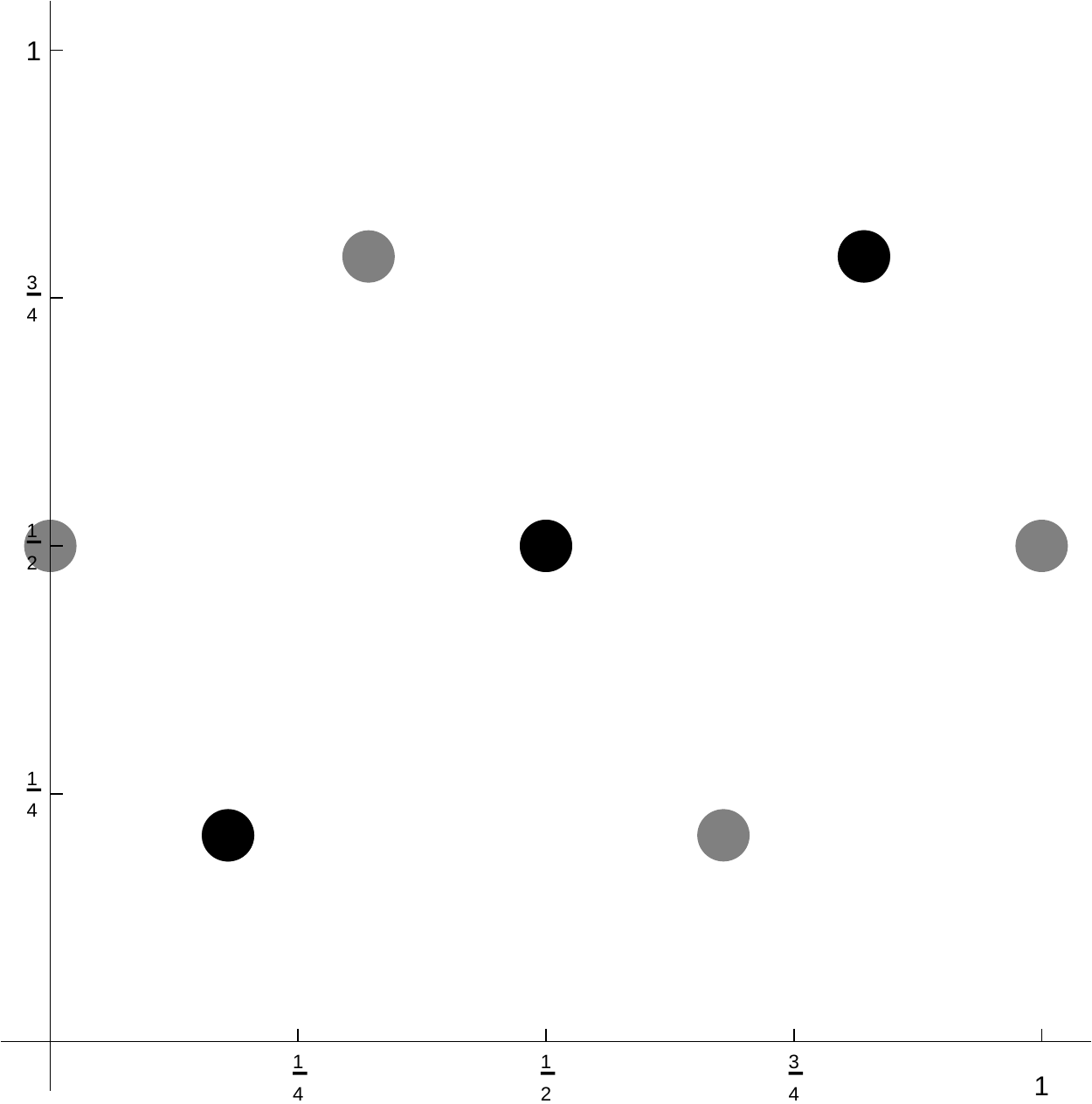}}%
}%

\begin{center}
Extremum 3
\end{center}

\\
{%
\setlength{\fboxsep}{8pt}%
\setlength{\fboxrule}{0pt}%
\fbox{\includegraphics[width=3.5cm]{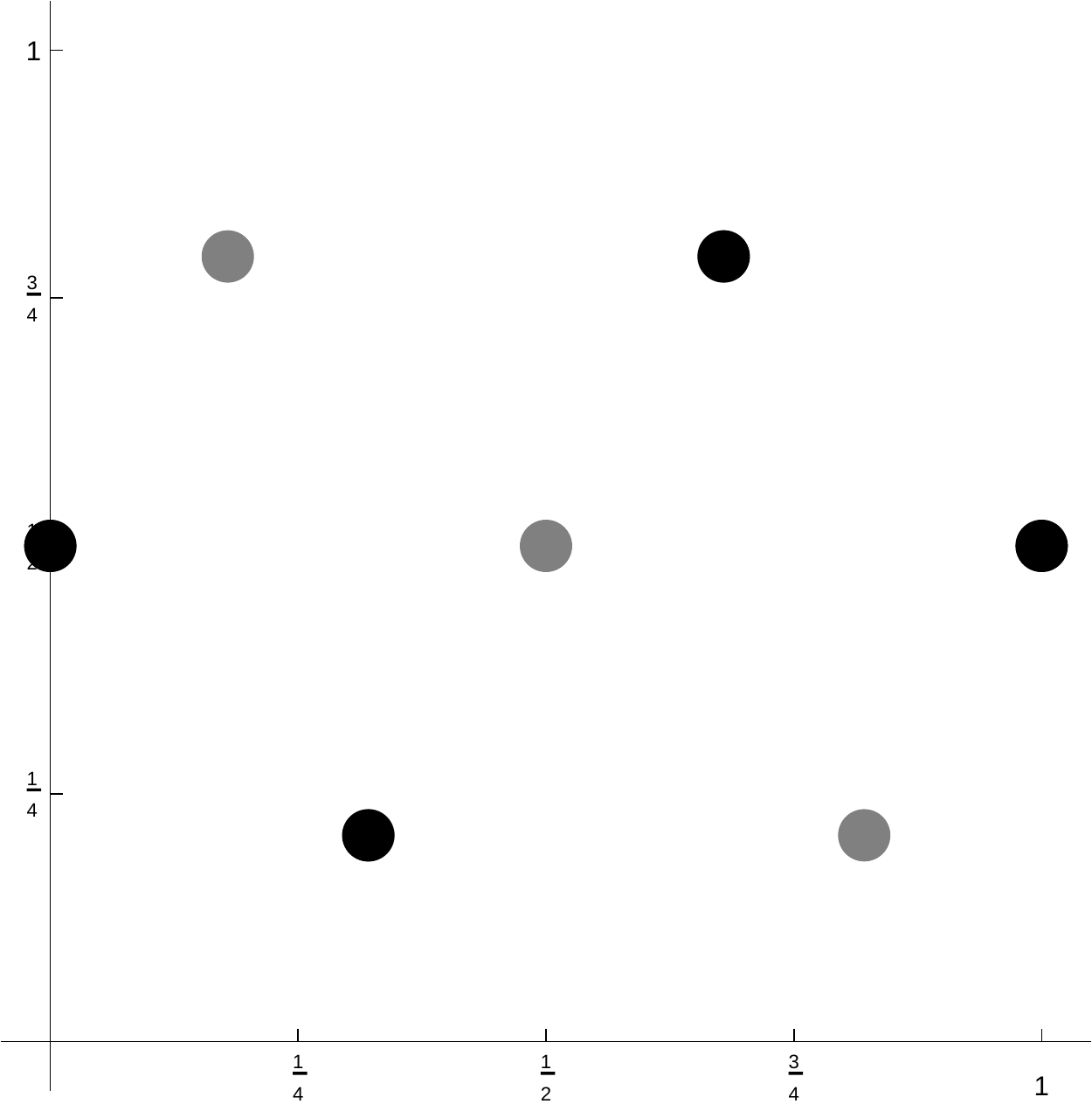}}%
}%

\begin{center}
Extremum 4
\end{center}

 & 
 {%
\setlength{\fboxsep}{8pt}%
\setlength{\fboxrule}{0pt}%
\fbox{\includegraphics[width=3.5cm]{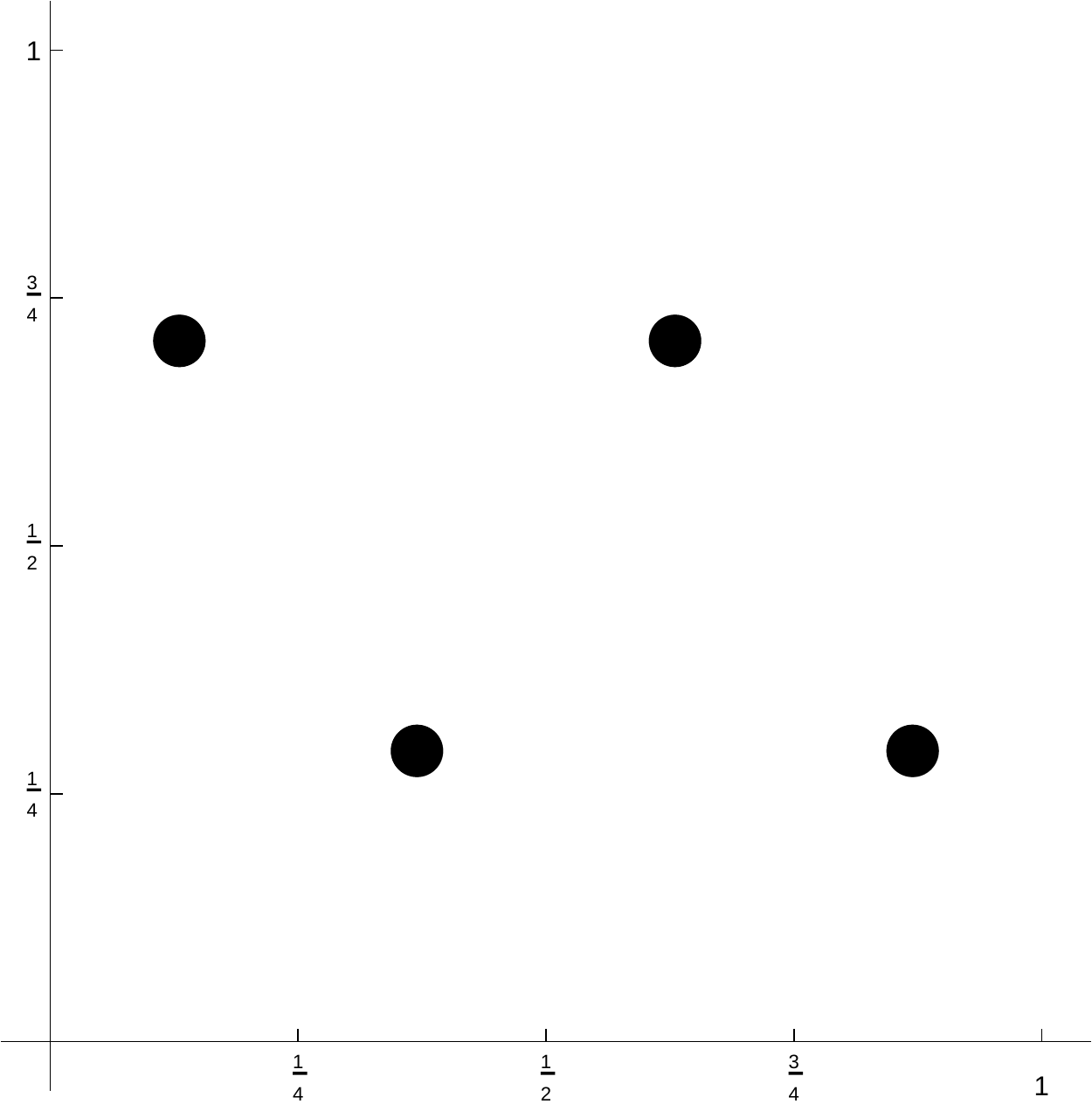}}%
}%

\begin{center}
Extremum 5
\end{center}
 & 
 {%
\setlength{\fboxsep}{8pt}%
\setlength{\fboxrule}{0pt}%
\fbox{\includegraphics[width=3.5cm]{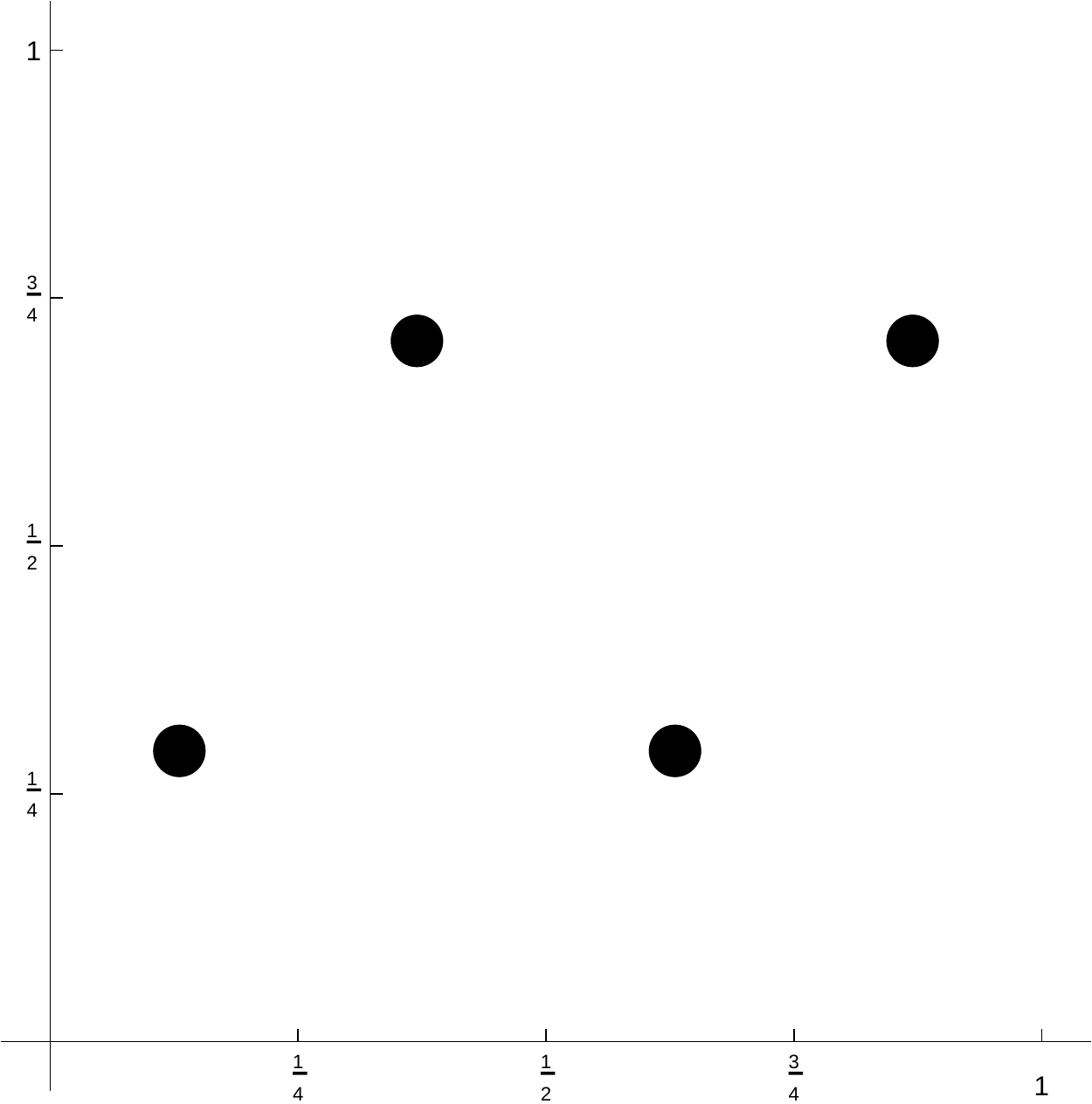}}%
}%

\begin{center}
Extremum 6
\end{center}

\\
{%
\setlength{\fboxsep}{8pt}%
\setlength{\fboxrule}{0pt}%
\fbox{\includegraphics[width=3.5cm]{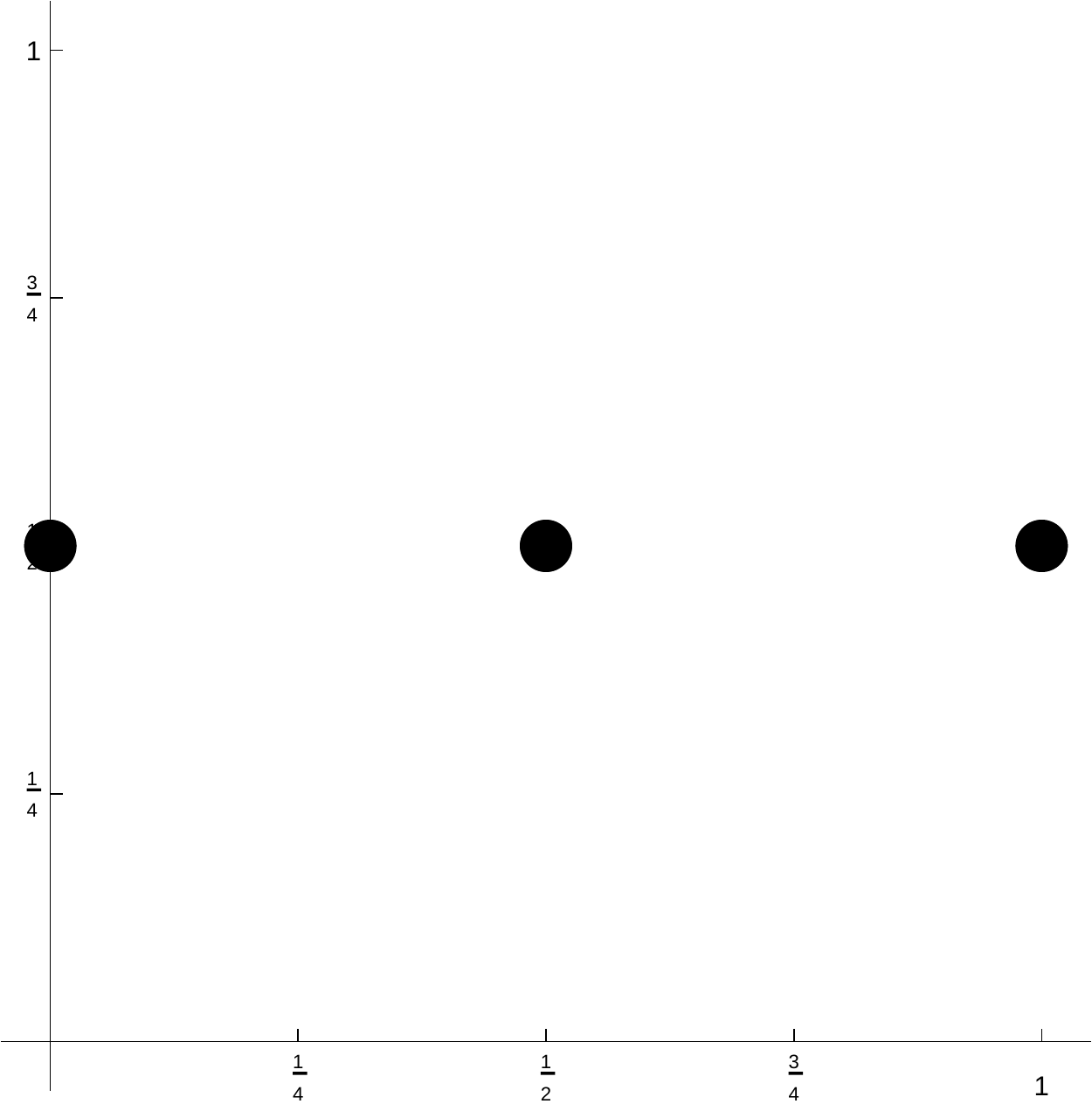}}%
}%

\begin{center}
Extremum 7
\end{center}

& & 
\end{tabular}

\end{minipage}
\caption{Extrema for the superpotential at coupling $\tau=i$ for the Lie algebra $so(5)$ are drawn in dark. 
Configurations obtained by translation by $\omega_1$ are drawn in light gray.}
\label{extremaso5}
\end{figure}
\begin{figure}
\centering
\includegraphics[width=0.5\textwidth]{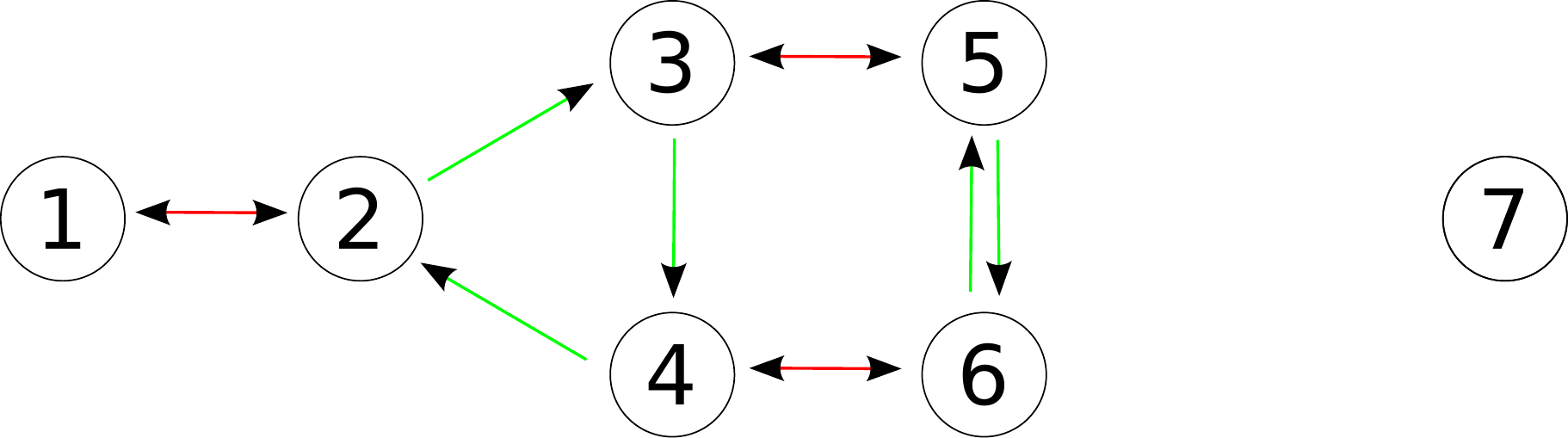}
\caption{The diagram of  dualities on the extrema of the integrable system for $so(5)$. In red, we draw the action of Langlands $S_2$-duality, and in green, $T$-duality, when the action on a given extremum is non-trivial.  }
\label{dualities_so5}
\end{figure}

The Dynkin diagram for the affine algebra $B_2^{(1)}$ (as well as its finite counterpart, upon deleting the zeroth node)
can be read off from figure
\ref{DynkinB2}.
The long simple root $\alpha_1$ of $B_2$ can be  parametrized as $\alpha_1 = \epsilon_1 - \epsilon_2$ 
and the short root $\alpha_2$ as $\alpha_2 = \epsilon_2$,
where the $\epsilon_i$ are orthonormal basis vectors in a two-dimensional Euclidean vector space.\footnote{
Let us recall a few Lie algebra data for future reference. The root lattice is generated
by $\epsilon_{1,2}$. 
The fundamental weights are $\pi_1=\epsilon_1$   and $\pi_2=(\epsilon_1+\epsilon_2)/2$.
The dual simple roots are $\alpha_1^\vee = \alpha_1$ and $\alpha_2^\vee = 2 \alpha_2= 2 \epsilon_2$.
The dual weight lattice is spanned by the $\epsilon_i$. The Weyl group allows for permutations of the $\epsilon_i$, and all sign changes.
We follow the conventions of \cite{OV}.
}
The superpotential of the twisted elliptic Calogero-Moser model with root system $B_2$ is  \cite{D'Hoker:1998yg}
\begin{eqnarray}
\label{B2superpotential}
\mathcal{W}_{B_2,tw} (Z) &=& \wp \left(z_1 + z_2 \right) + \wp \left(z_1 - z_2 \right) + \frac{1}{2} \left[ \wp_2 \left(z_1\right) + \wp_2 \left(z_2\right) \right]  \\
&=& \wp \left(Z_1 \right) + \wp \left(Z_1+Z_2\right) + \frac{1}{2} \left[ \wp_2 \left(\frac{Z_2}{2}\right) + \wp_2 \left(Z_1 + \frac{Z_2}{2}\right) \right] \, , 
\end{eqnarray}
where we combine the Wilson line $a$ and dual photon $\sigma$ of the low-energy theory in the Coulomb phase in a complex field $Z=\sigma + \tau a$
parametrized by 
\begin{equation}
Z= Z_1 \pi_1+Z_2 \pi_2 = z_1 \epsilon_1 + z_2 \epsilon_2 \, . 
\end{equation}
Throughout the paper we use capital letters to denote the components of an element of the dual Cartan space decomposed on the basis of fundamental weights, and small letters to denote its components in the basis $\epsilon_i$.
For $so(5)$, the relation is
\begin{equation}
\begin{cases}
Z_1 = z_1 - z_2 \\
Z_2 = 2z_2
\end{cases} 
\qquad \mbox{or} \qquad
\begin{cases}
z_1 = Z_1 + \frac{1}{2}Z_2 \\
z_2 = \frac{1}{2}Z_2 
\end{cases} \, .
\end{equation}
The superpotential ${\cal W}$ depends on the elliptic Weierstrass function $\wp$ with half-periods $\omega_1=1/2$ and $\omega_2=\frac{\tau}{2}$ (where the complexified gauge coupling is $\tau=\omega_2/\omega_1$) and its twisted cousin $\wp_2$ which is defined to have half the period in the $\omega_1$ direction, $\wp_2(z,\tau) = \wp (z,\tau) + \wp(z+\frac{1}{2} , \tau)$. The ratio of the coupling constants for short and long roots was fixed in \cite{Kumar:2001iu} and checked using Langlands duality in \cite{Bourget:2015cza}. In \cite{Bourget:2015cza}, we established the existence of seven massive vacua (up to a given equivalence relation to be discussed shortly), determined their positions numerically, and provided analytic expressions for the value of the superpotential in each of these massive vacua. The extremal positions at $\tau=i$ are rendered in figure \ref{extremaso5}. We moreover established the duality diagram in figure \ref{dualities_so5} between the seven massive vacua.

In the present section, we wish to add to the analysis presented in \cite{Bourget:2015cza} in several ways. We analyze the semi-classical
limits of the effective low-energy superpotential. We propose a list of such limits, and show that we
obtain an analytic handle on each of the seven vacua, and on the massless vacua as well. Moreover,
we will  carefully exhibit the differences between various global choices of gauge group and spectra
of line operators, and consequently a more refined duality diagram.

Importantly, our list of limits is based on table \ref{so5orbitsrefined}. Each nilpotent orbit and conjugacy class of the component group is associated, by Bala-Carter-Sommers theory to a choice of inequivalent\footnote{The equivalence relation is given precisely in \cite{Sommers}, and can be technical in some cases. For the gauge theories we are concerned with, it can be stated as follows. In algebras of type $A$ and $G$, two subsystems are equivalent if they have the same Lie algebra type and the same repartition of long and short roots. For type $B$, one should moreover distinguish between $A_1+A_1$ and $D_2$, and between $A_3$ and $D_3$, using the index of the subsystem. } subsystem of simple roots of the affine root system. To each such subsystem, we associate a limit of the integrable system as follows. We demand that for simple roots $\alpha_i$ in the subsystem, we have that the simple root is orthogonal to the vector of extremal positions $Z$, namely $(\alpha_i^\vee , Z ) =0$, to leading (linear) order in the complexified gauge coupling $\tau$ in the large imaginary $\tau$ limit. The simple root systems in the complement must have non-zero leading term.

We denote the part in $Z$ that is linear in $\tau$ by $Y \tau$. 
We moreover introduce the redundant coordinate $Y_0$ which we constrain by the equation $Y_0+Y_1+Y_2=1$. It is a coordinate that is natural in treating this problem governed by affine algebra symmetry as will become more manifest in section \ref{semiclassical}.
The list of semi-classical limits that we will consider are then labelled by the set $J_0$ which contains all $i$ for which $\alpha_i$ is in the chosen simple root subsystem indicated in table \ref{so5orbitsrefined}.
 We therefore distinguish five limits, which we treat one-by-one below. 
 \begin{itemize}
 \item 
The first limit corresponds to the empty set, $J_0 = \emptyset$. The arguments of the Weierstrass
functions will all contain  a linear term in $\tau$. Therefore all terms are well-approximated by
exponentials (see section \ref{semiclassical}, formulas (\ref{largetauP}) and (\ref{largetautwistedP})). 
The limiting procedure in this case is described in detail in \cite{D'Hoker:1998yh},
which in turn is a generalization of the Inozemtsev limit \cite{I}.
By demanding that all these exponentials have the same dependence
on the instanton counting parameter $q=e^{2 \pi i \tau}$, which is necessary in order to stabilize all variables, we determine that
the linear behavior of the coordinates in $\tau$ is $Y_0=Y_1=Y_2=1/3$. We obtain a (fractional instanton) 
$B_2$ affine Toda system in the limit, with 3 extrema. 
The solutions $(z_1,z_2)$ in the semi-classical limit are then $\left(\frac{\tau}{2} ,\frac{\tau}{6} \right)$, $\left( \frac{\tau}{2} , \frac{1}{3} + \frac{\tau}{6} \right)$ and $\left( \frac{\tau}{2} , \frac{2}{3} + \frac{\tau}{6} \right)$. One can check these solutions against the behavior of the numerical extrema labelled $2,3$ and $4$ in \cite{Bourget:2015cza}
(and figures \ref{extremaso5} and \ref{dualities_so5}), and they match in the semi-classical limit.
This codes the physics of the pure ${\cal N}=1$ gauge theory with gauge algebra $so(5)$.\footnote{We discuss the global choice of gauge group and line operators in
subsection \ref{globalso5}.} Indeed,
the partition $1+1+1+1+1$  leaves the whole of the gauge group unbroken.
\item
The second case is the choice of subroot system $J_0 = \{0\}$. Note that this is completely equivalent to the choice $J_0 = \{1\}$, since the corresponding marked Dynkin diagrams are the same in both cases, so we concentrate on the first of these sets. Then we have $Y_0=0$ as a consequence, and to match powers of $q$ in subleading terms, we choose $Y_1=Y_2=1/2$. In the semi-classical limit, we then obtain a trigonometric $A_1$ system at leading order (associated to the long root $\alpha_0$). At subleading order, we find a superpotential ${\cal W}(z_1,z_2)$ consisting of a sum of exponentials 
\begin{equation}
{\cal W}_{B_2,tw} \left(\frac{3 \tau}{4} + \frac{1}{4} + \frac{\delta x}{2}, \frac{\tau}{4}+\frac{1}{4}-
\frac{\delta x}{2} \right) =
\pi^2 ( 12 e^{-2 i \pi  \delta x}-4 e^{2 i \pi  \delta x}  )  q^{\frac{1}{2}}  +\dots \, .
\end{equation}
The two extrema at large $\tau$ are $\left( \frac{1}{8} + \frac{3 \tau}{4} , \frac{3}{8} +  \frac{\tau}{4} \right)$ and $\left( - \frac{1}{8} + \frac{3 \tau}{4} , \frac{5}{8} +  \frac{\tau}{4} \right)$.\footnote{The subleading behavior of the extrema in the large $\tau$ limit can easily be computed analytically as well. See later for more intricate explicit examples.} These match the behavior of the massive vacua number $5$ and $6$ in figures \ref{extremaso5} and \ref{dualities_so5} at large $\tau$. These are the two confining vacua of the unbroken pure ${\cal N}=1$ $su(2)$ gauge theory. Note how this limit is intermediate in that one coordinate is fixed at leading order in the $q^{\frac{1}{2}}$ expansion, while a second is fixed at subleading order.
\item Thirdly, we have the case $J_0 = \{1,2\}$, which by the same token is equivalent to $J_0 = \{0,2\}$. 
We find the  trigonometric potential $B_2$ with a real extremum, which can be characterized in terms of zeroes of orthogonal polynomials \cite{Corrigan:2002th,Odake:2002xm}. This corresponds to the fully Higgsed vacuum, with label 1 in figure \ref{extremaso5}. Importantly, there are other, complex extrema of the trigonometric integrable system.\footnote{Complex extrema of integrable systems are rarely discussed. The observation we make here on the trigonometric $B_2$ integrable system, for instance, appears to be new.} In the limit $\tau \rightarrow i \infty$, one example extremum is given by $(z_1,z_2) \sim (\frac{1}{4} + \frac{ \log(1+\sqrt{2})}{2 \pi}  i,\frac{1}{4} - \frac{ \log(1+\sqrt{2})}{2 \pi}  i)$. This is a massless extremum, 
part of a branch that we analyze in section \ref{so5massless}.
\item For our fourth pick, we take $J_0 = \{0,1\}$ and obtain two trigonometric potentials, corresponding to the root system $D_2$. We find the one extremum 
$\left( \frac{\tau}{2} , \frac{1}{2} + \frac{\tau}{2} \right)$. This corresponds to extremum number 7. This is a massless vacuum lifted by the presence of a $\mathbb{Z}_2$ Wilson line. It is thus semi-classically massive on $\mathbb{R}^{2,1} \times S^1$. The $\mathbb{Z}_2$ Wilson line sits inside the non-trivial conjugacy class $(12)$ of the component group $\mathbb{Z}_2$ of the unbroken gauge group. This is an occurrence of a general phenomenon that we analyze further in section \ref{discrete}. 
\item
Finally, we turn to the fifth possibility, $J_0 = \{2\}$. 
The leading $\tau$ behavior of the second coordinate is $Y_2=0$. As a first stab at the semi-classics in this regime, we choose the values $Y_0=Y_1=1/2$, which is a natural ansatz given the symmetry of the Dynkin diagram about $\alpha_2$. In any event, we obtain the trigonometric $\tilde{A}_1$ system (where the tilde stands for a short root) at leading order. The extremization of the superpotential at order 0 gives $Z_2 = \frac{1}{2}$. The value of $Z_2$ gets corrected non-perturbatively, namely at order $q^{\frac{1}{2}}$, order $q^{\frac{3}{2}}$
and higher strictly half-integer orders, by terms depending on $\delta Z_1$ exponentially. For the particular
value of $Y_1$ that we chose we find
\begin{eqnarray}
\delta Z_2 &=& \frac{1}{\pi} \left( \vphantom{\frac{1}{3}} -e^{-2 i \pi  \delta Z_1} q^{\frac{1}{2}}-e^{2 i \pi  \delta Z_1} q^{\frac{1}{2}} \right.  \nonumber \\
&& \left.  +\frac{1}{3} e^{-6 i \pi  \delta Z_1} q^{\frac{3}{2}}+5 e^{-2 i \pi  \delta Z_1} q^{\frac{3}{2}}+5 e^{2 i \pi  \delta Z_1} q^{\frac{3}{2}}+\frac{1}{3} e^{6 i \pi  \delta Z_1} q^{\frac{3}{2}} + \dots \right) \, .
\label{nonpertcor}
\end{eqnarray}
Injecting this value for $Z_2$ in the superpotential ${\cal W}(Z_1,Z_2)$ finally gives 
\begin{eqnarray}
\label{expansionMassless}
{\cal W}_{B_2,tw}
\left(\frac{\tau}{2} +\delta Z_1,\frac{1}{2}+\delta Z_2 (\delta Z_1) \right) &=& \pi^2 \left(\frac{2}{3}+16 q+16 q^2
+64 q^3+16 q^4 +  \dots \right) \nonumber \\
&=&- \frac{2 \pi^2}{3} E_{2,2}(q) \,  ,
\end{eqnarray}
to order $q^4$.
We observe that  non-perturbatively correcting the leading coordinate $Z_2$ leads to a vanishing potential for $Z_1$,
in perturbation theory in $q$. The value of the coordinate $Z_1$ determines the non-perturbative correction to
the leading coordinate $Z_2$.
For instance, for the special value $\delta Z_1=1/2$, the non-perturbative correction is zero. See equation (\ref{nonpertcor}). Thus,
we find a one-dimensional
complex branch of massless vacua to which we return in section \ref{so5massless}.
The value of the superpotential in these vacua can be determined by a combination
of numerics, and analytical expectations to be ${\cal W}= -\frac{2\pi^2}{3 } E_{2,2}(q)$. 
The Einstein series $E_{2,2}$ is the modular form of weight
2 of $\Gamma_0(2)$ that has a $q$-expansion that starts out with $-1$.\footnote{See \cite{Bourget:2015cza}
for more details on the combination of techniques used to determine these modular forms.}
\end{itemize}
We have made a list of semi-classical limits for the  $so(5)$ integrable system. In particular, we have analytically recuperated all the numerical results of \cite{Bourget:2015cza}, in the large imaginary $\tau$ limit. We have moreover
made inroads into extra vacua, which are massless. Before discussing the particular features of the $so(5)$ analysis
that we will concentrate on in the rest of the paper, we pause to discuss global aspects of the gauge theory at hand.

\subsection{Global Properties of the Gauge Group and Line Operators} 
\label{globalso5}
Up to now, we have implemented a concept of equivalence on the configuration space
in which we identify the variables proportional to $\omega_1$ by shifts in the weight lattice  and the
variables in the $\omega_2$ direction by shifts in the dual weight lattice \cite{Bourget:2015cza}. These are natural
identifications when one is concerned with analyzing the elliptic integrable potential. However,
from the gauge theory perspective, the global and local symmetries are fixed a priori, and  in this subsection
we will carefully
track how they influence both the counting of vacua and their duality relations.

In other words, we give an example of how to generalize the analysis
of the global choice of gauge group and the spectrum of line operators,
performed for pure ${\cal N}=1$ gauge theories and ${\cal N}=4$ theories
in \cite{Aharony:2013hda,Aharony:2013kma} to ${\cal N}=1^\ast$ theories.
Recall that ${\cal N}=4$ gauge theories with $so(5)$ gauge algebra come in three varieties which satisfy Dirac
quantization and maximality of the operator algebra.
We first distinguish between the choice of gauge group $SO(5)$ and $Spin(5)$.\footnote{The nomenclature is
fixed by demanding that a choice of electric gauge group implies that all possible purely electric charges for Wilson line operators corresponding to the electric gauge group must be realized.} The $Spin(5)$ theory is unique.
The $SO(5)$ theories come in two versions, depending on whether they include a 't Hooft operator which transforms
in the fundamental of the dual gauge group, or a Wilson-'t Hooft operator that transforms in the fundamental of
both the electric and the magnetic gauge group. The first can be denoted $SO(5)_+$ theory, and the second  $SO(5)_-$ theory.
The refined duality map of ${\cal N}=4$ theories described
in  \cite{Aharony:2013hda,Aharony:2013kma} states that the
$SO(5)_+$ theory is $S_2$-dual to the
$Spin(5)$ theory.\footnote{We denote by $S_2$ the Langlands duality transformation $\tau \rightarrow - \frac{1}{2 \tau}$.}
The $SO(5)_-$ theory is self-$S_2$-dual.
The goal of this subsection is to carefully examine the global electric and magnetic 
identifications of the extrema of the low-energy effective superpotential to show that the
refined classification of vacua of the ${\cal N}=1^\ast$ theory is consistent with the
duality imparted by the ${\cal N}=4$ theory.

To make contact with our set-up, we first analyze the 
periodicity of the Wilson line, which follows from the global choice of gauge
group and line operators. In the case where we work with the adjoint gauge group $Spin(5)/\mathbb{Z}_2=SO(5)$
and the spectrum of line operators corresponding to the $SO(5)_+$ theory,
we allow gauge parameters that close only up to 
an element in the center of the covering group. The Wilson line periodicity 
is then the dual weight lattice. The dual weight lattice if spanned by the $\epsilon_i$ and therefore
the two variables on the Coulomb branch will each have periodicity $2 \omega_2$.
When the gauge group is the covering group $Spin(5)$,
gauge parameters are strictly periodic, and the
periodicity of the Wilson line is the dual root lattice. In this case, 
Wilson lines are equivalent under shifts by $\epsilon_1-\epsilon_2$ and $2 \epsilon_2$. Thus, both coordinates are periodic with periodicity
$ 4 \omega_2$, and we can further divide by simultaneous shifts by $2 \omega_2$.

For the magnetic line operator spectrum for the $Spin(5)$ and $SO(5)_+$ theories, it suffices to 
Langlands $S_2$-dualize the above reasoning. We thus obtain that for  $SO(5)_+$  we can shift
by $ 2 \omega_1$ separately each coordinate (i.e. by the root lattice), and for $Spin(5)$ we add on top of this
the simultaneous shift by $\omega_1$ (i.e. the weight lattice). The factor of two difference in the lattice spacing is due
to the mechanics of the  Langlands $S_2$ duality. For the
$SO(5)_-$ theory, the story is more subtle. There is a 't Hooft-Wilson line operator in the spectrum
which is in the fundamental of both the dual gauge group and the ordinary gauge group.
We allow for the identifications common to $Spin(5)$ and $SO(5)_+$, and add the identification
that shifts an individual coordinate by $2 \omega_2$ and both coordinates simultaneously
by $\omega_1$. This is the diagonal $\mathbb{Z}_2$ in the magnetic and electric weight lattices
divided by the magnetic and electric root lattices respectively.

\begin{figure}
\centering
\includegraphics[width=0.45\textwidth]{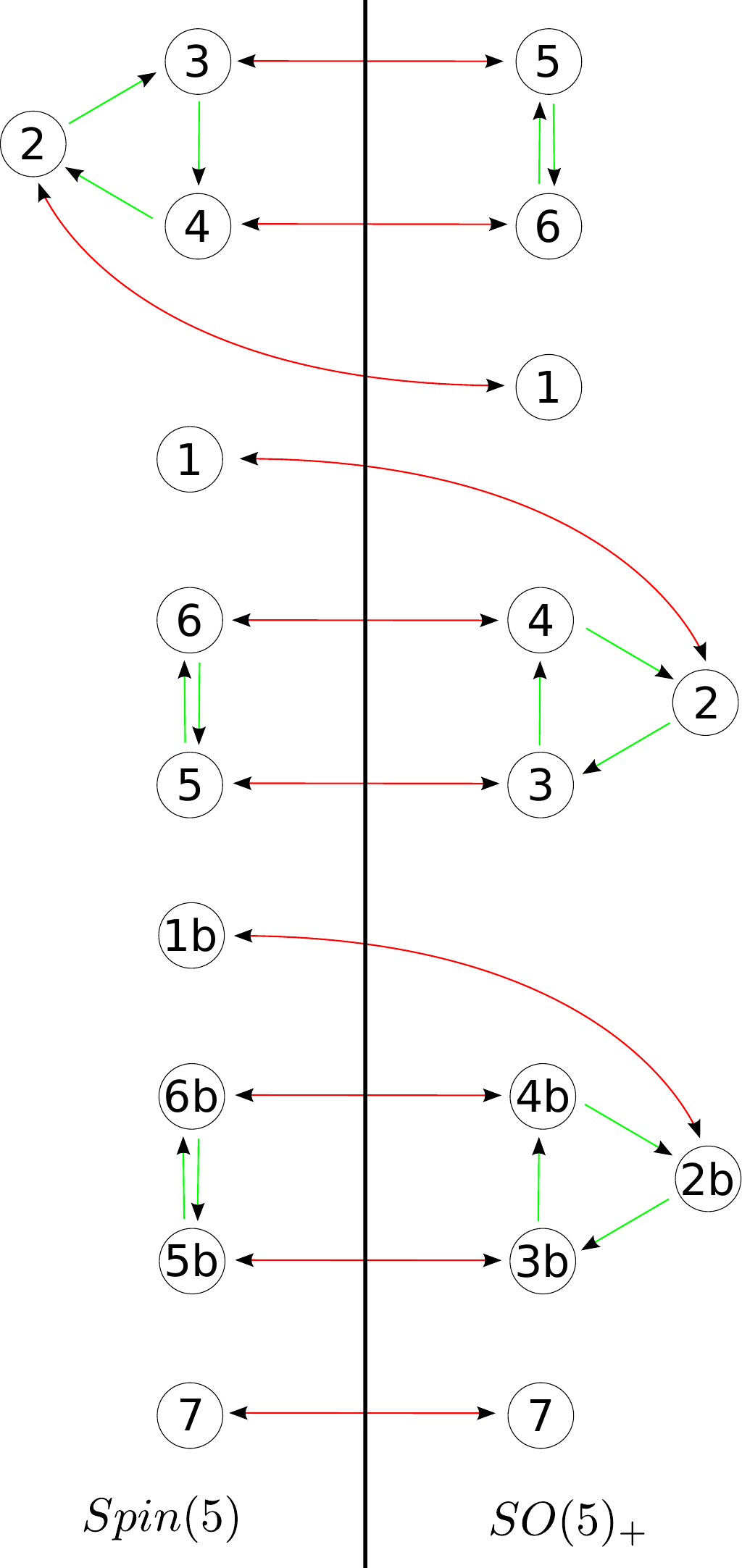}
\caption{The diagram of the action of dualities on the massive vacua for the different $B_2$ theories. In red, we draw the action of Langlands $S_2$-duality, and in green, $T$-duality (when the action is non-trivial). On the left we represent the 10 vacua of the $Spin(5)$ theory, and on the right the 10 vacua of the $SO(5)_+$ theory. The diagram of dualities for the self-dual $SO(5)_-$ theory is identical to figure \ref{dualities_so5}. }
\label{dualities_so5_extended}
\end{figure}

\subsubsection{$SO(5)_+$ vacua}
Given the more limited identifications above, we obtain a longer list of extrema. The list of massive extrema for the $SO(5)_+$ theory is $(1,2,3,4,5,6,7,2b,3b,4b)$ where the extrema $(2b,3b,4b)$ are obtained from $(2,3,4)$ by shifting by $\omega_1$ (see figure \ref{extremaso5plus} in the appendix). The doubling of the number of massive vacua arising from pure ${\cal N}=1$ super Yang-Mills theory with $SO(5)_+$ gauge group is as expected from \cite{Aharony:2013hda,Aharony:2013kma}.
We thus have ten massive vacua.

\subsubsection{$SO(5)_-$ vacua}
In this case, we remain with seven massive vacua. For the vacua $(2,3,4)$, this is as for the pure ${\cal N}=1$ theory. By self-$S_2$-duality, this is expected for the vacua $(1,5,6)$ as well.
\subsubsection{$Spin(5)$ vacua}
For the $Spin(5)$ theory, we again find ten massive vacua. 
The doubling of vacua is $S_2$-dual to the duplication for $SO(5)_+$, and extrema $(1,5,6)$ obtain partner vacua $(1b,5b,6b)$ (see figure \ref{extremaspin5}). The duality diagrams for the massive vacua are drawn in figure \ref{dualities_so5_extended}. 
The analysis in \cite{Aharony:2013hda,Aharony:2013kma} shows that the pure $\mathcal{N}=1$ $Spin(5)$ theory on $\mathbb{R}^{2,1} \times S^1$ has $3$ vacua, which is consistent with the one triplet under $T$-duality that we find on the left
in figure \ref{dualities_so5_extended}. 
To explain the doubling of the singlet and doublet in the $Spin(5)$ theory, we refine our analysis of
the unbroken gauge group further, and adapt it to include the differences between the adjoint
group $SO(5)$ and the covering group $Spin(5)$. The results are in table \ref{centralizerswithcentre}.
\begin{table}[H]
\centering
\begin{tabular}{|c|c|c|c|c|c|c|c|}
\hline 
{Partition} & \multicolumn{2}{c|}{Centralizers}  & \multicolumn{3}{c|}{Massive vacua on $\mathbb{R}^{2,1} \times S^1$} \\
\hline
$B_2$  
&  $SO(5)$  & $Spin(5)$  & $SO(5)_+$  & $SO(5)_-$ & $Spin(5)$  \\
\hline
$1+1+1+1+1$ 
& $SO(5)$ & $Spin(5)$ & 6 & 3 & 3 \\
$2+2+1$ 
& $SU(2)$ & $SU(2) \times \mathbb{Z}_2$ & 2 & 2 & 4 \\
 $3+1+1$ 
 & $\mathbb{Z}_2 \rtimes U(1)$ & $\mathbb{Z}_2 \rtimes U(1)  $ & 1 & 1 & 1 \\
$5$ 
& $1$ & $\mathbb{Z}_2$ & 1 & 1 & 2 \\
\hline
\multicolumn{3}{|c|}{Total} & 10 & 7 & 10 \\
\hline
\end{tabular}
\caption{For each $B_2$ partition we use the Springer-Steinberg theorem to compute the centralizer inside $SO(5)$ and $Spin(5)=Sp(4)$. 
Finally, we compute the number of massive vacua on $\mathbb{R}^{2,1} \times S^1$ in the different theories. }
\label{centralizerswithcentre}
\end{table}
\noindent
For the $Spin(5)$ gauge group, we find that the centralizer for the $2+2+1$ partition and the $5$ partition, contains an extra $\mathbb{Z}_2$ discrete
factor. We can turn on a Wilson line in this $\mathbb{Z}_2$ group, which doubles the number of massive vacua on $\mathbb{R}^{2,1} \times S^1$ corresponding to these partitions. This matches perfectly with the doubling of the T-duality doublet and singlet extrema  of the integrable
system that we witness on the left of figure \ref{dualities_so5_extended}.

\subsubsection*{Summary of the Global Analysis}
Thus, we have checked the duality inherited from ${\cal N}=4$, including the choice of the
center of the gauge group as well as the spectrum of line operators, in the case of
the Lie gauge algebra $so(5)$. The ${\cal N}=1^\ast$ theory neatly illustrates both the features
of the pure ${\cal N}=1$ theory as well as those of the ${\cal N}=4$ theory
discussed in \cite{Aharony:2013hda,Aharony:2013kma}.
The global refinement of the analysis of all vacua can be performed for 
${\cal N}=1^\ast$ theories with {\em any} gauge group, but we will refrain from belaboring this particular point
in the rest of our paper.

\subsection{Summary and Motivation}
\label{summaryso5}
By now, the reader may be convinced that the ${\cal N}=1^\ast$ theory, even in the rank two case of the $so(5)$ gauge algebra,
exhibits interesting elementary physical phenomena hiding in a maze  governed by modularity and ellipticity.
We will isolate a subset of these interesting phenomena, and clarify the mathematical structures relevant to each.
We will show that they are general, and that they can often be understood in algebraic, modular or elliptic terms. 
The points we will concentrate on are the following.
\begin{itemize}
\item We used semi-classical limits of elliptic integrable systems to render an analytic
exploration of the vacuum structure coded in the low-energy effective superpotential possible.
In the process, we uncovered limits of integrable systems that generalize the Inozemtsev limit \cite{I,D'Hoker:1998yh}. 
{From} the gauge theory perspective, these limits are intermediate between the confining and the Higgs regimes.
In section \ref{semiclassical} we describe these limits in more detail, and 
show that they are closely related to the semi-classical analysis of the ${\cal N}=1^*$ theory on $\mathbb{R}^{2,1} \times S^1$ with gauge algebra $\mathfrak{g}$.
\item We saw that a branch of massless vacua appeared as semi-classical limiting solutions, for the
gauge algebra $so(5)$. The appearance of massless vacua  as limiting solutions is again generic and also occurs for $su(N)$ theories, as we will  show in sections \ref{semiclassical}
and \ref{acase}. We will be able to analytically characterize  the manifold of massless vacua for the $su(3)$ theory, including its  duality properties. For the $su(4)$ theory, an analogous picture will be developed.
Finally, the massless manifold of the $so(5)$ gauge theory will be scrutinized in section \ref{so5massless}.
\item
We claimed that one vacuum of the $so(5)$ theory arises from turning on a $\mathbb{Z}_2$ Wilson line that breaks the abelian
gauge group factor such as to render the vacuum massive on $\mathbb{R}^{2,1} \times S^1$. 
We will show that this phenomenon as well is rather generic and that
we can characterize the discrete gauge group, and the Wilson line in terms
of the Lie algebra data associated to the corresponding semi-classical limit. This will 
be demonstrated in sections \ref{semiclassical} and \ref{discrete}.
\end{itemize}
The  clarification of these points will occupy us for the rest of this paper. There are  further open issues,
some of which are enumerated
in the concluding section \ref{conclusions}.

\section{Limits of Elliptic Integrable Systems and Nilpotent Orbit Theory}
\label{semiclassical}
In this section, we firstly propose new limits of elliptic integrable systems that generalize the
Inozemtsev limits performed in \cite{I,D'Hoker:1998yh}. We are motivated by the
fact that these limits describe semi-classical physics of supersymmetric gauge theories in four
dimensions. The existence of these limiting behaviors may also be of interest in the theory of integrable systems
\cite{I,D'Hoker:1998yh,Khastgir:1999pd}. Each limit is associated to a choice of subset of the set of simple
roots of (the dual of) the 
affine root system that enters the definition of the twisted elliptic integrable system.

In a second part of this section, we review how subsets of simple roots of affine root systems
enter in the theory of nilpotent orbits. Thus, we will be able to associate semi-classical
limits of the elliptic integrable system, and therefore the low-energy superpotential of
${\cal N}=1^\ast$ theory, to a detailed description of nilpotent orbits and the component
group of their centralizer. We will exploit this map in the following sections.

\subsection{Semi-Classical Limits of Elliptic Integrable Systems}
In this subsection, we study the (twisted) elliptic integrable systems which arise as the low-energy
effective superpotentials of ${\cal N}=1^\ast$ supersymmetric gauge theories compactified
on a circle \cite{Kumar:2001iu}. The derivation of the effective superpotential used the technique of compactification and
mass deformation \cite{Seiberg:1996nz} of ${\cal N}=2$ theories in four dimensions, as applied to the $su(N)$ theories in \cite{Dorey:1999sj}. The relevant elliptic integrable systems were described
in \cite{D'Hoker:1998yh}, where also the limits towards trigonometric and affine Toda integrable
systems were presented.\footnote{This analysis extended the one performed in \cite{I}. See also \cite{Khastgir:1999pd}.} This subsection is concerned with generalizing this analysis to include 
combinations of trigonometric and Toda integrable systems.
These limits  code possible symmetry
breaking patterns of the gauge theory. The limit we take can be described as a limit towards large imaginary modular parameter $\tau$, or as the semi-classical limit from the perspective of the ${\cal N}=1^\ast$ gauge theory where this parameter is identified with
the complex combination $\tau=\frac{4 \pi i }{g^2} + \frac{\theta}{2 \pi}$ of the gauge coupling $g$ and the $\theta$ angle. The procedure gives an
analytical handle on the extrema of the superpotential in the semi-classical regime.

\subsubsection{The Dual Affine Algebra and Non-Perturbative Contributions}
The large imaginary $\tau$ expansion of the (twisted) elliptic integrable potential is known
to be governed by affine algebras \cite{Lee:1998vu,Hanany:2001iy,Kumar:2001iu,Kim:2004xx}. Thus, it will be useful to introduce some affine algebra notation.\footnote{See e.g. \cite{Kac} for the theory
of affine Kac-Moody algebras.}
The (untwisted) affine algebra $\hat{\mathfrak{g}}=\mathfrak{g}^{(1)}$ is built from the loop algebra of $\mathfrak{g}$, the central extension $\hat{k}$ and the derivation 
$d$.
We build a Cartan subalgebra of $\hat{\mathfrak{g}}$ from a Cartan subalgebra of $\mathfrak{g}$ by adding the
generators $\hat{k}$ and $d$. Elements of the dual of the Cartan are denoted $(\lambda , k , n)$ with the Lorentzian scalar product $(\lambda , k , n) \cdot (\lambda' , k' , n') = \lambda \cdot \lambda' + kn' + k'n$.
If we define the imaginary root $\delta$ to be equal to $\delta = (0,0,1)$, 
the set of affine roots is
\begin{equation}
\hat{\Delta} = \{ \alpha + m \delta | m \in \mathbb{Z} \textrm{ and } \alpha \in \Delta \} \cup \{ m \delta | m \in \mathbb{Z} \textrm{ and } m \neq 0 \} \, ,
\end{equation}
and the set of positive affine roots is 
\begin{equation}
\hat{\Delta}^+ = \Delta ^+ \cup \{ \alpha + m \delta | m \in \mathbb{N}^* \textrm{ and } \alpha \in \Delta \} \cup \{ m \delta | m \in \mathbb{N}^* \}  \, .
\end{equation}
A set of positive simple roots is given by adjoining the affine root $\alpha_0=\delta - \vartheta$
(where $\vartheta$ is the highest root of $\mathfrak{g}$) to a simple root system of $\mathfrak{g}$. The theory of twisted affine algebras, their classification, their (simple,
positive) roots is also pertinent here, and can be looked up in \cite{Kac}.

Armed with this knowledge, let's analyze how the potential behaves in the large imaginary $\tau$ limit, and how the
low-energy effective superpotential codes non-perturbative corrections to
gauge theory on $\mathbb{R}^{2,1} \times S^1$. The low-energy
effective superpotential for the ${\cal N}=1^\ast$ gauge theory with gauge algebra $\mathfrak{g}$
is given by \cite{Kumar:2001iu}
\begin{equation}
\mathcal{W}_{tw} (Z) =  \sum\limits_{\alpha \in \Delta^+} g_{\nu (\alpha)} \wp_{\nu (\alpha)} (\alpha \cdot Z; \tau) 
\end{equation}
where the index $\nu(\alpha)$ is defined by
\begin{equation}
\nu(\alpha) = \frac{|\alpha_{long}|^2}{|\alpha|^2} = \frac{2}{|\alpha|^2}
\end{equation}
and the short and long root coupling constants are expressed in terms of a single constant $g$ by 
\begin{equation}
g_{\nu} = \frac{g}{\nu} \, . 
\end{equation}
We normalize the long roots to have length squared two.
The gauge coupling $\tau$  is given by the ratio of the periods of the torus
$\tau=\omega_2/\omega_1$.
To perform the semi-classical large imaginary $\tau$ expansion, we can exploit the result
\begin{eqnarray}
\label{largetauP}
\wp(2 \omega_1 x;\omega_1,\omega_2) &=&
-\frac{\pi^2}{12 \omega_1^2} E_2(q)
+ \frac{\pi^2}{4 \omega_1^2} \csc^2 \left( \pi x \right)
- \frac{2 \pi^2}{\omega_1^2} \sum_{n=1}^\infty \frac{n q^n}{1-q^n} \cos 2 \pi n x
\, ,
\end{eqnarray}
where $q=e^{2 \pi i \tau}$. This expansion is valid whenever the series is convergent, which requires $|q|<1$ or equivalently that $\tau \in \mathcal{H}$, and also $|\Im (x)|< \Im (\tau)$. The space $\mathcal{H}$ is the  upper-half plane of complex numbers with positive imaginary part. For $x \in \mathcal{H}$ or $x \in \mathbb{R} \setminus \mathbb{Z}$ we can use the further expansion:
\begin{equation}
-4 \sum\limits_{k=1}^{\infty}k e^{2 \pi i k x} = \csc^2 \pi x \, ,
\label{cosec}
\end{equation}
to find
\begin{eqnarray}
\wp(2 \omega_1 x;\omega_1,\omega_2) &=&
-\frac{\pi^2}{12 \omega_1^2} E_2(q)
-\frac{\pi^2}{\omega_1^2} \sum\limits_{n=1}^{\infty} n  \left[ e^{2 \pi i n  x} +  \sum\limits_{m=1}^{\infty} q^{nm} \left(e^{-2 \pi i n  x}+e^{2 \pi i n x} \right) \right]
\, .
\end{eqnarray}
For the twisted Weierstrass function $\wp_\nu$ defined for $\nu \in \mathbb{N}^\ast$ by
\begin{equation}
\wp_{\nu}( z ;\omega_1,\omega_2) = \sum\limits_{k=0}^{\nu -1} \wp \left( z + \frac{ k}{\nu} 2 \omega_1  ;\omega_1,\omega_2 \right) \, , 
\end{equation}
we have the counterpart
\begin{eqnarray}
\wp_\nu(2 \omega_1 x;\omega_1,\omega_2) &=&
-\frac{\nu \pi^2}{12 \omega_1^2} E_2(q)
+ \frac{\nu^2 \pi^2}{4 \omega_1^2} \csc^2 \left( \pi \nu x \right)
- \frac{2 \nu^2 \pi^2}{\omega_1^2} \sum_{n=1}^\infty \frac{n q^{n\nu}}{1-q^{n\nu}} \cos 2 \pi n \nu x \nonumber \\
 &=&
-\frac{\nu \pi^2}{12 \omega_1^2} E_2(q) -\frac{\nu ^2 \pi^2}{\omega_1^2} \sum\limits_{n=1}^{\infty} n  \left[ e^{2 \pi i n \nu x }  +  \sum\limits_{m=1}^{\infty} q^{nm \nu} \left( e^{-2 \pi i n \nu x}
 +e^{2 \pi i n \nu x} \right) \right] \, . 
 \label{largetautwistedP}
\end{eqnarray}
Again, this expansion is valid for $x \in \mathbb{C} \setminus \mathbb{Z}$ such that $0 \leq \Im (x) < \Im (\tau)$. 
It should be clear that the  part of the argument of the Weierstrass function proportional to $\tau$ plays a crucial role in the Taylor series in the large $\tau$ limit. This is illustrated by the fact that for any $0<a<1$ and any $b \in \mathbb{R}$,
\begin{equation}
\lim\limits_{\tau \rightarrow i \infty} \wp (a \tau + b ; \tau) = - \frac{\pi^2}{3} \, . 
\label{wp_asymptotic}
\end{equation}
It is therefore useful to separate the argument into  a part proportional to $\tau$ and a part that will not grow with $\tau$, by setting
\begin{equation}
 Z =  X + \tau Y   \, ,
\end{equation}
where $X$ and $Y$ are complex variables. At this stage, this decomposition is arbitrary. We have doubled the number of degrees of freedom, and we will use this redundancy in subsection \ref{section_semiclassical_limits} to impose the value of $Y$. Plugging this parametrization into the (twisted) Weierstrass function yields
\begin{eqnarray}
\wp_{\nu}(2 \omega_1 \alpha \cdot Z;\omega_1,\omega_2) &=&
-\frac{\nu \pi^2}{12 \omega_1^2} E_2(q) -\frac{\nu ^2 \pi^2}{\omega_1^2} \sum\limits_{n=1}^{\infty} n  \left[ q^{n \nu \alpha \cdot Y} e^{2in\pi \nu \alpha \cdot X } \vphantom{\sum\limits_{m=1}^{\infty}} \right. \\ 
 & & \left. +  \sum\limits_{m=1}^{\infty} q^{nm \nu} \left(q^{-n \nu \alpha \cdot Y}e^{-2in\pi \nu \alpha \cdot X}+q^{n \nu \alpha \cdot Y} e^{2in\pi \nu \alpha \cdot X} \right) \right]
\, .
\end{eqnarray}
Using these expansion formulas for the potential, we arrive at a sum of exponential terms, each associated to a positive affine root
\begin{eqnarray}
\mathcal{W}_{tw} (\hat{Z})
&=& -\frac{g \pi^2}{\omega_1^2}  \left( \frac{|\Delta ^+| }{12 } E_2(q) + \sum\limits_{n=1}^{\infty} n \left[ \sum\limits_{\hat{\alpha} \in \hat{\Delta}^+} \nu(\hat{\alpha}) q^{n \nu(\hat{\alpha}) \hat{\alpha} \cdot \hat{Y} } e^{2 \pi i n \nu(\hat{\alpha}) \hat{\alpha} \cdot X } \right] \right)  \, .
\label{exponents}
\end{eqnarray}
We have used the notations $\hat{X} = (X,0,0) = X$, $\hat{Y} = (Y,1,0)$ and $\hat{Z} = \hat{X} + \tau \hat{Y}$ so that for any affine root $\hat{\alpha} = \alpha + m \delta \in \hat{\Delta}$, we have the equality  $\hat{\alpha} \cdot \hat{Y} = \alpha \cdot Y + m$. We also define $\nu$ on affine roots with non-zero real part by $\nu(\hat{\alpha}) = \nu(\alpha + m \delta) = \nu(\alpha) $, so that 
\begin{equation}
\hat{\alpha}^{\vee} = \frac{2}{|\hat{\alpha}|^2}\hat{\alpha} = \frac{2}{|\alpha|^2}\hat{\alpha} = \nu(\hat{\alpha}) \hat{\alpha} \, .
\end{equation}
We have arbitrarily declared $\nu(m \delta) = 0$.\footnote{We note that low-energy effective superpotential is ambiguous up to a purely $q$-dependent term.} The form of the exponents in equation (\ref{exponents}) suggests switching from the affine root system to its dual
\begin{eqnarray}
\mathcal{W}_{tw} (\hat{Z})&=& -\frac{g \pi^2}{\omega_1^2}  \left( \frac{|\Delta ^+| }{12 } E_2(q) + \sum\limits_{n=1}^{\infty} n \left[ \sum\limits_{\hat{\alpha}^\vee \in \left( \hat{\Delta}^+ \right)^\vee} \frac{1}{\nu(\hat{\alpha}^\vee)} q^{n \hat{\alpha}^\vee \cdot \hat{Y}} e^{2 \pi i n \hat{\alpha}^\vee \cdot X } \right] \right) 
\, . 
\label{expansion}
\end{eqnarray}
In the sum, we again disregard the terms associated to purely imaginary roots.

In the gauge theory, the semi-classical expression (\ref{expansion}) has an interpretation as a sum over three-dimensional monopole-instanton 
contributions \cite{Kumar:2001iu}.\footnote{See \cite{Hanany:2001iy} for a graphical
representation of the non-perturbative states that contribute, in terms of
D-brane systems in string theory.} Note that the purely four-dimensional 
instanton terms associated to the imaginary roots contribute a $\tau$ dependent, but position independent term in the superpotential.
We have two forms for the final expression. One expression (namely (\ref{exponents}))
is in terms of the root system we started out with, the other (namely (\ref{expansion}))
in terms of co-roots. Both forms are equally canonical, due to the fact that both the electric Wilson line variable  and the dual photon variable are present in the potential and are interchanged under Langlands duality. This is a manifestation of 
the S-duality of the parent ${\cal N}=4$ theory. In a given semi-classical expansion (i.e. $\tau \rightarrow i \infty$), we may more easily read expression
(\ref{expansion}), which has an interpretation as a sum over magnetic monopole instantons in this limit.

\subsubsection{Semi-Classical Limits}
\label{section_semiclassical_limits}
Concretely, we take semi-classical limits as follows. We consider a particular isolated extremum whose positions $Z$ depend only on $\tau$ (up to discrete equivalences that depend on the gauge group). We assume that at weak coupling, the limit 
\begin{equation}
Y \equiv \lim\limits_{\tau \rightarrow i \infty} \frac{1}{\tau} Z(\tau)
\label{assumption_isolated}
\end{equation}
exists and we define $X (\tau)= Z(\tau) - \tau Y$. Note that for any $\tau \in \mathcal{H}$ we have $Z(\tau) = X(\tau) + \tau Y$ as before, and the parametrization $Y$ is a vector that is independent of $\tau$ and which characterizes the extremum (or several extrema) under consideration. It is a non-trivial task to enumerate the set of vectors $Y$ that give rise to isolated extrema. We will also deal with continuous branches of extrema, for which the definition (\ref{assumption_isolated}) has no intrinsic meaning. In this case we can nevertheless choose an arbitrary set of coordinates of the branch, and take the limit while keeping these coordinates fixed. Depending on the choice of parametrization, this may lead to a continuous set of values for the vector $Y$. {From} now on, when studying a given extremum, we trade the variable $Z(\tau)$ for the variable $X(\tau)$ which is finite in the limit we want to perform, and use the expansion (\ref{expansion}). 

Before doing so, let's choose a basis of simple roots $(\alpha_1 , \dots , \alpha_r)$ in the root system $\Delta$. Then $(\alpha_0, \alpha_1 , \dots , \alpha_r)$ are the simple roots of the affine root system $\hat{\Delta}$. The dual root system has a set of simple roots $((\alpha_0)^\vee,\alpha^\vee_1 , \dots , \alpha^\vee_r)$. To be more explicit about the semi-classical limit, we must distinguish between variables that sit on the boundary of the fundamental alcove, and those that reside inside. We therefore choose a vector $Y$ in the fundamental affine Weyl chamber (or fundamental alcove), which implies that $(\alpha_i)^\vee \cdot \hat{Y} \geq 0$ for $i=0,1,...,r$.
We decompose the positive roots in terms of simple roots of the dual of the affine algebra, and the vector $Y$ in the weight space in terms of affine fundamental weights $\hat{\pi}_i$:
\begin{eqnarray}
\hat{\alpha}^\vee = \sum_{i=0}^r n_i \alpha^\vee_i 
& \quad &
\hat{Y} = \sum_{i=0}^r Y_i \hat{\pi}_i 
\end{eqnarray}
where the $n_i$ are non-negative integers, 
\begin{eqnarray}
\hat{\pi}_i = (\pi_i;a_i^\vee;0)
& \quad &
\hat{\pi}_0 = (0;1;0)
\, ,
\end{eqnarray}
and $a_i^\vee$ denote the co-marks of the Lie algebra. The fundamental weights satisfy the orthonormality conditions $(\hat{\pi}_i,\alpha^\vee_j) = \delta_{ij}$, so that $Y_i = \alpha^\vee_i \cdot \hat{Y} \geq 0$ and 
\begin{equation}
\hat{\alpha}^\vee \cdot \hat{Y} = \sum_{i=0}^r n_i Y_i \, . 
\end{equation}
Note that the definition of $\hat{Y}=(Y,1,0)$ 
gives a linear relation between the $r+1$ coordinates $Y_i$, 
\begin{equation}
1 =Y_0  + \sum_{i=1}^r Y_i a_i^\vee = \sum_{i=0}^r Y_i a_i^\vee \, . 
\label{constraintequation}
\end{equation}
Similarly we define $X_i = \alpha^\vee_i \cdot X$, and have the constraint $\sum_{i=0}^r X_i a_i^\vee =0$. The  distinction we now make is between those variables $Y_i$ that lie on the boundary of the fundamental alcove, and those that lie inside. 
This will fix the leading behavior of the extrema that we focus on. For $Y_i=0$, we note that there is an infinite set of non-perturbative contributions that needs to be taken into account in the semi-classical limit, and in particular, we need to resum them to the trigonometric term (as in equation (\ref{cosec})). The set of roots $\alpha^\vee$ for which this phenomenon occurs will again 
form a root system.
Thus, to leading order in the modular parameter
$q=e^{2 \pi i \tau}$, we will have a trigonometric integrable
system corresponding to a choice of  subset of simple
roots inside the affine simple root system. 
In a second step, by assumption, we have the remaining coordinates
$Y_j$ that do not vanish to leading order in $\tau$.
As a consequence (of formula (\ref{expansion})), these directions $Y_j$ lead to subleading exponential terms. 

More in detail, let's group positive roots by their inner products  with $\hat{Y}$ and form the sets:
\begin{eqnarray}
( \hat{\Delta}^+_{t} (Y))^\vee &=&  \left\{ \hat{\alpha}^\vee \in (\hat{\Delta}^+)^\vee | \hat{\alpha}^\vee \cdot \hat{Y} = 2 \omega_1 t \right\} 
\, , 
\end{eqnarray}
and also the spectrum $S$ of such inner products
\begin{eqnarray}
S(Y) &=&  \left\{t \in \mathbb{R}  | ( \hat{\Delta}^+_{t} (Y) )^\vee \neq \emptyset \right\} 
\, . 
\end{eqnarray}
The spectrum of inner products without zero will be denoted $S(Y)^\ast = S(Y) - \{0 \}$. The set of roots with zero inner product is finite while the full spectrum $S(Y)$ is generically infinite, due to the infinite nature of the affine root system. The superpotential 
\begin{equation}
\mathcal{W}_{tw} (\hat{Z}) =-\frac{g \pi^2}{\omega_1^2}  \left(C(\tau) + \sum\limits_{n=1}^{\infty} n \left[\sum\limits_{t \in S(Y)} q^{n t} \sum\limits_{\hat{\alpha}^\vee \in (\hat{\Delta}^+_t)^\vee} \frac{1}{\nu(\hat{\alpha}^\vee)}  e^{2 \pi i n\hat{\alpha}^\vee \cdot X } \right] \right) \, , 
\end{equation}
will split into two sets of terms. Note that the exponents of $q$ are non-negative, so that the expression remains finite when we take the limit $q \rightarrow 0$. As mentioned previously, the first split happens between terms with zero inner product and non-zero inner product:
\begin{eqnarray}
\mathcal{W}_{tw} (\hat{Z}) &=& -\frac{g \pi^2}{\omega_1^2}  \left( C(\tau) -\frac{1}{4} \sum\limits_{\hat{\alpha}^\vee \in (\hat{\Delta}^+_0(Y))^\vee} \frac{1}{\nu(\hat{\alpha}^\vee)} \csc^2 \left( \pi  \hat{\alpha}^\vee \cdot X  \right) \right. \\
 & & \left. + \sum\limits_{t \in S(Y)^\ast} \sum\limits_{n=1}^{\infty} n  q^{n t} \left[ \sum\limits_{\hat{\alpha}^\vee \in (\hat{\Delta}^+_t(Y))^\vee} \frac{1}{\nu(\hat{\alpha}^\vee)}  e^{2 \pi i n\hat{\alpha}^\vee \cdot X } \right] \right) \, . 
\end{eqnarray}
We obtain  a sum of a trigonometric and an exponential system
\begin{equation}
\label{expansionTrigToda}
\mathcal{W}_{tw} (\hat{Z}) =-\frac{g \pi^2}{\omega_1^2}  \left( C(\tau)  - \frac{1}{4} \mathcal{W}_{trig}^{Y}(X)  + \sum\limits_{t \in S(Y)^\ast} \sum\limits_{n=1}^{\infty} n  q^{n t} \mathcal{W}_{exp}^{(n,t,Y)}(X) \right) \, ,
\end{equation}
where
\begin{eqnarray}
\mathcal{W}_{trig}^{Y}(X) &=& \sum\limits_{\hat{\alpha}^\vee \in \left( \hat{\Delta}^+_0(Y) \right)^\vee} \frac{1}{\nu(\hat{\alpha}^\vee)} \csc^2 \left( \pi  \hat{\alpha}^\vee \cdot X \right) \\
\mathcal{W}_{exp}^{(n,t,Y)}(X) &=& \sum\limits_{\hat{\alpha}^\vee \in \left( \hat{\Delta}^+_t(Y) \right)^\vee} \frac{1}{\nu(\hat{\alpha}^\vee)}  e^{2 \pi i n\hat{\alpha}^\vee \cdot X } \label{expSystem}
\, .
 \end{eqnarray}
The behavior of the subdominant system is intricate. 
A first stab at the subdominant system consists in realizing that the remaining
variables (indexed by the set $\bar{J}_0=\{0,1, \dots, r \} \setminus J_0$ where $J_0$ is the set
of coordinates with zero inner product) will all have a leading exponential
term. These exponentials, combined with the constraint equation (\ref{constraintequation}),
may give rise to exponential interactions, stabilized by an exponential interaction
of opposite sign. The affine Toda potential is an example of this type of subdominant 
potential. Roughly speaking,
this reasoning goes through, but the devil is in the details. 
The first complicating factor is the
influence of the dominant terms on the subdominant terms when searching for an equilibrium position.
In particular, corrections to equilibrium positions for leading coordinates may 
strongly influence subdominant contributions. Particular equilibrium configurations for the leading
trigonometric system can also
give rise to subtle and persistent cancellations in the coefficients of subdominant exponential terms.
There may also be a staircase of subdominant terms, each with its own limiting behavior. Even a continuous
set of limiting behaviors can occur.
Moreover, the solutions
to the trigonometric system are only known as zeroes of orthogonal polynomials, making this process hard to carry through analytically in full generality. 

Therefore, we develop only a partial picture of the integrable systems that result in the limit.
Still, we provide a generalization of the limit discussed in  \cite{D'Hoker:1998yh} in the following subsection,
and useful heuristics based on the examples in sections \ref{so5}, \ref{acase} and \ref{discrete}. 

\subsubsection{The Trigonometric, Affine Toda and Intermediate Limits}
Here we will treat the special case in which no cancellation of (sub)leading exponentials occurs, and in which the subleading exponential integrable system stabilizes all the remaining coordinates and leads to an isolated extremum. We can then analytically solve for the remaining variables. 
Due to the constraint equation we have that the set $J_0 \subsetneq \{0,1,...,r\}$ is a true subset of the set of simple roots (we identify a simple root $\alpha_i$ with its index $i$). We obtain a trigonometric integrable system for the root system corresponding to the simple roots in $J_0$. This system gives solutions for $|J_0|$ of the $r+1$ variables $X_i$. 
Let $t_1$ be the smallest non-zero element of the spectrum $S(Y)$. At the next level in the $q$-expansion, we find contributions corresponding to the set $( \hat{\Delta}_{t_{1}}^+ (Y))^\vee$, which is equal to a set of positive  roots.

 The final Toda integrable system is a sum over 
 $|\bar{J}_0|$ vectors where $\bar{J}_0$ is the complement of the set of affine
 simple roots that enter the trigonometric system, by assumption. 
We then obtain
\begin{eqnarray*}
\mathcal{W}_{Toda}^{(n=1,t_{1},Y)}(X) 
&=& 
C 
\sum\limits_{i \in \bar{J}_0
} \frac{1}{\tilde{\nu}(\alpha^\vee_i)}  e^{2 \pi i \alpha^\vee_i\cdot X } \, .
\end{eqnarray*}
In the last equality we have indicated the fact that
for each individual index $i$, there may be a renormalization of the constant $\tilde{\nu}$
in front of the exponential term, due to various roots contributing to the same exponential behavior.
The constraint equation then gives 
\begin{equation}
1  = \sum_{i \in \bar{J}_0} Y_i a_i^\vee = 2 \omega_1 t_1 \sum_{i \in \bar{J}_0} a_i^\vee
\end{equation}
from which we extract $t_1$ and finally 
\begin{equation}
\hat{Y}= \frac{\sum\limits_{i \in \bar{J}_0} \hat{\omega}_i}{\sum\limits_{i \in \bar{J}_0} a_i^\vee} \, . 
\end{equation}
After projection on the finite part, we find: 
\begin{equation}
Y= \frac{\sum\limits_{i \in \bar{J}_0} \omega_i}{\sum\limits_{i \in \bar{J}_0} a_i^\vee} \, . 
\end{equation}
where we define $\omega_0=0$. Here the dependence on $t_1$ has disappeared. We can simply use $Y_{\bar{J}_0}$ as an ansatz,
for every non-empty set $\bar{J}_0 \subset \{0,1,...,r\}$, where we have defined 
\begin{equation}
Y_J = \frac{\sum\limits_{i \in J } \omega_i}{\sum\limits_{i \in J} a_i^\vee} \, .
\end{equation}
On the condition that the subleading exponentials have non-vanishing coefficients, this gives the
semi-classical (linear order in $\tau$) values for the $Y$ coordinates of the integrable system. Namely, a first set sits at an extremum of the trigonometric integrable system, and  a second set
at the extrema of the affine Toda system. Known applications of this ansatz are the following.
A first  extreme case is
$\bar{J}_0=\emptyset$ and $Y=0$. Then $\Delta^+_0 (Y) = \Delta^+$ and we recover the trigonometric potential only.  The other extreme case is $\bar{J}_0=\{0,1,...,r\}$ and $Y=\rho / h^{\vee}$ (where $\rho$ is the Weyl vector and $h^\vee$ the dual
Coxeter number of the gauge algebra). We then obtain the affine Toda potential for the algebra $(\mathfrak{g}^{(1)})^{\vee}$, as described in \cite{D'Hoker:1998yh}.
There are many intermediate cases that follow the above pattern, or an even more intricate
one.\footnote{In the  gauge theory, these cases correspond respectively to a fully Higgsed vacuum,
confining pure ${\cal N}=1$ dynamics, and partial Higgsing.} Examples are provided in 
sections \ref{acase} and \ref{discrete}, and we already saw some
in section \ref{so5}.
It would be desirable to have a full classification of semi-classical limits. The ${\cal N}=1^\ast$
gauge theory
provides intuition in the case of the (twisted) elliptic Calogero-Moser system with 
particular coupling constants -- the question in the integrable system context is even more
general.

\subsection{The Nilpotent Orbit Theory of Bala-Carter and Sommers}
 {From} the previous subsection, we conclude that we can associate  semi-classical 
extrema of the elliptic integrable system
to subsets of the (dual) affine simple root system. In this section, we show that there is another
 way to understand the relevance of these subsets, in terms of nilpotent orbit theory and
 the physics of ${\cal N}=1^\ast$ theory on $\mathbb{R}^{2,1} \times S^1$.
 
 Firstly, let us briefly  review  highlights of nilpotent orbit theory. See e.g. 
 the textbooks  \cite{C,CM,LT,LS} for a gentler introduction.
The Bala-Carter classification of nilpotent orbits of simple algebraic groups
goes as follows. Each nilpotent orbit of a 
Lie algebra $\mathfrak{g}$ of a connected, simple algebraic group $G$ is a distinguished nilpotent orbit
of a Levi subalgebra \cite{BC1,BC2}. Levi subalgebras of $\mathfrak{g}$ correspond to subsets of simple roots of $\mathfrak{g}$
(up to conjugation by the Weyl group). Distinguished nilpotent orbits are those for which
the nilpotent element does not commute with a non-central semi-simple element.\footnote{
We illustrate the application of these concepts more concretely in the case of $B_2$, $A_{N-1}$ and $G_2$ in sections
\ref{so5}, \ref{acase} and \ref{discrete}, and the mathematics literature contains much more detail.}

Furthermore, there is  generalization of the Bala-Carter correspondence by Sommers \cite{Sommers}. There is a one-to-one correspondence
between nilpotent elements $n$ and conjugacy classes of the component group $Comp(n)$
of the centralizer
on the one hand, and pairs $(\mathfrak{l},n)$ of pseudo-Levi subalgebras $\mathfrak{l}$ and distinguished nilpotent
elements $n$ in $\mathfrak{l}$ on the other hand. The correspondence is up to group conjugacy. A pseudo-Levi
subalgebra corresponds, by definition, to a subset of the simple root system 
of $\mathfrak{g}$ extended by the lowest root $-\theta$. This classification  allows for the 
unified calculation of all the component groups of nilpotent orbits of simple Lie algebras \cite{Sommers}. The relevance of these  results can be gleaned from the
$so(5)$ example discussed in section \ref{so5}, from the semi-classical limits analyzed
above, and  can also generically be argued for, as follows.

\subsection{The Bridge between Gauge Theory and Integrable System}

Semi-classical solutions to the F-term equations of motion for ${\cal N}=1^\ast$ theory
on $\mathbb{R}^4$ are classified by matching them onto nilpotent orbits \cite{Bourget:2015lua}. When we compactify
the gauge theory on $S^1$, there are further aspects of nilpotent orbits that
come into play. In particular, we will allow for Wilson lines in the unbroken gauge group.
If the latter contains topologically non-trivial conjugacy classes, i.e. conjugacy classes
in the component group of the centralizer, then we need to consider each of these configurations
separately. 

As we saw,
there is a one-to-one correspondence between the pair (nilpotent orbit, conjugacy class of component group)
and pseudo-Levi subalgebras. The trivial conjugacy classes will correspond to a collection
of non-affine simple roots. Each inequivalent choice of subset that necessarily includes  the affine
root will correspond to a non-trivial conjugacy class of a component group.
These are classified by Bala-Carter-Sommers theory, which therefore is crucial in classifying semi-classical configurations for ${\cal N}=1^\ast$ theory compactified on a circle.
The example of the ${\cal N}=1^\ast$ gauge theory with $G_2$ gauge group discussed in section \ref{discrete}
will neatly illustrate our reasoning. 

Before we turn to this application, we demonstrate the use of the semi-classical limit in example systems.
In particular, the  techniques developed in this section
allow for the analysis of the physics and duality properties of the massless
vacua of $su(N)$ theories.

 \section{The ${\cal N}=1^\ast$ Theory with Gauge Algebra $A_{N-1}$}
\label{sun}
\label{an}
\label{acase}
In this section, we concentrate on the ${\cal N}=1^\ast$ theory with $su(N)$
gauge algebra. This theory has many simplifying features. In particular,
the unbroken gauge group in all semi-classical vacua of the $SU(N)/\mathbb{Z}_N$ 
theory is connected, so that the component group (in the adjoint group) is trivial.\footnote{The center of the gauge group would play the leading
role in the discussion of the global aspects of the gauge theory. See \cite{Aharony:2013hda,Aharony:2013kma},  and our
subsection \ref{globalso5}.}
Indeed, from the mathematical point of view we have that
Bala-Carter theory coincides with Bala-Carter-Sommers
theory. In the $A$-type case, all pseudo-Levi subalgebras are equivalent to Levi
subalgebras, since the lowest root is Weyl equivalent to any other simple root.
Thus, in this theory, we can isolate new semi-classical limits of the integrable system, and the corresponding gauge theory physics from other interesting features of ${\cal N}=1^\ast$ theories
when compactified on $S^1$.  We will find branches of massless vacua for low rank, characterize  their equilibrium positions,
analyze their superpotential and study how these vacua behave under duality.\footnote{A preliminary discussion of massless vacua can be found in \cite{Aharony:2000nt}.}

 To understand the fate of semi-classically massless and massive vacua in $su(N)$ ${\cal N}=1^\ast$ theory,
 we again take the elliptic Calogero-Moser Hamiltonian as our starting point \cite{Dorey:1999sj}.
 For starters, we analyze this effective superpotential in the semi-classical regime
 $\tau \rightarrow i \infty$ and classify extrema of the integrable system using the technique
 laid out in section \ref{semiclassical}. We will be able to promote parts of our limiting knowledge
 to exact statements at finite coupling.
 
 \subsection{Semi-Classical Preliminaries}
 As argued previously, a classification of extrema is governed by pseudo-Levi subalgebras, in turn determined by  Weyl (i.e. permutation) inequivalent subsets of the affine root system (whose Dynkin diagram  is a circle with $N$ nodes). The number of inequivalent subsets of roots (except all of them) is  the number of partitions of $N$. In more detail, we let $\Delta = \{\alpha_1 , ... , \alpha_{N-1}\}$ be a set of simple roots of $A_{N-1}$. For any subset $\Theta \subset \Delta$ we construct a partition of $N$. We can write $\Theta$ uniquely  as a disjoint union of sets of the form $\Delta_{k_i,d_i} = \{\alpha_{k_i}, ... , \alpha_{k_i+d_i-2}\}$ where $d_i \geq 2$. If our choice of subset $\Theta$ is 
 \begin{equation}
 \Theta = \bigcup\limits_{i} \Delta_{k_i , d_i}
 \end{equation}
then the partition is $1+...+1+\sum_i d_i=N$ with as many $1$'s as necessary to obtain a partition of $N$.

 For each choice of subsystem, we know the corresponding centralizer subgroup in the complexification of $SU(N)$. We denote the latter by $GL(N)$, the group of size $N$ invertible matrices with complex entries. The algebra of the centralizer is given in \cite{C}. With the notation $r_i = |\{j | d_j=i\}|$ for the number of times a representation of dimension $i$ occurs in the $sl(2)$ representation spanned by the adjoint scalars, so that 
 \begin{equation}
 \sum\limits_{i} i r_i = N \, , 
 \end{equation}
 the centralizer algebra is
 \begin{eqnarray}
 \left(\prod_{i} A_{r_i-1} \right) \times u(1)^k  \, ,
 \label{acentralizer}
 \end{eqnarray}
 where $k = | \{i | r_i > 0 \} | -1$ (i.e. the number of distinct dimensions minus one). Then the global structure of the centralizer group is \cite{CM}
 \begin{equation}
 S \left( \prod GL(r_i)^i_\Delta \right) \, , 
 \end{equation} 
 where the $\Delta$ denotes the diagonal copy of $GL(r_i)$ inside $GL(r_i)^i$, and the $S$ in front means that we keep only the matrices with unit determinant. This is the centralizer group in the complexification of $SU(N)$. The counting of the abelian factors in this group goes as follows: there is one abelian factor for each term in the product 
  and the constraint of unit determinant reduces the total number of abelian factors
 by one. In terms of pseudo-Levi subalgebras, a group with no abelian factor is obtained from a  set of roots $\Theta$ containing all the simple roots except $d$ of them (where $d$ divides $N$) equally spaced on the cyclic affine Dynkin diagram.
 In other words, one takes disconnected groups of $d-1$ roots on the affine Dynkin diagram,
 where $d$ is a divisor. These give rise to the massive vacua of the ${\cal N}=1^\ast$
 theory that were described in \cite{Donagi:1995cf,Dorey:1999sj}. The corresponding
 semi-classical limits of the integrable system are well-understood. We wish to advance the more general case.
 We will do this on a case-by-case basis, working our way up in rank.
 
In the next subsections, we  use the semi-classical limiting technique to gain information on the massless vacua of the first non-trivial low rank cases. For the $su(3)$ theory, we will complete the picture at finite coupling, while for the $su(4)$ 
algebra, we present a few features that will be typical of higher rank.

 \subsection{The Gauge Algebra $su(3)$, the Massless Branch and the Singularity}
 
We remind the reader that the superpotential for the $su(3)$ algebra \cite{Dorey:1999sj} can be parametrized in terms of the coordinates $z_{i}$ with $i = 1, 2, 3$ where we can use a shift symmetry to put $z_3=0$:
 \begin{eqnarray}
 \mathcal{W}_{A_2} &=& \wp (z_1-z_2)+\wp(z_2-z_3)+\wp(z_3-z_1)
 \nonumber \\
 &=& \wp(z_1-z_2) + \wp (z_2) +\wp(z_1)
 \nonumber \\
 &=& \wp(Z_1) + \wp(Z_2) +\wp(Z_1+Z_2) \, .
 \end{eqnarray}
 In the last line, we have used  the more intrinsic parametrization in terms of the coordinates $Z_i$ associated to the fundamental weights.
 
\subsubsection*{Semi-classical analysis}

When we apply our program of identifying vacua in the semi-classical limit to the case of $su(3)$, we recuperate the known results for the massive vacua, and find new results for massless vacua. 
 \begin{itemize}
 \item For the choice $J_0 = \emptyset$, which corresponds to the partition $1+1+1$, we set the leading behavior $Y_0=Y_1=Y_2=\frac{1}{3}$. One finds three confined massive vacua (with $k=0,1,2$) at 
 \begin{equation}
  (z_1,z_2,z_3) = \left( \frac{k}{3} + \frac{2}{3} \tau , \frac{2k}{3} + \frac{1}{3} \tau , 0 \right) \, . 
 \end{equation}
 These extremal positions are exact and the superpotential in these vacua is known \cite{Dorey:1999sj}. 
 \item For the pick $J_0 = \{1\}$, namely the partition $3=1+2$, we choose $Y_0=Y_2=\frac{1}{2}$, and at first order the $A_1$ trigonometric system fixes $X_1=\frac{1}{2}$. We analyze the potential near this equilibrium by expanding ${\cal W}_{A_2} (\frac{1}{2} + \delta X_1, \frac{\tau}{2} + X_2)$ in perturbation theory in $\delta X_1$, and as a function of $X_2$. We find that the first coordinate is corrected as follows
\begin{eqnarray}
\delta X_1 &=& -\frac{4 i \left(e^{2 i \pi  X_2}-e^{-2 i \pi  X_2} \right) \sqrt{q}}{\pi }-\frac{16 i  \left(e^{4 i \pi  X_2}-e^{-4 i \pi  X_2}\right) q}{\pi } + \dots
\end{eqnarray}
Plugging this correction into the superpotential leads to a superpotential which to the relevant order no longer depends on $X_2$, and in fact, is equal to zero. We have checked this to order $q^4$. These facts point towards the existence of a branch of massless vacua, with zero superpotential along the whole branch. We will obtain full analytic control of this branch below. 
 
 \item Finally, for the choice $J_0=\{1,2\}$, namely the partition $3$, one obtains the $A_2$ trigonometric potential. This potential has a real extremum,  the fully Higgsed vacuum
 \begin{equation}
 (z_1,z_2,z_3) =  \left( \frac{2}{3}, \frac{1}{3}  , 0 \right) \, ,
 \end{equation}
 as well as complex massless extrema
which form a portion of the same branch of vacua with zero superpotential just mentioned.  
 \end{itemize}

\subsubsection*{The Massless Branch and the Singularity}
Semi-classically, we have found evidence for the existence of a massless branch of vacua with zero superpotential. In the following, we will concentrate on describing the properties of this branch analytically, at any finite coupling $\tau$. Together with the known results about massive vacua that our analysis also recovers, we thus obtain all the vacua of the ${\cal N}=1^\ast$ theory with $su(3)$ gauge algebra exactly.

Firstly, we introduce some notation. We will denote the elliptic curve variables as 
\begin{eqnarray}
\label{gauge_inv_var}
\mathcal{X}_i &=& \wp(Z_i)
\nonumber \\
\mathcal{Y}_i &=& \wp'(Z_i) \,  
\end{eqnarray}
for $i=1,2,3$, where $Z_3 = -Z_1 -Z_2$ by convention. The points $(\mathcal{X}_i,\mathcal{Y}_i)$ all lie on the same elliptic curve, parametrized by $\tau$, and described by the equation
\begin{eqnarray}
\mathcal{Y}^2 &=& 4 \mathcal{X}^3 - g_2 \mathcal{X} - g_3 \, .
\end{eqnarray}
The equations for extremality of the superpotential then read
\begin{eqnarray}
\mathcal{Y}_1 = \mathcal{Y}_2 =
\mathcal{Y}_3 \, . & &
\end{eqnarray}
Moreover, the addition theorem for the elliptic Weierstrass function  implies
\begin{eqnarray}
\mathcal{X}_i &=& \frac{1}{4} \left( \frac{\mathcal{Y}_j-\mathcal{Y}_k}{\mathcal{X}_j-\mathcal{X}_k}  \right)^2 - \mathcal{X}_j - \mathcal{X}_k \, ,
\end{eqnarray}
where $i,j,k$ take three distinct values in the set $\{ 1,2,3 \}$. Thus, we see that there are two possibilities: either the superpotential is zero 
\begin{eqnarray}
\mathcal{X}_1+\mathcal{X}_2+\mathcal{X}_3 &=& 0 \, ,
\label{su3zero}
\end{eqnarray}
or we must have that 
\begin{eqnarray}
\mathcal{X}_1 = \mathcal{X}_2 = \mathcal{X}_3 \, . 
\label{su3oneistwoisthree}
\end{eqnarray}
We split the analysis of the extrema according to these two cases. Firstly, we consider the case in which we have the equality (\ref{su3oneistwoisthree}). This equation, together with extremality 
shows that $Z_1 \equiv Z_2 \equiv Z_3$ modulo a period. This implies that all $Z_i$ equal a non-trivial third of a period of the torus, and gives rise to $4$ inequivalent vacuum solutions, which are the known massive vacua \cite{Dorey:1999sj}. The superpotential is three times the Weierstrass function evaluated at a third period. 

Let us return then to the first possibility, which is that the superpotential is zero, equation (\ref{su3zero}).
By eliminating the variables $\mathcal{Y}_i$ through the curve equation and extremality, we obtain two
equations characterizing the massless branch
\begin{eqnarray}
\label{massless_branch}
\mathcal{X}_1+\mathcal{X}_2+\mathcal{X}_3 &=& 0
\nonumber \\
\mathcal{X}_1^2+\mathcal{X}_2^2+\mathcal{X}_3^2 &=& \frac{g_2}{2} \, .
\end{eqnarray}
These equations are gauge invariant. 
Solving for the variables $\mathcal{Y}_i$ will provide a further double cover
of this space. 
Moreover, we mod out the space by the discrete gauge symmetry $S_3$,
which exchanges the three indices $\{ 1,2,3 \}$ of the variables $\mathcal{X}_i$
(and flips the sign of the variables $\mathcal{Y}_i$ if the permutation is odd, exchanging
the two sheets of the cover).
We can parametrize the curve
more explicitly by eliminating more variables. A description of the curve
in terms of two variables is
\begin{eqnarray}
\mathcal{X}_1^2 + \mathcal{X}_1 \mathcal{X}_2 + \mathcal{X}_2^2 &=&  \frac{g_2}{4} 
\, .
\label{constraint}
\end{eqnarray}
This equation parametrizes a complex line.

Note that at the values $\tau_0$ of the complexified gauge coupling where the fourth Eisenstein series $g_2$ is zero, the complex line has a singular point at $\mathcal{X}_1=\mathcal{X}_2=0$. 
The singularity is a crucial feature of the massless branch. The zeros of $g_2$ in the $\tau$ upper half-plane are exactly the $SL(2,\mathbb{Z})$ images of $\tau_0=e^{2\pi i/3}$, which is the only zero of $g_2$ in the fundamental domain.\footnote{It is easy to show that $\tau_0=e^{2\pi i/3}$ is a zero of $E_4 = \frac{3}{4 \pi^4} g_2$ using $E_4 (\tau_0) = E_4 (\tau_0+1)=E_4(-1/\tau_0)= \tau_0^4 E_4(\tau_0) = \tau_0 E_4(\tau_0)$. There remains to show that there is no other zero. We use the formula $\textrm{ord}_{i\infty} + \frac{1}{2} \textrm{ord}_{i} + \frac{1}{3} \textrm{ord}_{\tau_0} + \sum \textrm{ord}_{\tau} = k/6$, valid for any modular form of weight $2k$. At weight $2k=4$ the formula gives $\textrm{ord}_{\tau_0} = 1$ and there can be no other zero.} Thus, at these couplings the massless branch develops a singularity. These are elliptic points of order three.

Finally, we note that the conditions that all $\mathcal{X}_i$ be equal (which is valid for the 4 vacua associated
to third periods), and that the superpotential vanish can both be satisfied at the singular points. More precisely, for each given singular coupling in the $SL(2,\mathbb{Z})$ orbit of $\tau_0$, one of the four formerly massive vacua becomes massless and joins the massless branch. The fact that a massive vacuum becomes massless at this coupling may indicate
a higher order critical point, and the existence of an
interacting ${\cal N}=1$ superconformal field theory. The value of the critical coupling points towards a natural candidate for this theory, which is the Argyres-Douglas  theory \cite{Argyres:1995jj} broken to ${\cal N}=1$  \cite{Terashima:1996pn}. 

In fact, the analysis of $\mathcal{N}=2$ $SU(3)$ theory with an adjoint hypermultiplet reveals that the Seiberg-Witten curve has eight 
cusps \cite{Donagi:1995cf}. When we analyze the cusps at values of the moduli such that they coincide with vacua that would be massive at generic coupling, we find that the number of cusps reduces to four.\footnote{The operation $S^2$
discussed in \cite{Donagi:1995cf} acts
trivially in this circumstance.} Of these four cusps, one is associated to a pure $SU(3)$ theory, and the other three correspond to a $SU(2)$ theory with a massless fundamental hypermultiplet at a (generalized) Argyres-Douglas point \cite{Argyres:1995xn}.  
Since the ${\cal N}=1^\ast$
massive vacua are invariant under $\Gamma_0(3)$, we can classify the singular couplings $\tau$ into $\Gamma_0(3)$ cosets of
$SL(2,\mathbb{Z})$ according to which massive vacuum becomes massless at the given singular coupling. We find that 
at $\tau = \frac{-1}{\tau_0 +2}$  the Higgs vacuum becomes massless,
while at $\tau = \tau_0 +2$ the confined vacuum situated on the imaginary axis becomes massless,
at $\tau = \tau_0 +1 = e^{ \pi i/3}$, its T-dual and at $\tau_0 = e^{2 \pi i/3}$ the third confined vacuum.
Thus, the  $SL(2,\mathbb{Z})$ action on these Argyres-Douglas singularities coincides with the 
action of the duality group on the four massive vacua of the $\mathcal{N}=1^\ast$ theory. From the action of the T-transformation,  we can identify the confined vacua with the $SU(2)$ cusps
and the Higgs vacuum with the pure $SU(3)$ cusp
\cite{Donagi:1995cf}. Our analysis provides a concrete picture for how the transformation properties of the massive phases are locked with the duality properties of the cusps.

At generic coupling $\tau$, the duality properties of the massive vacua  are well-known. We find that the massless branch, in the description in terms of
elliptic curve variables, is invariant under the action of the
T-transformation, since the fourth Eisenstein series is. Moreover, under the S-transformation, the variables
$\mathcal{X}_i$ transform with weight two, as one expects from their definition in terms of the elliptic Weierstrass function.
Thus, the branch is self-dual under the full modular group (or more precisely, is mapped to an equivalent, scaled branch at dual coupling).

\subsubsection*{The Massless Branch in the Toroidal Variables}
The description of the massless branch was straightforward in terms of gauge invariant polynomials of the variables
$\mathcal{X}_i$. Still, we can ask for the description of the massless branch of vacua in terms of the extrema
of the integrable system, parametrized by the coordinates $Z_i$ (namely, the complexified Wilson lines), at finite
coupling $\tau$. That description too can be obtained, but it demands further effort. We can for instance work with the following parameterization of the massless branch   
\begin{eqnarray}
\label{lambda_param}
\mathcal{X}_1 &=& \frac{i\sqrt{g_2}}{2 \sqrt{3}} \left( \lambda - \frac{1}{\lambda}\right) \nonumber  \\
\mathcal{X}_2 &=& \frac{\sqrt{g_2}}{4 } \left[ \left( \lambda + \frac{1}{\lambda}\right) - \frac{i}{\sqrt{3}} \left( \lambda - \frac{1}{\lambda}\right) \right] \nonumber \\
\mathcal{X}_3 &=&\frac{\sqrt{g_2}}{4 } \left[ -\left( \lambda + \frac{1}{\lambda}\right) - \frac{i}{\sqrt{3}} \left( \lambda - \frac{1}{\lambda}\right) \right] \, ,
\end{eqnarray}
for $\lambda \in \mathbb{C}^\ast$. 
However, we still have to take into account both the fact that we have a double cover (when we solve for $\mathcal{Y}_i$)
as well as the action of the 
Weyl group to faithfully describe  the branch of vacua. The Weyl group has generators that exchange two distinct coordinates, $Z_i \leftrightarrow -Z_j$ (while also changing the sign of the third coordinate).  This translates into identifications on our parameter space:
\begin{itemize}
\item $Z_2 \leftrightarrow -Z_3$ corresponds to $\lambda \leftrightarrow - \frac{1}{\lambda}$. 
\item $Z_1 \leftrightarrow -Z_2$ corresponds to $\lambda \leftrightarrow e^{-2\pi i /3} \lambda$. 
\item $Z_1 \leftrightarrow -Z_3$ corresponds to $\lambda \leftrightarrow e^{2\pi i /3} \lambda$ \, ,
\end{itemize}
and each transformation exchanges the two sheets of the $\mathcal{Y}$ cover.
Hence the massless branch is a double cover of the sphere parametrized by $\lambda$. 
We excise the points $\lambda=0$ as well as the point $\lambda=\infty$, because the superpotential blows up in these
points. This indicates the enhancement of gauge symmetry, and the breakdown of the
effective superpotential description at these $\mathbb{Z}_3$ fixed points.
A fundamental domain for $\lambda$ is given by the following region: $|\lambda| \leq 1$ and $\pi/6 \leq \arg \lambda \leq 5\pi/6$ with a $\mathbb{Z}_2$ identification of the borders of the unit disk as well as of the two rays on the boundary. 

We now wish to distinguish between two physically distinct sets of configurations. 
They are characterized by the way they behave under the charge
conjugation symmetry of the gauge theory. Conjugation acts by exchanging $Z_1 \leftrightarrow Z_2$, which is a global
symmetry of the gauge theory, inherited by the low-energy effective superpotential. When we have the equality
$\mathcal{X}_1=\mathcal{X}_2$ (or a permutation thereof), we can either have $Z_1=-Z_2$ or $Z_1=Z_2$, modulo 
the periodicity of these variables. The first case corresponds to a fixed point of the local Weyl symmetry group,
and it leads to a singular term in the effective superpotential, indicating the enhancement of the gauge group (i.e. the
fact that we leave the Coulomb branch). We exclude this singular configuration from our analysis.
The second case indicates a fixed point of the charge conjugation symmetry. This occurs  when $\lambda^6=-1$. When there is no equality between any
of the variables $\mathcal{X}_i$, we are at a less symmetric point on the massless branch.
These two regimes will lead to a qualitatively different solution for the variables $Z_i$ as we show in detail below.

We would like to solve equation (\ref{gauge_inv_var}) for the complexified Wilson lines $Z_i$.
The solution relies on inverting the Weierstrass function. The techniques for performing this inversion 
were presented in \cite{EZ} by Eichler and Zagier in their analysis of the zeros of the Weierstrass $\wp$ function. These authors also study the solutions to 
 the equation $\wp(Z) =\mathcal{X}(\tau)$ where $\mathcal{X}(\tau)$ is a (e.g. meromorphic) 
modular form of weight $2$. Our equation does not  fit this mold -- the Weierstrass function is equal to the 
square root of a modular form of weight 4. Still, we can apply the bulk of the Eichler-Zagier methods.
The Eichler-Zagier technique  for inverting the Weierstrass function consists of two parts. On the one
hand, since the argument $Z$ is multi-valued due to the periodicity of the Weierstrass function, it is useful to derive
with respect to the modular parameter $\tau$ twice, to eliminate this ambiguity. The two integration constants
that one subsequently needs can be determined by matching the semi-classical limits.
On the other hand, one inverts the equation through integration of the defining equation for the elliptic curve
\begin{equation}
\wp (Z ; \tau) = \mathcal{X}(\tau) \qquad \Longleftrightarrow \qquad Z = \pm \frac{\sqrt{3}}{2 \pi} \int\limits_{\frac{3}{\pi^2} \mathcal{X}(\tau)}^{\infty} \frac{dt}{\sqrt{t^3 - 3 E_4 (\tau) t - 2 E_6 (\tau)}} \, . 
\label{directintegration}
\end{equation}
{From} this equation, we determine the second derivative with respect to $\tau$, by multiple application of the
Ramanujan identities for the derivatives of the Eisenstein series. The calculation is presented in pedagogical
detail in \cite{EZ} and results in the equality
\begin{eqnarray}
\pm \frac{\mathrm{d}^2 Z}{\mathrm{d} \tau^2} &=&  \Big( 4 \pi^2 (g_3 - 4 \mathcal{X}^3 +g_2 \mathcal{X}) D_3 D_6 \mathcal{X} + 2 \pi^2 (12 \mathcal{X}^2-g_2) (D_6 \mathcal{X})^2 +
(6 g_3 \mathcal{X} +g_2^2/3) D_6 \mathcal{X}
\nonumber \\
& & + \frac{1}{72 \pi^2} (12 g_2 \mathcal{X}^4+3 g_2^2 \mathcal{X}^2+
6 g_2 g_3 \mathcal{X} - g_2^3 + 27 g_3^2) \Big) /  (4 \mathcal{X}^3-g_2 \mathcal{X}-g_3)^{\frac{3}{2}} \, ,
\label{expansionderivative}
\end{eqnarray}
where the function $\mathcal{X}$ acts as a seed, 
and the modular covariant derivative is given by $D_n = q \partial_q - \frac{1}{n} E_2$.
The integration constants are fixed by taking the semi-classical limit of the formula (\ref{directintegration}).
Here, we will add a point to the analysis in \cite{EZ}, by exhibiting a special case
of the limiting formula, which is also physically distinct.
We define the variable $\mathcal{X}_{i \infty} = \mathcal{X} (\tau \rightarrow i \infty)$. If $\mathcal{X}_{i \infty} \neq - \frac{\pi^2}{3}$, the semi-classical limit is given by  \cite{EZ}
\begin{equation}
Z (\tau \rightarrow i \infty) = \frac{1}{2} \pm \frac{1}{2 \pi i } \log \frac{1+\sqrt{\frac{2}{3}-\frac{1}{\pi^2}  \mathcal{X}_{i \infty} }}{1-\sqrt{\frac{2}{3}-\frac{1}{\pi^2} \mathcal{X}_{i \infty}}}
\label{EZasymptotic1}
\end{equation}
while for the case $\mathcal{X}_{i \infty} = - \frac{\pi^2}{3}$  it is
\begin{equation}
Z (\tau \rightarrow i \infty) = \pm \frac{1}{4} + \frac{\tau}{2} \, . 
\label{EZasymptotic2}
\end{equation}
The latter case occurs when the $\mathcal{X}_i$ are at a charge conjugation fixed point, i.e. a fixed point of the global $\mathbb{Z}_2$ symmetry. Note that the limit formula (\ref{wp_asymptotic}) shows that this case is common.
Let us nevertheless first concentrate on the case in which all the variables $\mathcal{X}_i$ are different, and construct
the solution for the variables $Z_i$. We then come back to the global $\mathbb{Z}_2$ fixed point.

\subsubsection*{The points $\lambda^6 \neq -1$ }

The point $\lambda = 1$, for instance, is representative for  all $\lambda$
not at a $\mathbb{Z}_2$ fixed point. In this case, we have 
\begin{eqnarray}
\mathcal{X}_1 &=& 0 \nonumber  \\
\mathcal{X}_2 &=& \frac{\sqrt{g_2}}{2}  \nonumber  \\
\mathcal{X}_3 &=& - \frac{\sqrt{g_2}}{2} \nonumber  \, ,   
\end{eqnarray}
and the formulas (\ref{EZasymptotic1}) and (\ref{expansionderivative}) from \cite{EZ} apply. We can for instance
write the solution as a series expansion $q$ at large imaginary $\tau$
\begin{eqnarray}
\pm \pi Z_1 &=& \frac{1}{2} \left(\pi -i \cosh ^{-1}(5)\right)+36 i \sqrt{6} q+6588 i \sqrt{6} q^2 + ...
\label{explicitseries}
\\
\pm \pi Z_2 &=& \left(\frac{\pi }{2}-i \tanh^{-1}\left(\frac{1}{\sqrt{2}}-\frac{1}{\sqrt{6}}\right)\right)+\left(-36 i \sqrt{2}-18 i \sqrt{6}\right) q+\left(1188 i \sqrt{2}-3294 i \sqrt{6}\right) q^2 + ... \nonumber \\
\pm \pi Z_3 &=& - \pi -\frac{i}{2} \log \left(1-\frac{6}{3+\sqrt{6+3 \sqrt{3}}}\right)+\left(36 i \sqrt{2}-18 i \sqrt{6}\right) q+\left(-1188 i \sqrt{2}-3294 i \sqrt{6}\right) q^2 +... \nonumber
\end{eqnarray}
The series that we obtain has a finite radius of convergence.
The integration formula (\ref{directintegration}) is valid at any modular parameter $\tau$.
In this explicit solution (\ref{explicitseries}), we can choose a sign for each $Z_i$, consistently  with the constraint $Z_1+Z_2+Z_3 \equiv 0$. Thus, we see that we must pick the same sign for all $Z_i$ -- there are two solutions. The solutions are invariant under $T$-duality. This implies that they are also $S$-invariant, since $1=(ST)^3=S^3=S$. 
The semi-classical limit of these vacua lies in the class $J_0 = \{1,2\}$. The semi-classical limit of the massless
branch that contains these vacua can be obtained by setting $Y_{1,2}=0$ and taking the corresponding limit on the 
equations (\ref{massless_branch}) parameterizing the branch.

\subsubsection*{The $\mathbb{Z}_2$ symmetric points }
We return to the $\mathbb{Z}_2$ symmetric values of $\lambda$ which lie at $\lambda^6=-1$.
Let us further concentrate on the case where the equality $\mathcal{X}_2=\mathcal{X}_3$ holds. Note that the condition we impose is duality invariant. The solutions will therefore transform into each other under the $SL(2,\mathbb{Z})$ action of the duality group. We solve for the coordinates $Z_i$ at these particular points. {From} equation (\ref{lambda_param}) we read that the equality $\mathcal{X}_2=\mathcal{X}_3$ translates into $\lambda^2 = -1$, which implies that we can focus on the  two points  $\lambda = \pm i$. 

At the value $\lambda = -i$, we have to solve the equations:
\begin{eqnarray}
\mathcal{X}_1 &=& \sqrt{\frac{g_2}{3}} \label{lambda-i1} \\
\mathcal{X}_2 &=& - \frac{1}{2}\sqrt{\frac{g_2}{3}} \label{lambda-i2}   \\
\mathcal{X}_3 &=& - \frac{1}{2}\sqrt{\frac{g_2}{3}} \label{lambda-i3} \, .
\end{eqnarray}
We begin with the second equation (\ref{lambda-i2}), for which the equality (\ref{EZasymptotic2}) gives the asymptotic value $Z_2 (i \infty)= \pm \frac{1}{4} + \frac{\tau}{2}$. We define \begin{equation}
Z_2 = \frac{1}{4} + \frac{\tau}{2} - \frac{1}{2}\alpha(q)
\end{equation}
to be the solution of (\ref{lambda-i2}) such that $\alpha(q)$ has semi-classical behavior 
\begin{equation}
\alpha (q) = \frac{8}{\pi} q^{\frac{1}{2}} + O\left(q^{3/2} \right)
\label{alphaexpansion}
\end{equation}
and is analytic along the line $i \mathbb{R}_+^*$. Note that from the equation, we can compute the Fourier expansion to arbitrary order.\footnote{We have 
 \begin{eqnarray}
\pi \alpha (q) &=& 8 q^{\frac{1}{2}}-
\frac{1088             q^{  \frac{3}{2}  } }{3}
+\frac{198288          q^{  \frac{5}{2}  } }{5}
-\frac{39006080        q^{ \frac{7}{2}   } }{7}
+\frac{7975383560      q^{ \frac{9}{2}  } }{9} \nonumber \\
& & 
-\frac{1669600216512   q^{ \frac{11}{2}  } }{11}
+\frac{355119960987280 q^{ \frac{13}{2}  } }{13} + \dots
 \end{eqnarray}
} 
Next,  we consider the first equation (\ref{lambda-i1}). The asymptotic behavior of its solutions $\pm Z_1$ is now given by equation (\ref{EZasymptotic1}), and it is $Z_1 (i \infty) = \frac{1}{2}$. The exact solution involves the function $\alpha (q)$ just defined, $Z_1 = \frac{1}{2} \pm \alpha (q)$ as a consequence of the doubling formula
\begin{equation}
\wp \left(\frac{1}{2} \pm \alpha \right) = -2 \wp  \left( \frac{1}{4} + \frac{\tau}{2} - \frac{1}{2}\alpha(q)\right) +  \frac{6 \wp  \left( \frac{1}{4} + \frac{\tau}{2} - \frac{1}{2}\alpha(q)\right)^2 - \frac{1}{2} g_2}{4 \wp ' \left( \frac{1}{4} + \frac{\tau}{2} - \frac{1}{2}\alpha(q)\right)^2} = \sqrt{\frac{g_2}{3}} \, . 
\end{equation}
The relative sign is determined by the requirement that $Z_3=-Z_1-Z_2$ be a solution of (\ref{lambda-i3}). Therefore we have found two inequivalent vacua at $\lambda = -i$:
 \begin{equation}
 (Z_1,Z_2) = \left(\frac{1}{2} + \alpha , \frac{1}{4} + \frac{\tau}{2} - \frac{\alpha}{2} \right) \qquad (z_1,z_2,z_3) = \left(  \frac{3}{4} + \frac{\tau}{2}  + \frac{\alpha}{2} , \frac{1}{4} + \frac{\tau}{2}  - \frac{\alpha}{2}, 0 \right)  
 \label{vacuumlambda-i1}
 \end{equation}
 and 
  \begin{equation}
 (Z_1,Z_2) =  \left(\frac{1}{2} - \alpha , \frac{1}{4} + \frac{\tau}{2} + \frac{\alpha}{2}\right) \qquad (z_1,z_2,z_3) = \left(  \frac{3}{4} + \frac{\tau}{2}  - \frac{\alpha}{2} , \frac{1}{4} + \frac{\tau}{2}  + \frac{\alpha}{2}, 0 \right)  \, . 
  \label{vacuumlambda-i2}
 \end{equation}
 We now turn to the value $\lambda = i$ and proceed similarly. Our task is to solve 
 \begin{eqnarray}
\mathcal{X}_1 &=& - \sqrt{\frac{g_2}{3}} \label{lambdai1} \\
\mathcal{X}_2 &=&  \frac{1}{2}\sqrt{\frac{g_2}{3}} \label{lambdai2}   \\
\mathcal{X}_3 &=&  \frac{1}{2}\sqrt{\frac{g_2}{3}} \, .  \label{lambdai3} 
\end{eqnarray}
We define $\frac{1}{2} - \beta(\tau)$ to be the solution of (\ref{lambdai2}) with semi-classical behavior
\begin{equation}
\beta(q) = \frac{i}{2\pi} \log \left( 2 + \sqrt{3} \right) + O (q)
\end{equation}
and demand analyticity on $i \mathbb{R}_+^*$,\footnote{We have the further expansion 
\begin{eqnarray*}
i \pi \beta &=& - \log \left( \frac{1 + \sqrt{3}}{\sqrt{2}}\right) + 12 \sqrt{3} (q-87 q^2+11080 q^3-1671095 q^4+\frac{1384694994 q^5}{5} \\
 & & -48732765432 q^6+\frac{62575601740112 q^7}{7}-1690589139219255
   q^8+327268705474374265 q^9+...) \, .
\end{eqnarray*} } 
and again we find that $2 \beta (q)$ is a solution of (\ref{lambdai1}) using the duplication formula for the Weierstrass function. The signs are determined as previously, and we conclude that a solution is
 \begin{equation}
\label{masslessSU3_03}
(Z_1,Z_2)=(2\beta , \frac{1}{2} - \beta) \qquad(z_1,z_2,z_3) = \left(\frac{1}{2} + \beta , \frac{1}{2} - \beta ,0\right)  \, . 
\end{equation}
As before, we could flip the sign in front of $\beta$ in this expression, but this would lead to an equivalent vacuum.
We have only one vacuum at $\lambda = i$. 

While for generic $\lambda$ the action of $T$-duality and as a consequence $SL(2,\mathbb{Z})$ duality on the vacua was trivial, here we see, e.g. from the expansion (\ref{alphaexpansion}), that $T$-duality exchanges the two vacua (\ref{vacuumlambda-i1}) and (\ref{vacuumlambda-i2}). As a consequence $S$-duality will act as well. We devote the next paragraph to a detailed study of these dualities.

\subsubsection*{Dualities at the $\mathbb{Z}_2$ symmetric points}
In the course of our analysis we have found the four solutions of the equation 
\begin{equation}
\wp (z)^2 = \frac{g_2}{2} 
\label{ellipticbeauty}
\end{equation}
that we can gather in a vector 
\begin{equation}
V (\tau) = \begin{pmatrix}
\frac{1}{2} + \alpha (\tau)  \\
\frac{1}{2} - \alpha  (\tau) \\
2 \beta  (\tau) \\
- 2 \beta  (\tau) 
\end{pmatrix} \, ,
\end{equation}
which can be interpreted as a vector-valued and multi-valued modular form \cite{EZ}. The word multi-valued  here refers to the fact that these quantities are defined up to periods of the Weierstrass function. This vector transforms under $SL(2,\mathbb{Z})$ according to 
\begin{equation}
V(\tau) \xrightarrow{T} V(\tau +1) = 
\begin{pmatrix}
0&1&0&0\\
1&0&0&0\\
0&0&1&0\\
0&0&0&1
\end{pmatrix}
V (\tau) \, ,
\end{equation}
and 
\begin{equation}
V(\tau) \xrightarrow{S} V \left( \frac{-1}{\tau} \right) =  \frac{-1}{\tau} \left[
\begin{pmatrix}
0&0&1&0\\
0&0&0&1\\
0&1&0&0\\
1&0&0&0
\end{pmatrix}
V (\tau) 
-
\begin{pmatrix}
\tau\\
0\\
0\\
1
\end{pmatrix}
\right] \, ,
\end{equation}
where we have used the analyticity along $i \mathbb{R}_+^*$ to fix the periodic dependence. Thus, we have a weight $-1$ modular form up to periodicity. Let's call $T$ and $S$ the matrices that appear in these equations and that are associated to the two generators of $SL(2,\mathbb{Z})$. The periodicity is linear in the modular parameter $\tau$, such that again,
if we take two derivatives with respect to $\tau$, this ambiguity drops out, and
we find a vector valued modular form of weight 3:
\begin{equation}
 V''(\tau +1) = T V'' (\tau) \, ,  \qquad
  V '' \left( \frac{-1}{\tau} \right) = (-\tau)^3 S V'' (\tau)  \, .
\end{equation}
The method of \cite{EZ} gives the explicit solution
\begin{equation}
\psi (\tau) = V''(\tau) \times \left[ \left(\frac{g_2(\tau)}{3}\right)^{\frac{3}{2}} - g_3 (\tau) \right] ^{\frac{3}{2}}
\end{equation}
and each component of the vector $V$ is given as
\begin{eqnarray}
\pm V'' =  \frac{4 \pi^2 (g_3 -\phi_\pm^3) D_3 D_6 \phi_\pm + 18 \pi^2  \phi_\pm^2 (D_6 \phi_\pm)^2 +3 \phi_\pm (2g_3+\phi_\pm^3)D_6 \phi + \frac{1}{8 \pi^2} (4 \phi_\pm^6 + 2 g_3 \phi_\pm^3 + 3g_3^2 ) }{  ( \phi_\pm^3-g_3)^{\frac{3}{2}}}
\nonumber 
\end{eqnarray}
where the seed $\phi_\pm$ is a branch of the square root of the Eisenstein series $\phi_\pm = \pm \sqrt{\frac{g_2}{3}}$, 
and the 4 components of $V''$ correspond to the 4 possible choices of signs (on the left, and on the right hand side independently). 
We give the first few terms (the first line is obtained from $\phi_+$ and the second line from $\phi_-$): 
\begin{equation}
V''(\tau) = \pm 8 \pi
\begin{pmatrix}
q^{\frac{1}{2}}-408 q^{\frac{3}{2}}+123930 q^{5/2}-34130320 q^{7/2} \\
12 i \sqrt{3} (q-348 q^2+99720 q^3-26737520 q^4+...)
\end{pmatrix}
= \pm
\begin{pmatrix}
\alpha '' \\
2 \beta ''
\end{pmatrix}
\end{equation}
and
\begin{equation}
\psi (\tau) \propto -q^2+336 q^3-94824 q^4+25238080 q^5-6506938620 q^6 +... \, .
\end{equation}
After double integration, this characterizes the $q$ expansion of $\alpha,\beta$, and therefore analytically completes the series we obtained previously.
We further analytically continue the functions $\alpha$ and $\beta$  in the double cover of the upper half plane. The triplet of solutions to the equation becomes degenerate at the zeros of $E_4$.
Note that we can switch branch for the seed by rotating around the zero $\tau_0=e^{2 \pi i/3}$ of the weight
4 Eisenstein series. As a consequence, this operation flips $\alpha$ and $\beta$, and this introduces
a monodromy amongst the sheets of massless vacua in the elliptic integrable system parameterization.

\subsubsection*{Summary Remarks}
We recapitulate the duality diagram for both the massive and massless extrema of the $su(3)$ integrable system.\footnote{
We repeat that the global aspects of the gauge group  can  be taken into account  by carefully
treating the subgroup of $\mathbb{Z}_3$ which one chooses as center, and the possible electro-magnetic line operators in the theory, which have consequences on the periodic identifications of variables.}
We have four massive vacua, of which two are self-$S$-dual, and
two are mutually $S$-dual. They form a singlet and a triplet under $T$-duality.
We have one massless branch which is duality invariant in the elliptic curve variables.\footnote{There is a point this branch which is S-duality and
T-duality invariant. It is given by $(z_1,z_2,z_3)=(1/2,\tau/2,(1+\tau)/2)$, and is mentioned in \cite{Aharony:2000nt}.
} 
We note that the semi-classical limit that allows for the Higgs vacuum, also sees the massless branch.
  \begin{table}[H]
\centering
\begin{tabular}{|c|c|c|c|}
\hline
Partition & $J_0$ & Unbroken & Vacua \\
\hline 
$1+1+1$ & $\emptyset$ & $su(3)$ &  3 confining vacua of pure $\mathcal{N}=1$\\
$2+1$ & $\{1\}$ or $\{2\}$ & $u(1)$ & the massless branch  \\
$3$ & $\{1,2\}$ & 1 & 1 Higgs vacuum + the massless branch \\
\hline
\end{tabular}
\caption{Summary of vacua for $su(3)$}
\end{table} 
\noindent
There is a more intricate description of the massless branch in terms of the elliptic integrable system variables, which allows to follow the duality map on the massless vacua point by point. For the extremal positions of the
 massless vacua in terms of the complexified Wilson lines,
 we have exhibited a point of monodromy on the boundary of the fundamental domain, and in particular, the elliptic point of order 3 of the $SL(2,\mathbb{Z})$ action on the upper half plane. This point is a singular point for the manifold of
 massless vacua.  
 It is reminiscent of the point
 of monodromy in the interior of the fundamental domain for two massive vacua of the $so(8)$ theory \cite{Bourget:2015cza}.

 \subsection{The Gauge Algebra $su(4)$}
We have obtained a complete picture of the massive and massless vacua of the $su(3)$ theory.
In this subsection, dedicated to  the gauge algebra $su(4)$, we will only perform a partial analysis. 
 Recall that for $su(4)$, the partition $1+1+1+1$ gives rise to an affine Toda limit with
 four solutions, which correspond to the four confining vacua of pure ${\cal N}=1$.
 The partition $2+2$ corresponds to the choice $J_0=\{\alpha_1,\alpha_3 \}$ which 
 gives rise to two trigonometric $A_1$ systems with one  solution,
 and the two remaining variables then form an $A_2$ affine Toda system which has
 two solutions, corresponding to the two confining vacua of the unbroken $su(2)$ gauge
 algebra.
 Finally, we have the partition $4$ which corresponds to the 
 trigonometric $A_3$ system. This gives rise to a real extremum which represents
the fully Higgsed vacuum. We have a total of seven massive vacua.\footnote{There are other complex extrema of the trigonometric integrable systems.} Our focus in the following are
massless vacua.
A natural way to generate massless vacua is by exploring 
the partitions $2+1+1$ and $3+1$ which leave unbroken abelian gauge group factors. We will consider them in turn.
Let us first remind the reader that the superpotential for the $su(4)$ gauge algebra is
 \begin{equation}
 \label{superpotA3}
 \mathcal{W}_{A_3} (Z_1,Z_2,Z_3) = \wp (Z_1) + \wp (Z_2) + \wp (Z_3) + \wp (Z_1+Z_2)+ \wp (Z_2+Z_3)+ \wp (Z_1+Z_2+Z_3) 
 \, ,
 \end{equation}
 in variables $Z_i$ which are coefficients of fundamental weights.

 \subsubsection*{The Partition $2+1+1$}
The partition $2+1+1$ corresponds to a choice of simple root system $J_0 = \{ 1 \}$. 
The centralizer algebra is $su(2) \oplus u(1)$ in this case. 
We may intuit the existence of two massless branches on the basis of this centralizer algebra. We will approach
them through the semi-classical limit.

\begin{figure}
 \centering
 \includegraphics[width=0.2\textwidth]{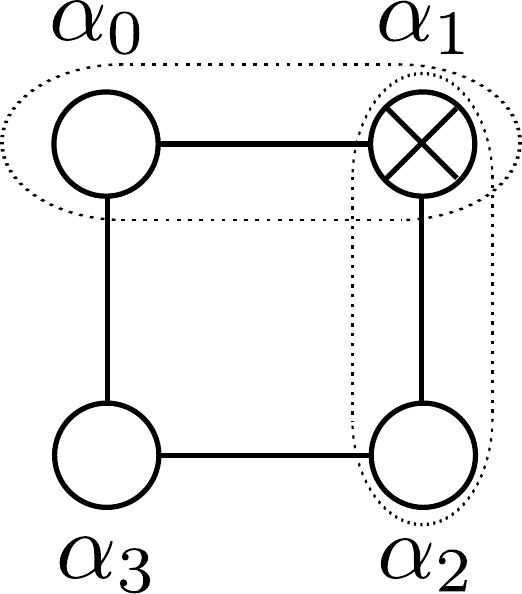}
 \caption{Affine Dynkin diagram for algebra $A_3$ and partition $2+1+1$. The crossed simple root corresponds to the set $J_0 =\{1\}$, and the dotted lines encircle roots that cancel the contribution of the simple roots $\alpha_0$ and $\alpha_2$. } 
 \label{DynkinA3Affine}
 \end{figure}

In this limit, we have one leading trigonometric root that sets $Y_1=0$. To find the other shifts, we use a heuristic argument based on cancellations that happen in the superpotential at first order in perturbation theory, where $X_1=\frac{1}{2}$. Such cancellations occur at level $n=1$ in (\ref{expansionTrigToda}) in the sum over the roots that have a non-vanishing scalar product with $\hat{Y}$. As illustrated on the affine Dynkin diagram on figure \ref{DynkinA3Affine}, the contributions of $\alpha_0$ and $\alpha_0 + \alpha_1$ will cancel each other in (\ref{expSystem}), as well as $\alpha_2$ and $\alpha_1 + \alpha_2$, and all other roots involving $\alpha_0$ and $\alpha_2$ are suppressed in the semi-classical limit. Therefore in order to stabilize the system we use the next level $n=2$ for these roots, which then contribute with factors of $q^{2Y_0}$ and $q^{2Y_2}$. On the other hand $\alpha_3$ contributes with a factor $q^{Y_3}$. Stabilization at leading order requires that these powers of $q$ be equal, and we therefore propose the following ansatz: 
\begin{equation}
\label{Y221}
\begin{cases}
Y_1 = 0 \\
2Y_0=2Y_2=Y_3 \\
Y_0+Y_1+Y_2+Y_3=1
\end{cases}
\implies
\begin{cases}
Y_0=\frac{1}{4} \\ Y_2=\frac{1}{4} \\ Y_3=\frac{1}{2} 
\end{cases} \, .
\end{equation}
To obtain the subleading Toda potential, we need to take into account the non-perturbative corrections to the value $X_1=\frac{1}{2}$. Firstly, we expand the superpotential (\ref{superpotA3}) around the leading order values (\ref{Y221}), assuming that the variation $\delta X_1$ of $X_1$ behaves as a power of $q$. The dominant terms are 
\begin{eqnarray}
\frac{1}{\pi^2} \mathcal{W} \left( \frac{1}{2} + \delta X_1 , \frac{\tau}{4} + X_2 , \frac{\tau}{2} + X_3 \right) &=& -1+\pi ^2 \delta X_1^2 + 8 i \pi \delta X_1 q^{\frac{1}{4}} \left(  e^{2 i \pi  X_2} -  e^{-2 i \pi  X_2-2 i \pi  X_3} \right)
\, .
\end{eqnarray}
There is a linear term in the non-perturbative correction $\delta X_1$ which determines its value at order $q^{\frac{1}{4}}$: 
\begin{equation}
(\delta X_1)_{\frac{1}{4}} = \frac{4i}{\pi} q^{\frac{1}{4}} \left( e^{- 2 i \pi (X_2 + X_3)} - e^{2 i \pi X_2}  \right) \, .
\end{equation}
This confirms that the value $X_1=1/2$ has to be corrected, and that the superpotential should be expanded around the point
shifted by $(\delta X_1)_{\frac{1}{4}}$: 
\begin{eqnarray}
\label{expansion_su4_211}
\frac{1}{\pi^2} \mathcal{W} \left( \frac{1}{2} + (\delta X_1)_{\frac{1}{4}} + \delta X_1 , \frac{\tau}{4} + X_2 , \frac{\tau}{2} + X_3 \right) &=& -1 - 4q^{\frac{1}{2}} \left( 9 e^{-2 \pi i X_3} +e^{2 \pi i X_3}  \right) \, . 
\end{eqnarray}
We conclude that $X_3$ can be determined at this step,
and we find 
\begin{equation}
X_3 = - \frac{i \log 3}{2\pi} + 
\frac{1}{2} \mathbb{Z} \, . \label{delta3leading}
\end{equation}
A longer calculation at higher order shows that $X_3$ in turn receives non-perturbative corrections, starting
at order $q^{\frac{1}{2}}$. Taking into account this second step in our non-perturbative staircase, we find that the 
superpotential becomes independent of $X_2$, and equal to
$-1 \mp 24 \sqrt{q} - 24 q + \dots = \pi^2  E_{2,2} (\pm q^{\frac{1}{2}}) $ where the upper sign
is for the choice of an integer in equation (\ref{delta3leading}) and the lower sign for
a strictly  half-integer choice. Thus, we have  found semi-classical evidence for two
one-dimensional complex manifolds of massless vacua characterized by these superpotentials.
Again, numerical and analytical evidence can be amassed to argue that the superpotentials
are exact.\footnote{One extra technique compared to those presented elsewhere is to find a special
point on the branch, and then prove that at that point the superpotential takes the claimed value.
For the case at hand, for instance, we can concentrate on the point 
\begin{equation}
(Z_1,Z_2,Z_3) = \left( \frac{1}{2} , \frac{\tau}{4} + \frac{\gamma}{2} , \frac{\tau}{2} - \gamma \right) \,  .
\end{equation} 
One then shows that these positions are indeed extremal provided the function $\gamma(\tau)$ 
satisfies the equation 
\begin{equation}
\wp'(2A)+\wp'_2(A)=0 \qquad \textrm{with} \qquad A=\frac{\tau}{4}+\frac{\gamma}{2} \, .
\end{equation} One can then also analytically prove that this vacuum is massless and has the claimed
superpotential. The result is then valid along the whole branch.
}

 \subsubsection*{The Partition $3+1$}
The partition $3+1$ corresponds to the choice of simple roots $J_0 = \{ 1,2 \}$. The unbroken gauge algebra is $u(1)$, and we expect one massless branch. Our ansatz for the linear behavior in $\tau$ is dictated by the choice of the partition which gives $Y_1=Y_2=0$ and by the symmetry of the affine Dynkin diagram which leads to $Y_3=Y_0$. Together with the normalization (\ref{constraintequation}) we obtain $Y_3=\frac{1}{2}$. The trigonometric $A_2$ system arises, and we consider the standard Higgs solution of this system. We thus have $X_1=1/3$ and $X_2=1/3$ to leading order. We obtain a series of non-perturbative corrections to both leading coordinates and find that when we take those into account, the third variable $X_3$ parametrizes a massless branch. We calculate the superpotential to order $q^2$ and it is consistent with the exact value we propose, namely ${\cal W}=-2\pi ^2 E_{2,2}(q)$.

 \subsubsection*{The Duality Diagram}
We have gathered semi-classical and exact data on the $su(4)$ ${\cal N}=1^\ast$ theory.  The duality diagram for the massive states is essentially known, 
with our without the refinement due to the global choice of gauge group and line operator spectrum.
The massless branches fit into the following scheme: we have two massless branches that arise from the partition
$2+1+1$ and they are T-dual. This is consistent with the confining dynamics of the summand $su(2)$ in the unbroken
gauge group.
The branch that we found for partition $3+1$ is self-T-dual. Moreover, the branch with
superpotential  
$\pi^2 E_{2,2} (q^{ \frac{1}{2}}) = -\pi^2 \left( \theta_2 (q )^4+\theta_3 (q)^4 \right)$
 is S-dual to the manifold with superpotential $\pi^2 ( \theta_4(q)^4 + \theta_3(q)^4) = -2 \pi^2 E_{2,2}(q)$.
Similarly, the branch with superpotential $\pi^2 E_{2,2} (-q^{ \frac{1}{2}})$ is self-S-dual.
This is a familiar three-node permutation representation of the $SL(2,\mathbb{Z})$ duality group.
The table below gives a summary of some of the data we laid bare.

 \begin{table}[H]
\centering
\begin{tabular}{|c|c|c|c|}
\hline
Partition & $J_0$ & Unbroken & Vacua \\
\hline 
$1+1+1+1$ & $\emptyset$& $su(4)$ & 4 confining vacua of pure $\mathcal{N}=1$\\
$2+1+1$ & $\{1\}$, $\{2\}$ or $\{3\}$& $su(2) \oplus u(1)$ & branches \quad $\pi^2 E_{2,2}(\pm q^{ \frac{1}{2}})$ \\
$3+1$ & $\{1,2\}$ or $\{2,3\}$ & $u(1)$ & branch \quad $-2 \pi^2 E_{2,2}(q)$  \\
$2+2$ & $\{1,3\}$ & $su(2)$ & 2 massive vacua\\
$4$ & $\{1,2,3\}$ & $1$ & 1 Higgs massive vacuum  \\
\hline
\end{tabular}
\caption{Summary of vacua for $su(4)$}
\end{table}

\subsubsection*{Summary Remarks}
We again found new features in the $su(4)$ analysis. These included a staircase structure for determining the positions,
with each step corresponding to non-perturbative corrections of a given order.
We also discovered an example in which  massless manifolds fit into a non-trivial duality diagram.
These features are expected to be generic. We moreover are bound to find higher dimensional vacuum manifolds when higher
dimensional abelian centralizers are present.

It would be interesting to fully complete the analysis of the vacua of the $su(4)$ theory, in the spirit of the analysis
we performed for $su(3)$. In particular, one can exploit the algebraic approach, and parametrize the extrema in terms
of  algebraic equations. This will allow to determine for instance potential singularities, and possible intersections of manifolds
of massless vacua for given values of the complexified coupling $\tau$.

 \subsection{A Word on the $su(N)$ Theory}
 It should be clear that  in the $su(N)$ case there will be many
 extra solutions compared to the known sublattices of order $N$ of the torus that represent the massive
 vacua. Below, we offer only one rudimentary observation on the massless vacua.
 
The number of massless directions at $\tau \rightarrow i \infty$ in the integrable system equals the number of $U(1)$ factors for the semi-classical vacuum in the Coulomb phase.
 Indeed, all directions that are stabilized by terms with leading behavior a power of $q$ will become
 untethered when we take the semi-classical limit. In our problem, these are all directions
 associated to the (affine) Toda system(s). Thus, in the semi-classical limit, we obtain $|\bar{J}_0|-1$
 flat directions. We can check that this matches the dimension of the semi-classically unbroken abelian factors, when we go to the Coulomb branch.
 
 Recall that a partition $(d_i)$ (which satisfied $\sum d_i=N$) corresponds to strands $(d_i-1)$ in the set of simple
 roots of $A_{N-1}$. The number of coordinates we fix at leading order, using the trigonometric
 integrable system potential, is equal to $\sum_{i} (d_i-1)$. The number of coordinates
 that is unfixed then, in the semi-classical limit (taken on the low-energy effective action) is
 equal to $N-1-\sum_i (d_i-1)= |i|-1$ where $|i|$ counts the number of (non-zero) terms in the partition.
 
 The number of abelian factors in the Coulomb phase is given by the rank of the centralizer of the nilpotent orbit. The rank is equal to 
$\sum_j (r_{j}-1) + k$ where the $r_j$ are defined as the number of times 
the summand $j$ appears in the partition 
and $k= | \{i | r_i > 0 \} | -1$ as in the discussion around (\ref{acentralizer}). If we compute this sum, using $\sum_j r_j = |i|$ and $\sum_j 1=k+1$, we find $|i|-1$, which matches the number of massless modes in the semi-classical limit.

\section{ Discrete Gauge Groups, Wilson Lines and the $G_2$ Theory}
\label{discrete}

In section \ref{acase} we described features of semi-classical limits and massless vacua for the ${\cal N}=1^\ast$ theory with $su(N)$ gauge algebra. We now wish to examine more closely another feature that we encountered in the example of $so(5)$ put forward in section \ref{so5}. We study the appearance of extra massive vacua that occur on $\mathbb{R}^{2,1} \times S^1$, coded in advanced nilpotent orbit theory.
We classified classical vacua of ${\cal N}=1^\ast$ theory on $\mathbb{R}^4$ using nilpotent orbit theory in \cite{Bourget:2015lua}. In this subsection, we wish to stress an important extra feature that comes into play after compactifying the theory on a circle, namely the multiplication of massive vacua through the existence of discrete gauge groups and Wilson lines.

\subsection{Discrete Gauge Groups and Wilson Lines}
We have a space-time equal to  $\mathbb{R}^{2,1} \times S^1$, and parametrize the circle by the coordinate $\theta$ with period $2 \pi$, and we denote by $R$ the radius of the circle. Suppose we fix constant vacuum expectation values $X^j (0)$ for the three ${\cal N}=1$ adjoint chiral multiplets ($j=1,2,3$). Moreover, we have them satisfy $su(2)$ commutation relations, as required for constant scalar field configurations to obey the F-term equations of motion. Let us further suppose that the chosen $su(2)$ algebra has a discrete  centralizer (equal by assumption to the component group of the centralizer). We therefore have a discrete unbroken gauge group.

It should be clear that a discrete component group permits discrete Wilson lines upon compactification. Suppose that a  discrete centralizer of the $sl(2)$ triple contains a non-trivial element $e^{2 \pi i a}$ with $a$ an element in the Lie algebra $\mathfrak{g}$ of the gauge group. Then we can propose semi-classical configurations that are new to the theory compactified on the circle, compared to the theory on $\mathbb{R}^4$. These configurations are:\footnote{Note that we satisfy ordinary boundary conditions. Interesting boundary conditions twisted by outer automorphisms can be imposed for gauge algebras of type $A$, $D$ and $E_6$. See e.g. \cite{Hanany:2001iy,Kim:2004xx}.}
 \begin{eqnarray}
X^j (\theta) &=& \exp(i a \theta) X^j (0) \exp(-i a \theta)\, .
 \end{eqnarray}
 The gauge field component along the circle is fixed to be the constant $A_\theta = \frac{1}{R} a$. These configurations are
 covariantly constant, since the gauge covariant derivative is given by
 $D_\theta X^j = \partial_{\theta} X^j - i [a , X^j ]$, and they are periodic by the fact that the group element $e^{2 \pi i a}$ belongs to
the centralizer of $X^j(0)$:
\begin{eqnarray}
D_\theta X^j (\theta) &=& 0
 \nonumber \\
 X^j(\theta+2 \pi) &=& X^j (\theta) \, .
\end{eqnarray}
By construction these configurations satisfy the F-term equations of motion.
Thus, on $\mathbb{R}^{2,1} \times S^1$, the unbroken discrete gauge group gives rise to a larger set of semi-classical configurations. A formal, non-periodic gauge transformation transforms the 
solutions $X^j(\theta)$ with non-zero Wilson line into the configurations
$X^j(0)$ with zero Wilson line. Needless to say, these 
configurations remain physically distinct on $S^1$. Furthermore, true gauge transformations with constant parameter transform the constant gauge field
$a$ within
a given conjugacy class. Thus, for each non-trivial conjugacy class in the discrete gauge group,
we find a new semi-classical configuration on the circle.

For a purely discrete centralizer, the above discussion is complete. When there are both continuous
identity components and a discrete component group, the analysis requires more care.
Note for instance that the role of the component group can also be to exchange continuous factors in the centralizer,
as discussed in detail in \cite{Bourget:2015lua}, or to break an abelian factor in the centralizer as illustrated
in  \cite{Bourget:2015lua} and section \ref{so5}.

\subsection{The Semi-Classical Vacua for $G_2$}
\label{g2section}
 We will discuss  in greater detail an example theory that illustrates the above configurations neatly,
 namely ${\cal N}=1^\ast$ theory with gauge algebra $G_2$.
We start out with a description of our semi-classical expectations. We will see that $G_2$ is a good testing
ground for the above general discussion. We perform semi-classical limits on the low-energy effective potential, and compare
the results to our semi-classical expectations for the gauge theory. The extra configurations described above will indeed appear as solutions.
We conclude with a duality diagram for the vacua, a  point of monodromy,
and other findings on the gauge theory and integrable system that are of interest.

Firstly, let's recall the classification of semi-classical configurations for ${\cal N}=1^\ast$ theory with gauge group $G_2$ on $\mathbb{R}^4$. The group $G_2$ is both connected and simply-connected. For the ${\cal N}=1^\ast$ theory on $\mathbb{R}^4$, we classify semi-classical configurations by enumerating embeddings $X^j : \mathfrak{sl}_2 \rightarrow G_2$, which are in one-to-one correspondence with nilpotent orbits of the Lie algebra $G_2$. Again, we apply the classification theory of Bala-Carter \cite{BC1,BC2} and Sommers \cite{Sommers}. We pause for a while to explain how this classification is obtained. 

\subsubsection*{Bala-Carter Theory for Nilpotent Orbits}
Suppose we want to find the nilpotent orbits of a Lie algebra $\mathfrak{g}$. The Bala-Carter theorem states that this is equivalent to finding the pairs $(\mathfrak{l},\mathfrak{p_l})$ where $\mathfrak{l}$ is a Levi subalgebra of $\mathfrak{g}$ and $\mathfrak{p_l}$ is a distinguished parabolic subalgebra of $[\mathfrak{l} , \mathfrak{l}]$. In order to fully understand this statement, we recall three useful definitions and properties: 
\begin{itemize}
\item A parabolic subalgebra of $\mathfrak{g}$ is a subalgebra which is conjugate to a subalgebra of the form $\mathfrak{p}_\Theta$ where $\Theta$ is a set of simple roots, and where $\mathfrak{p}_\Theta$ is generated by 
\begin{enumerate}
\item[(a)] The Cartan subalgebra ; 
\item[(b)] The root spaces corresponding to the root system $\langle \Theta \rangle$ created by $\Theta$ ; 
\item[(c)] The root spaces corresponding to all other positive roots. 
\end{enumerate}
We have that $\mathfrak{p}_{\Theta}$ and $\mathfrak{p}_{\Theta ' }$ are conjugate if and only if $\Theta = \Theta '$. 
\item We can decompose a parabolic subalgebra $\mathfrak{p}_{\Theta} = \mathfrak{l}_{\Theta} \oplus \mathfrak{n}_\Theta$, where the part generated by points (a) and (b) above is the Levi subalgebra $\mathfrak{l}_\Theta$, and the part generated by point (c) is the nilradical $\mathfrak{n}_\Theta$. The algebras $\mathfrak{l}_{\Theta}$ and $\mathfrak{l}_{\Theta ' }$ are conjugate if and only if $\langle \Theta \rangle$ and $\langle \Theta ' \rangle$ are Weyl-conjugate. 
\item A parabolic subalgebra $\mathfrak{p}_\Theta = \mathfrak{l}_\Theta \oplus \mathfrak{n}_\Theta$ is \emph{distinguished} if and only if $\dim \mathfrak{l}_\Theta = \dim  \mathfrak{n}_\Theta / [\mathfrak{n}_\Theta , \mathfrak{n}_\Theta ]$. 
\end{itemize} 

Let's apply this to $G_2$. In table \ref{G2parabolic} we compute, for the 4 conjugacy classes of parabolic subalgebras, the dimensions of the corresponding Levi subalgebra and of the nilradical. This gives the list of Levi subalgebras, which is the first step of the classification. The second step is to find, for each Levi subalgebra, the distinguished parabolic subalgebras of $[\mathfrak{l}_\Theta,\mathfrak{l}_\Theta]$. This is trivial for $0$, $\tilde{A}_1$ and $A_1$, in which one can check that there is exactly one distinguished parabolic subalgebra, while for $G_2$ we use again table \ref{G2parabolic} in which we read that there are two distinguished parabolic subalgebras. 
\begin{table}[H]
\centering
\begin{tabular}{|c|c|c|c|c|}
\hline
$\Theta$ & dim $\mathfrak{l}_\Theta$ & dim $\mathfrak{n}_\Theta$ & dim $\mathfrak{n}_\Theta/[\mathfrak{n}_\Theta,\mathfrak{n}_\Theta]$& $[\mathfrak{l}_\Theta,\mathfrak{l}_\Theta]$
\\
\hline
$\emptyset$  &  2 & 6 & 2 & $0 $\\
$\{ \alpha_1 \}$ & 4 & 5 & 4 & $\tilde{A}_1$\\
$\{ \alpha_2 \}$ &  4 & 5 & 2 & $A_1 $ \\
$\{ \alpha_1,\alpha_2 \}$ & 14 & 0 & 0 & $G_2 $ \\
\hline
\end{tabular}
\caption{The 4 (conjugacy classes of) parabolic subalgebras of $G_2$, which are in one-to-one correspondence with subsets of the set of simple roots. We read that a given parabolic subalgebra is distinguished if and only if the numbers in the second and fourth columns are equal. 
}
\label{G2parabolic}
\end{table}
\noindent
We conclude that there are 5 nilpotent orbits in the $G_2$ Lie algebra, which are summarized in table \ref{G2BalaCarter}. 
\begin{table}[H]
\centering
\begin{tabular}{|c|c|c|}
\hline
$\Theta$  & Name and Number\\
\hline
$\emptyset$  &  $0 \rightarrow 1$ orbit \\
$\{ \alpha_1 \}$ &   $\tilde{A}_1 \rightarrow 1$ orbit \\
$\{ \alpha_2 \}$ & $A_1 \rightarrow 1$ orbit \\
$\{ \alpha_1,\alpha_2 \}$ & $G_2 \rightarrow 2$ orbits called $G_2$ and $G_2(a_1)$\\
\hline
\end{tabular}
\caption{The 4 (conjugacy classes of) parabolic subalgebras of $G_2$, which are in one-to-one correspondence with subsets of the set of simple roots. }
\label{G2BalaCarter}
\end{table}
\noindent
Sommers generalizes this classification by allowing $\Theta$ to be a proper subset of the set of affine simple root, and calling \emph{pseudo-Levi} subalgebras the additional generated algebras \cite{Sommers}. The main theorem states that there is a bijection between 
\begin{enumerate}
\item[(i)] Conjugacy classes of pairs $(X,C)$, where $X$ is a nilpotent element and $C$ is a conjugacy class in the component group of the centralizer and 
\item[(ii)] Conjugacy classes of pairs $(\mathfrak{l},\mathfrak{p_l})$ where $\mathfrak{l}$ is a pseudo-Levi subalgebra and $\mathfrak{p_l}$ is a distinguished parabolic subalgebra of $[\mathfrak{l},\mathfrak{l}]$. 
\end{enumerate} 
As a result, we find that for $G_2$ there are 7 such conjugacy classes, captured in table \ref{G2BCS}. The centralizer of orbit $G_2(a_1)$ (which is a discrete group) has 3 conjugacy classes, and is in fact $S_3$, while the other orbits have trivial component group. 
\begin{table}[H]
\centering
\begin{tabular}{|c|c|c|c|c|c|c|}
\hline
W-classes of $\Theta$ & $[\mathfrak{l}_\Theta,\mathfrak{l}_\Theta]$ & Distinguished & Orbit & C. C. & Comm.\\
\hline
$\emptyset$  & 0 & 1 & 1 & 1 & $G_2$ \\
$\{ \alpha_0 \}$ , $\{ \alpha_2 \}$  & $A_1$ & Principal $A_1$ & $A_1$ & 1 & $\tilde{A}_1  $ \\
$\{ \alpha_1 \}$  & $\tilde{A}_1$   & Principal $\tilde{A}_1$ & $\tilde{A}_1  $ & 1 & $A_1  $  \\
$\{ \alpha_0, \alpha_1 \}$  & $A_1 + \tilde{A}_1$ & Principal $A_1 + \tilde{A}_1$ & $G_2(a_1)$ & (12) & 1 \\
$\{ \alpha_0, \alpha_2 \}$  & $A_2$ & Principal $A_2$ & $G_2(a_1)$ & (123) & 1 \\
$\{ \alpha_1, \alpha_2 \}$  & $G_2$ & $G_2$ & $G_2$ & 1& 1 \\
 &  & $G_2(a_1)$ & $G_2(a_1)$  & 1 & 1 \\
\hline
\end{tabular}
\caption{ The 7 classes of pairs $(X,C)$ where $O$ is a nilpotent orbit and $C$ is a conjugacy class in the component group of the centralizer of $O$. We tabulate the (derived algebra of the) pseudo-Levi subalgebra, its distinguished orbits, their name, the conjugacy class and the reductive part of the Lie algebra commutant. The discrete centralizer for the orbit $G_2(a_1)$ is the group $S_3$.}
\label{G2BCS}
\end{table}
\noindent
Finally, we can apply our reasoning on the multiplication of semi-classical vacua when we compactify the ${\cal N}=1^\ast$ theory on $\mathbb{R}^{2,1} \times S^1$
with gauge group $G_2$. For each nilpotent orbit for which we have a single conjugacy class in the component group, we apply the same
reasoning as on $\mathbb{R}^4$ based on the idea that we obtain pure ${\cal N}=1$ super Yang-Mills theories with a number of massive
vacua equal to the dual Coxeter number of the gauge group. We find $4+2+2+1=9$ massive vacua in this manner. Moreover, for the nilpotent orbit
$G_2(a_1)$, we have three conjugacy classes in the discrete centralizer $S_3$, and we therefore expect $3$ vacua. 
We therefore find a total of
\begin{equation}
4+2+2+1+3=12
\end{equation}
massive vacua for $G_2$. 

Similarly, for the semi-classical limits, we have the following expectations. The trigonometric
system will be determined by the gauge group breaking pattern, and more specifically by the root system
associated to the breaking.  We therefore may expect a
trigonometric system of type $G_2,A_2, A_1 + \tilde{A}_1,\tilde{A}_1, A_1$ and none at all in the above
cases (read from bottom to top). 
The $A_1$ and $\tilde{A}_1$ cases are cases of oblique confinement, and the case of trivial orbit, where the full $G_2$ gauge group remains unbroken, corresponds to confining vacua.
The first three cases are Higgsed vacua, possibly with non-trivial Wilson lines corresponding to the non-trivial conjugacy classes in the component group. We distinguish three different cases, namely zero Wilson line, a Wilson line 2-cycle and a Wilson line 3-cycle since these are the conjugacy classes of the $S_3$ component group. Note that the two Higgs vacua associated to the $G_2$ orbit and $G_2(a_1)$ orbit with trivial conjugacy class share the same symmetry breaking pattern. Although the vacuum expectation values $X^j$ are different, the integrable system may not distinguish them. Taking this last subtlety into account, we can predict that the integrable system has 11 extrema, which are recovered from the semi-classical limit in section \ref{ellint} and found numerically in subsection \ref{G2num}. 

\subsection{The Elliptic Integrable System and the Semi-classical Limits}
\label{ellint}
 In this subsection, we explicitly calculate the semi-classical limits on the  
 low-energy effective superpotential and compare the results to our expectations.
  The effective superpotential that we work with is 
\begin{equation}
\mathcal{W}_{G_2,tw} (Z) =  \sum\limits_{\alpha \in \Delta^+_{\text{long}}} \wp (\alpha \cdot Z) + \frac{1}{3}\sum\limits_{\alpha \in \Delta^+_{\text{short}}} \wp_3 (\alpha \cdot Z) \, .
\end{equation}
We refer to appendix \ref{g2app} for our conventions for the root system, weight system, and other Lie algebra data. We can parametrize  $Z=3 (z_1 \epsilon_1 + z_2 \epsilon_2)$ such that 
 \begin{eqnarray}
\mathcal{W}_{G_2,tw}(z_1,z_2) &=& \wp \left( 3z_1 - 3z_2 \right) + \wp \left( 3z_1 \right) + \wp \left( 3z_2 \right) \nonumber \\
 & & + \frac{1}{3}  \left[ \wp_3 \left( z_1+z_2 \right) + \wp_3 \left( 2z_1-z_2 \right) + \wp_3 \left( -z_1+2z_2 \right) \right] \, ,
\end{eqnarray}
or use the alternative parameterization $Z=Z_1 \pi_1 + Z_2 \pi_2 =(Z_1+ 2 Z_2) \epsilon_1 + Z_2 \epsilon_2$. The link is
\begin{eqnarray}
\label{OldNewG2variables}
\begin{cases}
 z_1 = \frac{1}{3} ( Z_1+2 Z_2 ) \\
 z_2 =  \frac{1}{3} Z_2
\end{cases}
\qquad 
\begin{cases}
 Z_1 = 3z_1 - 6z_2 \\
 Z_2 = 3z_2 \, ,
\end{cases}
\end{eqnarray}
and the explicit form of the superpotential is then
\begin{eqnarray}
\mathcal{W}_{g2,tw}(Z_1,Z_2) &=& \wp (Z_2) + \wp (Z_1 + Z_2) +  \wp (Z_1 + 2 Z_2)  \nonumber \\
& & + \frac{1}{3} \left[ \wp_3 (Z_1/3) +\wp_3 (Z_1/3+Z_2)+\wp_3 (2Z_1/3+Z_2) \right] \, ,
\end{eqnarray}
with $Z_i = X_i + \tau Y_i$. We still have to specify the periodicities and identifications. In the $\omega_1$ direction, we identify by shifts by the  weight lattice, and in the $\omega_2$ direction by the co-weight lattice. This implies : 
 \begin{equation}
 (z_1,z_2) \sim \left( z_1 + \frac{2}{3}\omega_1 ,z_2 \right) \sim \left( z_1  ,z_2 + \frac{2}{3}\omega_1  \right)
 \end{equation}
 and
 \begin{equation}
 (z_1,z_2) \sim \left( z_1 + 2\omega_2 ,z_2 \right) \sim \left( z_1  ,z_2 + 2\omega_2 \right) \sim \left( z_1  -\frac{2}{3}\omega_2 ,z_2 + \frac{2}{3}\omega_2 \right) \, .
 \end{equation}
The Weyl group action yields the further equivalences 
 \begin{equation}
 (z_1,z_2)  \sim (z_2,z_1) \sim (-z_1,-z_2) \sim \left( z_1  ,z_1 - z_2 \right) \, .
 \end{equation}
 We note that the group $G_2$ has trivial center.

Below, we distinguish between the trigonometric or Higgs limits, in which the leading trigonometric system is of rank two,
the oblique limits, in which it is of rank one, and the affine Toda, or confining limit.

\subsubsection{The Higgsed Limits}

Firstly, we describe the limit for the Higgs vacuum, the 2-cycle vacuum and the 3-cycle vacuum. 
The superpotential becomes the trigonometric system corresponding to the  pseudo-Levi subalgebra.

\begin{itemize}
\item  \subsubsection*{The trigonometric $G_2$ limit}
In the first $\tau \rightarrow i \infty$ limit, where we take the Wilson line to be $a=(0,0)$ and consequently
$Y_1=0=Y_2$, we find the trigonometric $G_2$ model for the choice of simple roots $J_0=\{ \alpha_1,\alpha_2 \}$: 
\begin{equation}
\mathcal{W}_{G_2,tw} (X) \rightarrow \mathcal{W}_{\mbox{trig},G_2} \, .
\end{equation}
We find a real extremum. It can be described through zeroes of an
orthogonal polynomial \cite{Corrigan:2002th,Odake:2002xm}.

\item  \subsubsection*{The trigonometric $A_2$ limit}
In the second limit, corresponding to the 3-cycle conjugacy class and Wilson line
$a=(1/3,0)$ we find the trigonometric $A_2$ system corresponding to the choice
of simple root system $J_0= \{ \alpha_0,\alpha_2 \}$. 
The co-marks give the constraint $1=Y_0+2Y_2+Y_1$.
If we  impose $Y_0=0=Y_2$, we find $Y_1=1$.
We are left with a trigonometric $A_2$ system corresponding to all the long roots.
The extremal positions are therefore given by the equilibria of the trigonometric $A_2$ integrable system. 
There is a massive extremum at $(z_1,z_2)=(1/3+\tau/3 , 2/9)$.\footnote{The $A_2$ trigonometric model also allows for massless complexified extrema at zeroth order. 
However, these extrema do not survive the order $q$ perturbation.}
\item  \subsubsection*{Trigonometric $A_1 + \tilde{A}_1$}
The third Higgs vacuum is associated to the Wilson line $a=(1/2,0)$, with the choice $J_0=\{ \alpha_0, \alpha_1 \}$,
and gives rise to
the trigonometric $A_1+ \tilde{A}_1$ system (with differing coupling constants).
One finds a unique extremum up to equivalences, namely $(z_1,z_2)=(\frac{1}{6}+\frac{\tau}{2},\frac{1}{6})$. 
\end{itemize}
\subsubsection*{Remark}

We remark that it is the centralizer of the Wilson line group element that determines the non-perturbative contributions to the superpotential in the semi-classical limit. Namely, the allowed monopole charges corresponds precisely to positive roots of the pseudo-Levi subalgebra. We recall that we have two configurations in which the full gauge group is broken,
namely the orbit labelled $G_2$ and the orbit labelled $G_2(a_1)$ with zero Wilson line. In the elliptic integrable system,
we only identified one real extremum. The two orbits are distinguished through their scalar adjoint vacuum expectation
values.

\subsubsection{The Confining Limit}
If we pick zero simple roots, we obtain the affine Toda potential for the algebra $D_4^{(3)}=(G_2^{(1)})^\vee$
 \begin{equation}
\mathcal{W}_{G_2,tw} ( x_1 + \frac{\tau}{4},x_2 + \frac{\tau}{12}) = q^{\frac{1}{4}} \left( e^{-6\pi i x_1} + e^{6\pi i  x_2} + 3 e^{6 \pi i (x_1-2x_2)}\right) + \dots \, .
\end{equation}
The associated simple roots are $\alpha_0$, $\alpha_2$ and $3\alpha_1$. The extrema of the
affine Toda potential can be obtained analytically (see e.g. \cite{Davies:2000nw}).

\subsubsection{The Oblique Limits}
Let's turn to the limits with partial breaking of the gauge algebra through adjoint vacuum expectation values.

\begin{itemize}
\item  \subsubsection*{The Oblique Limit $J_0=\{1\}$}

The limit $J_0=\{ 1 \}$ corresponds to the orbit $\tilde{A}_1$ with unbroken gauge group $A_1$.
We first determine the non-perturbative corrections to the leading coordinate $Z_1$, and find
\begin{equation}
(Z_1,Z_2)=\left(\frac{1}{2}- \frac{4i}{3 \pi} e^{2 i \pi \delta_2} q^{\frac{1}{4}}+\delta_1,\frac{\tau}{4}+\delta_2 \right) 
\, ,
\end{equation}
and a
final stabilized potential (at $\delta_1=0$)
\begin{eqnarray}
{\cal W}_{G_2,tw}(\delta_2) &=& \pi^2( 1 
+4 e^{-4 i \pi  \delta _2} q^{\frac{1}{2}}-\frac{20}{3} e^{4 i \pi  \delta _2} q^{\frac{1}{2}}
+O(q) ) \, .
 \end{eqnarray}
We can solve for the fluctuation $\delta_2$ using this superpotential, and then find the
superpotential at equilibrium to order $q^{\frac{1}{2}}$
\begin{eqnarray}
{\cal W}_{G_2,tw} &=& \pi^2( 1 -8 i \sqrt{\frac{5}{3}} \sqrt{q} + \dots)\, .
\end{eqnarray}

\item \subsubsection*{The Oblique Limit $J_0=\{2\}$}
If we put $Y_1=\frac{3}{4}, Y_2=0$, we get stabilization at level $q^{\frac{3}{2}}$. The first orders in the expansion of the coordinate $Z_2$ are given by
\begin{eqnarray}
Z_2 &=& \frac{1}{2}-\frac{4 i e^{-2 i \pi  \delta X_1} \sqrt[4]{q}}{\pi }+\frac{48 i e^{-4 i \pi  \delta X_1} q^{\frac{1}{2}}}{\pi } + \dots \, ,
\end{eqnarray}
to finally find stabilization for the coordinate $Z_1$ at
\begin{eqnarray}
Z_1 &=& \frac{3\tau}{4} - \frac{i \log 3}{4\pi} \quad \mbox{or} \quad \frac{3\tau}{4} - \frac{1}{4} - \frac{i \log 3}{4\pi} \, .
\end{eqnarray}
The resulting superpotentials in the two inequivalent vacua are
\begin{eqnarray}
{\cal W}_{G_2,tw} &=& \pi^2 ( -1 + 312 q \pm 5832 q^{\frac{3}{2}} + \dots) \, .
\label{hardestsuperpot}
\end{eqnarray}
The stabilizing potential for $Z_1$ arises at sixth order in the non-perturbative expansion parameter $q^{\frac{1}{4}}$.
\end{itemize}

\subsubsection*{Remark}
One can ask about the oblique limit $J_0=\{0\}$. We have found no choice of $Y_i$ consistent with the condition $Y_0=0$ such that the second 
coordinate stabilizes.
We note that the choices $J_0=\{ 0 \}$ and $J_0 = \{ 2 \}$ are Weyl equivalent in the horizontal
algebra,  but inequivalent in the affine algebra. They also are inequivalent as limiting choices. In this example, using the pseudo-Levi subalgebra classification scheme as a starting point for the semi-classical limits works,  if only because another, inequivalent limit, does not stabilize.

\subsection{Results Based on Numerics}
\label{G2num}

In this subsection, we present results based on numerical analyses performed at finite coupling $\tau$. The main strategy is to combine a random exploration of the parameter space with the requirement that vacua should form closed multiplets under $S_3$ and $T$ dualities. Our numerics is in essence based on the 
FindMinimum procedure of Mathematica, applied to the logarithm of the positive definite real potential of the gauge theory. Let us first explain how these dualities can be implemented numerically on a vacuum that we know at large $\tau$ (by which we always mean the semi-classical regime $\tau \rightarrow i \infty$): 
\begin{itemize}
\item $T$-duality is performed by taking the vacuum at large $\tau$ and changing continuously $\tau \mapsto \tau +1$ on a straight line. 
\item For $S_3$-duality, we first track the vacuum to the self-dual point $\tau_{sd}=i/\sqrt{3}$, then use the exact Langlands $S_3$-duality formula (see later, equation (\ref{G2Sduality})) to $S_3$-dualize it, and finally bring it back to large $\tau$. 
\end{itemize}
Note that it is crucial that $\tau$ be large to $T$-dualize, because of potential points of monodromy at finite gauge coupling. 

\subsubsection*{Finding the Vacua}

Using our numerical procedure, it is easy to find the Higgs vacuum on the real axis; we label it $H$. Taking its $S_3$-dual as explained above, one obtains the confining vacuum dubbed $C_0$. When we $T$-dualize the confining vacua we obtain three more vacua, $C_1,C_4,C_5$, for a quadruplet of confining vacua at large $\tau$. More subtle is the following fact.
Consider these vacua brought down to the self-dual value of the gauge coupling $\tau_{sd}$. We call $T_{sd}$ duality the operation 
\begin{equation}
T_{sd} : \tau_{sd}=i/\sqrt{3} \mapsto \tau_{sd} +1
\end{equation}
continuously along a straight line in the upper-half plane $\mathcal{H}$. If we apply this transformation to the confining vacuum $C_0$, we find that we need to repeat it six times before falling on this confining vacuum once more. We thus find a sextuplet of $T_{sd}$-duality that we denote $(C_0,C_1,C_2,C_3,C_4,C_5)$. This indicates a point of monodromy that lies above the self-dual point.\footnote{See \cite{Bourget:2015cza} for a more gentle introduction to points of monodromy.} The point of monodromy is located around $\tau_M \sim 1.440672920416 i$, and all of these digits are significant. At the self-dual point, we can analytically check that $S_3$-duality acts as $S_3(C_1)=C_4, S_3(C_2)=C_5$ and $S_3(C_3)=C_3$. Moreover, if we bring up the two extra vacua $(C_2,C_3)$ that complete the sextuplet to larger $\tau$, they behave as a doublet under $T$-duality. These seven vacua obtained from the Higgs are represented on the right of figure \ref{dualities_g2}. 

In addition to these, we also find two extrema which are $S_3$-duality and $T$-duality singlets, and also two $S_3$-singlets (labelled $J_1,J_2$) which are $T$-dual (and $T_{sd}$-dual) to each other. This completes the duality web summarized in figure \ref{dualities_g2}.

\subsubsection*{Identification with the Semi-classical Limits}

We have obtained a total of eleven extrema, as expected from section \ref{ellint}. We can be more precise and match each $T$-multiplet with its corresponding limiting integrable system, using the value of the superpotential when necessary.

The singlets correspond to the 2- and 3-cycle semi-classical vacua, while the doublet of the duality group matches the semi-classical $J_0=\{1 \}$ extrema. The confining quadruplet is easily matched to the semi-classical solutions. The semi-classical origin of $(C_2,C_3)$ is the choice $J_0=\{ 2 \}$. The numerical evidence we obtained for this last identification is limited to the first two coefficients in the superpotential (\ref{hardestsuperpot}).

\subsubsection*{Numerical values}

Finally, let us provide a few concrete numbers of our simulations for easier reproducibility.
The $(z_1,z_2)$ positions  of the numerical extrema are approximately given in the tables below,
where the first entry is real part of $z_1$ and the second entry is the imaginary part of $z_1$ 
expressed in units of the purely imaginary value of $\tau$. 
\begin{equation}
\begin{array}{|c|c|c|}
\hline
 \text{Vacuum} & \text{Positions at } \frac{i}{\sqrt{3}} & \text{Positions at } \frac{5i}{2}  \\
 \hline
 \text{H} & \{0.22754,0.,0.03944,0.\} & \{0.22738,0.,0.03954,0.\} \\
 \text{H2} & \{0.16667,0.5,0.16667,0.\} & \{0.16667,0.5,0.16667,0.\} \\
 \text{H3} & \{0.11111,0.33333,0.22222,0.33333\} &
   \{0.11111,0.33333,0.22222,0.33333\} \\
 \text{C0} & \{0.,0.26698,0.,0.41565\} & \{0.,0.26222,0.,0.41795\} \\
 \text{C1} & \{0.2727,0.2679,0.4226,0.4103\} &
   \{0.25257,0.2631,0.41731,0.41817\} \\
 \text{C2} & \{0.5594,0.3047,0.8509,0.4275\} &
   \{0.5,0.33139,0.83333,0.42738\} \\
 \text{C3} & \{0.86305,0.33333,1.26486,0.46124\} &
   \{0.83812,0.33333,1.25239,0.42829\}  \\
 \text{C4} & \{0.22607,0.36197,0.7085,0.45614\} &
   \{0.16667,0.40236,0.66667,0.4875\} \\
 \text{C5} & \{0.60603,0.39877,0.18344,0.47574\} &
   \{0.58591,0.40356,0.1686,0.4884\} \\
 \text{J1} & \{0.89497,0.60134,0.64682,0.45821\} &
   \{0.87587,0.58097,0.62587,0.41972\}  \\
 \text{J2} & \{0.98015,0.20845,0.56164,0.73199\} &
   \{0.9592,0.24694,0.54253,0.75236\}  \\
   \hline
\end{array}
\end{equation}
The superpotentials in these vacua are
\begin{equation}
\begin{array}{|c|c|c|}
\hline
 \text{Vacuum} & \text{Superpotential at } \frac{i}{\sqrt{3}} & \text{Superpotential at } \frac{5i}{2} \\
 \hline
 \text{H} & 271.5202972 & 256.6097930 \\
 \text{H2} & 26.54254786 & 19.73924450 \\
 \text{H3} & 26.54254786 & 19.73924450 \\
 \text{C0} & -218.4352014 & -22.81452733 \\
 \text{C1} & 42.47856497-33.32941024 i & -19.56246124-2.892724428 i \\
 \text{C2} & 10.60653076+33.32941024 i & -9.869136924 \\
 \text{C3} & 26.54254786 & -9.869143660 \\
 \text{C4} & 10.60653076-33.32941024 i & -17.01826338 \\
 \text{C5} & 42.47856497+33.32941024 i & -19.56246124+2.892724428 i \\
 \text{J1} & 26.54254786+13.36027231 i & 9.869700650+0.03957060700 i \\
 \text{J2} & 26.54254786-13.36027231 i & 9.869700650-0.03957060700 i \\
 \hline
\end{array}
\label{superpotentialvalues}
\end{equation}

\subsection{Langlands Duality  and the Duality Diagram}

 Aside from the simply laced Lie algebras of $A,D$ and $E$ type, there are three more algebras that are mapped to  themselves  under Langlands duality. These are $B_2$, $G_2$ and $F_4$.
 The twisted elliptic integrable systems with appropriate couplings are indeed Langlands self-dual \cite{Bourget:2015cza}, 
 namely, they permit the symmetry $S_{\alpha} : \tau \rightarrow - \frac{1}{\alpha \tau}$, where $\alpha$ is the ratio of the length
 squared of the long versus the short roots. 
The invariance under $S_{\alpha}$ translates into a relation involving the superpotentials evaluated at  different positions $X_i$, including a shift. Explicitly, the fact that $G_2$ is invariant under $S_3 : \tau \rightarrow  - \frac{1}{3 \tau}$ duality reads
 \begin{equation}
 \label{G2Sduality}
 \mathcal{W}_{G_2,tw}(X_1 , X_2 ; \tau) = \frac{1}{3 \tau^2} \mathcal{W}_{G_2,tw} \left( \frac{X_1 + X_2}{3 \tau} , \frac{2X_1 - X_2}{3 \tau} ; -  \frac{1}{3 \tau} \right) +2 \pi ^2 \left[  3E_2(3\tau) - E_2(\tau) \right] \, .
 \end{equation}
 As was the case for the $so(5)$ integrable system (see \cite{Bourget:2015cza}), the shift resulting from the  $S_\alpha$ duality transformation can be identified with the superpotential in one of the vacua. The latter property allows for the realization of duality symmetries as permutations on the
 list of extremal superpotential values.

\begin{figure}[H]
 \centering
 \includegraphics[width=0.6\textwidth]{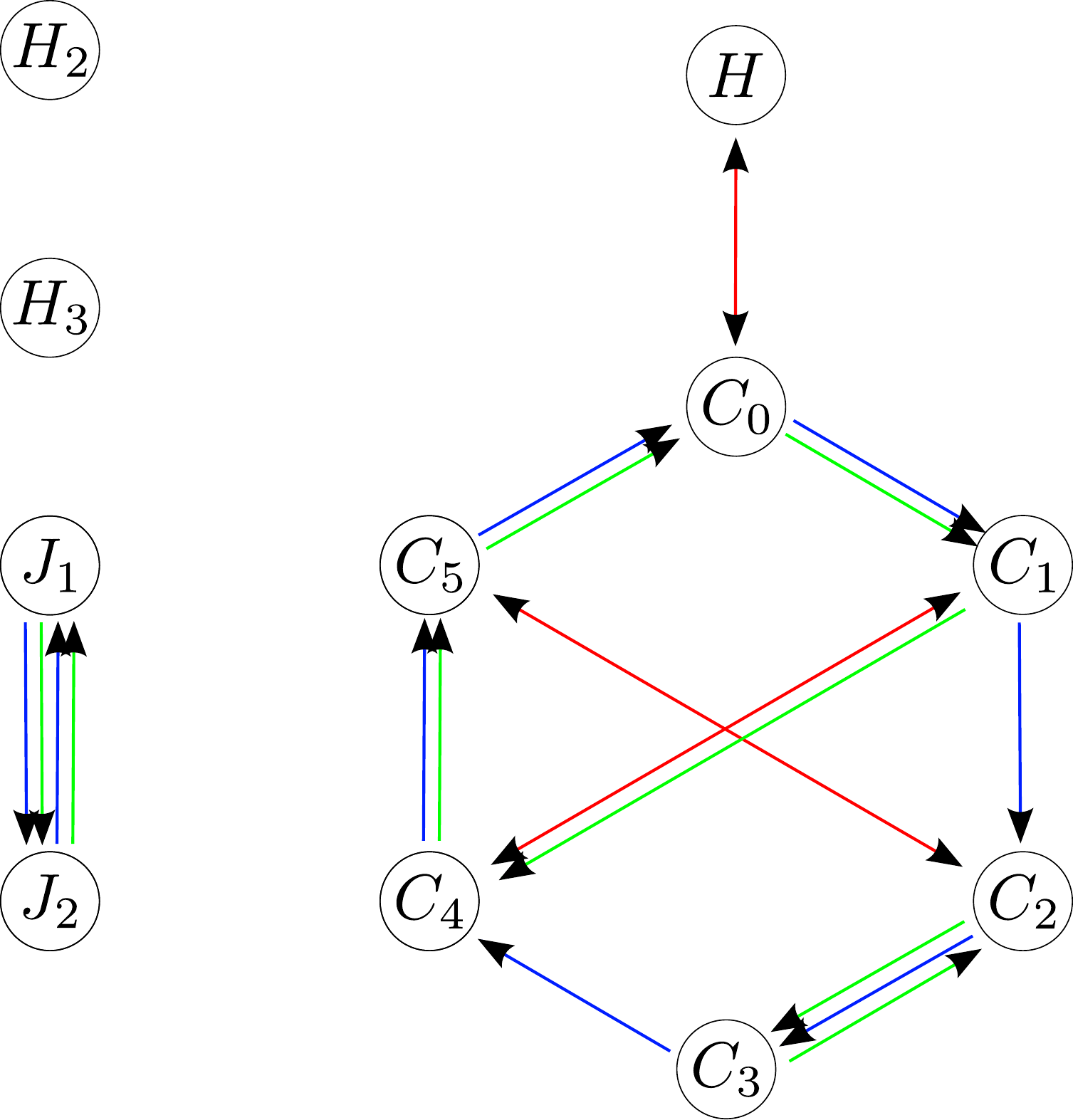}
 \caption{A diagram of dualities of $G_2$, below the point of monodromy. The blue arrows represent the transformation $T_{sd} : \tau \rightarrow \tau +1$ below the point of monodromy, while the green arrows represent the transformation $T : \tau \rightarrow \tau +1$ above the point of monodromy. To identify vacua at different $\tau$ we use the convention that  the branch cut is on the left of the monodromy point ($\mathbb{R_-}$ direction). The red arrows indicate the action of $S_3$-duality (\ref{G2Sduality}) at the self-dual point $\tau = \frac{i}{\sqrt{3}}$. The absence of a given arrow indicates invariance of the vacuum under the corresponding transformation. Note that the order of $S_3 T_{sd}$ is 7 while the order of $S_3 T$ is 6. } 
 \label{dualities_g2}
 \end{figure}

We have determined these permutations numerically (as reviewed above), and sum up the action of $S_3$, $T$ and $T_{sd}$ in the diagram shown in figure \ref{dualities_g2}. 
 This diagram demonstrates the importance of specifying the path followed in the moduli space while performing a duality: note for instance that in the diagram, $S_3 T_{sd}$ has order 7 while the order of the more standard operation $S_3 T$ is 6, as a consequence of monodromies. In \cite{Bourget:2015cza} one can find other examples of generalized duality groups that are generated by points of monodromy.

 Finally, we make a few remarks on the exact values of the superpotential in a number of vacua.
The superpotentials in the Higgs vacua with non-zero Wilson line are identical.
They are equal to 
\begin{eqnarray}
{\cal W}_{H2} &=& 2 \pi^2 (\theta_3(q^2) \theta_3(q^6) + \theta_2(q^2) \theta_2(q^6))^2 \, .
\end{eqnarray}
This is the theta series of the direct sum of 2 copies of a hexagonal lattice. It generates the (1-dimensional) space of modular forms of weight 2 for the congruence subgroup $\Gamma_0 (3)$. 
Many  further analytical statements can be made about the exact values
of the superpotential. As an example, we have the closed form 
expression 
\begin{equation}
26.54254786... = \frac{9 \Gamma \left(\frac{1}{3}\right)^6}{8 \times 2^{\frac{2}{3}} \pi ^2} \, ,
\end{equation}
for this particular entry in table (\ref{superpotentialvalues}) of values of the superpotential.
It will be interesting to classify the superpotential values into (vector valued) modular forms (potentially
with non-analyticity in the upper half plane) of $\Gamma_0(3)$ or the full Hecke group.

 \section{The $so(5)$ Massless Branch}
 \label{so5revisited}
 \label{so5massless}
In this section, we tie up a loose end. In section \ref{so5} we analyzed semi-classical limits for the $B_2$
twisted elliptic integrable system, and we found a single massless branch of complex dimension one. We wish to characterize this branch more precisely, including at finite coupling $\tau$. We also exhibit its duality and global properties in the different theories associated to the gauge algebra $so(5)$.

\subsection{The Local Description of the Massless Branch}
We show that a massless branch exists at all couplings by a brute force analysis. We postulate that the superpotential on the massless branch is equal to $e_1(q)$ (as we found in perturbation theory in section \ref{so5}). We will also consider the two equations that follow from the fact that we are studying an extremum of the superpotential. These equations give rise to three constraint equations in terms of two unknowns, for a single massless branch of complex dimension one. This doubly overdetermined system will have a simple solution which is the description of the massless branch. Before we get to the simple end result, we plough through some
elliptic function identities. Firstly, we recall the definition of the Weierstrass function evaluated at half-periods
\begin{equation}
\wp (\omega_i ; \tau) = e_i (q) \, 
\end{equation}
and note that we have the equality
\begin{equation}
e_1 (q) = -\frac{2\pi^2}{3} E_{2,2} (q) = -\frac{2\pi^2}{3} (E_2(q)-2E_2(q^2)) \, ,
\end{equation}
as well as the identities
\begin{eqnarray}
\wp_2 (z ; \tau) &=& 4 \wp (2z ; 2\tau) +e_1 \nonumber \\
&=& -e_1 + \frac{1}{4} \left( \frac{\wp' (z;\tau)}{\wp (z;\tau) - e_1}\right)^2
\nonumber \\
\wp'_2 (z ; \tau) &=& 8 \wp' (2z ; 2\tau) \, .
\end{eqnarray}
We again describe the superpotential and its derivatives algebraically using the variables\footnote{These variables describe faithfully the vacua of the $SO(5)_+$ theory, by which we mean that for any $(\mathcal{X}_1,\mathcal{X}_2,\mathcal{Y}_1,\mathcal{Y}_2) \in \mathbb{C}^4$ there is exactly one vacuum of the $SO(5)_+$ theory that satisfies (\ref{ellipticso5variables}). In the $Spin(5)$ and $SO(5)_-$ theories there are two such vacua, namely $(z_1,z_2)$ and $(z_1+2\omega_2,z_2+2\omega_2)$. Moreover, both in the $Spin(5)$ and in the $SO(5)_-$ theory (\ref{ellipticso5variables}) is not a well defined functional of a given vacuum because of the identification $(z_1,z_2) \sim (z_1+\omega_1 , z_2 + \omega_1)$ in $Spin(5)$ and $(z_1,z_2) \sim (z_1+\omega_1 + 2\omega_2 , z_2 + \omega_1)$ in $SO(5)_-$. These subtleties will be taken care of in subsection \ref{moduliSpaceso5}. } 
\begin{equation}
\mathcal{X}_i=\wp (z_i ; \tau) \qquad\textrm{and}  \qquad \mathcal{Y}_i=\wp' (z_i ; \tau) \, , 
\label{ellipticso5variables}
\end{equation}
for $i=1,2$. The value of the superpotential,  $\mathcal{W}_{B_2,tw}(Z ; \tau ) = e_1 (\tau)$, translates into 
the equation
\begin{equation}
2 \left( \frac{\mathcal{Y}_1-\mathcal{Y}_2}{\mathcal{X}_1-\mathcal{X}_2}\right)^2 + 2 \left( \frac{\mathcal{Y}_1+\mathcal{Y}_2}{\mathcal{X}_1-\mathcal{X}_2}\right)^2  +\left( \frac{\mathcal{Y}_1}{\mathcal{X}_1-e_1}\right)^2  + \left( \frac{\mathcal{Y}_2}{\mathcal{X}_2-e_1}\right)^2 =16(e_1 +\mathcal{X}_1+\mathcal{X}_2) \, .
\end{equation}
We also  need the addition formula for the derivative of the Weierstrass function:
\begin{eqnarray}
\wp'(u+v) = \frac{ \left[ \frac{1}{2} g_3 (\wp'(u)-\wp'(v)) 
   +  ( \wp'(v) \wp(u)^2 +\frac{1}{4}g_2 \wp'(u) )(\wp(u)+3 \wp(v)) \right] - [u \leftrightarrow v]}{ (\wp(u)-\wp(v))^3} 
   \, .
\end{eqnarray}
It is convenient to write this in a more symmetric way, where all the derivatives are isolated on the left-hand side : 
\begin{equation}
\frac{\wp'(u+v) + \wp'(u-v)}{\wp'(u)} = \frac{  g_2 (\wp(u)+3 \wp(v))+4 g_3 -4 \wp(v)^2 (3 \wp(u)+\wp(v)) }{2 (\wp(u)-\wp(v))^3} \, ,
\end{equation}
which we can use to express the derivative of the twisted Weierstrass function as
\begin{equation}
\frac{\wp'_2(u)}{\wp'(u)} = \frac{8 e_1^3-3 e_1 \left(g_2-4 \wp(u)^2\right)-\wp(u) \left(g_2+4 \wp(u)^2\right)-4   g_3}{4 (e_1-\wp(u))^3} \, . 
\end{equation}
Using these relations, the three equations describing the massless branch become
\begin{eqnarray}
\frac{4(4\mathcal{X}_1^3+4\mathcal{X}_2^3 - g_2 (\mathcal{X}_1+\mathcal{X}_2)-2g_3)}{(\mathcal{X}_1-\mathcal{X}_2)^2}  +\frac{4\mathcal{X}_1^3-g_2 \mathcal{X}_1 - g_3}{(\mathcal{X}_1-e_1)^2}  +\frac{4\mathcal{X}_2^3-g_2 \mathcal{X}_2 - g_3}{(\mathcal{X}_2-e_1)^2} - 16(e_1 +\mathcal{X}_1+\mathcal{X}_2)&=& 0 \nonumber \\
\frac{  g_2 (\mathcal{X}_1+3 \mathcal{X}_2)+4 g_3 -4
   \mathcal{X}_2^2 (3 \mathcal{X}_1+\mathcal{X}_2) }{ (\mathcal{X}_1-\mathcal{X}_2)^3} + \frac{8 e_1^3-3 e_1 \left(g_2-4
  \mathcal{X}_1^2\right)-\mathcal{X}_1 \left(g_2+4 \mathcal{X}_1^2\right)-4
   g_3}{4 (e_1-\mathcal{X}_1)^3} &=& 0 \nonumber\\
   \frac{  g_2 (\mathcal{X}_2+3 \mathcal{X}_1)+4 g_3 -4
   \mathcal{X}_1^2 (3 \mathcal{X}_2+\mathcal{X}_1) }{ (\mathcal{X}_2-\mathcal{X}_1)^3} + \frac{8 e_1^3-3 e_1 \left(g_2-4
  \mathcal{X}_2^2\right)-\mathcal{X}_2 \left(g_2+4 \mathcal{X}_2^2\right)-4
   g_3}{4 (e_1-\mathcal{X}_2)^3} &=& 0 \, . \nonumber
\end{eqnarray}
Finally, we express the Eisenstein series $g_2$ and $g_3$ of weight 4 and 6 in terms of the half-period values $e_i$ using the  relations $g_2 = 2(e_1^2+e_2^2+e_3^2)$ and $g_3=4e_1e_2e_3$ to obtain
\begin{eqnarray}
 \frac{ (2 e_1-\mathcal{X}_1-\mathcal{X}_2)^2 }{(e_1-\mathcal{X}_1) (e_1-\mathcal{X}_2) (\mathcal{X}_1-\mathcal{X}_2)^2} P_3(\mathcal{X}_1,\mathcal{X}_2,e_1,e_2) = 0
   \\
 \frac{(2 e_1-\mathcal{X}_1-\mathcal{X}_2) }{(e_1-\mathcal{X}_1)^2 (\mathcal{X}_1-\mathcal{X}_2)^3} P_4(\mathcal{X}_1,\mathcal{X}_2,e_1,e_2) = 0 \\
 \frac{(2 e_1-\mathcal{X}_1-\mathcal{X}_2) }{(e_1-\mathcal{X}_2)^2 (\mathcal{X}_2-\mathcal{X}_1)^3} P_4(\mathcal{X}_2,\mathcal{X}_1,e_1,e_2) = 0 \, ,
\end{eqnarray}
where $P_3$ and $P_4$ are  homogeneous polynomials of degree 3 and 4 respectively. We  see that $\mathcal{X}_1+\mathcal{X}_2=2e_1$ is a sufficient condition to be on the massless branch of vacua. Restricting to these solutions  -- except at special points in the space of couplings, these are the only solutions --, we can parametrize the line by a single complex number $\lambda \in \mathbb{C}^\ast$ as 
\begin{equation}
\mathcal{X}_1 = e_1(\tau) + \lambda \, ,  \qquad \qquad \mathcal{X}_2 = e_1(\tau) - \lambda \, .
\label{ellipticso5branch}
\end{equation}

\subsection{Duality and the Massless Branch}
T-duality manifestly leaves the description of the massless branch in terms of the elliptic curve variables invariant,
as can be seen from equation (\ref{ellipticso5branch}). We analyze Langlands $S_2$ duality next. 
In the notation of equation (\ref{B2superpotential}), the duality formula for $so(5)$
reads \cite{Bourget:2015cza}
\begin{equation}
\mathcal{W}_{B_2,tw} (z_1,z_2,\tau) = \frac{1}{2 \tau^2}  \mathcal{W}_{B_2,tw}
\left(\frac{z_1+z_2}{2 \tau},\frac{z_1-z_2}{2 \tau},-\frac{1}{2 \tau}\right) + 2e_1 (\tau)  \, .
\end{equation}
Using the identity
\begin{equation}
e_1 \left( - \frac{1}{2\tau}\right) = -2 \tau^2 e_1 (\tau) \, ,
\end{equation}
it can be written in the more symmetric form
\begin{equation}
\mathcal{W}_{B_2,tw} (z_1,z_2,\tau) -e_1(\tau) 
= \frac{1}{2 \tau^2} \left[ \mathcal{W}_{B_2,tw}
\left(\frac{z_1+z_2}{2 \tau},\frac{z_1-z_2}{2 \tau},-\frac{1}{2 \tau}\right) - e_1\left(-\frac{1}{2 \tau} \right) \right] \, .
\end{equation}
We define the dual elliptic curve variables
\begin{eqnarray}
\mathcal{X}'_1 &=& \wp \left( \frac{z_1+z_2}{2\tau} ; - \frac{1}{2\tau}\right) \\
\mathcal{X}'_2 &=& \wp \left( \frac{z_1-z_2}{2\tau} ; - \frac{1}{2\tau}\right) \\
\mathcal{Y}'_1 &=& \wp' \left( \frac{z_1+z_2}{2\tau} ; - \frac{1}{2\tau}\right) \\
\mathcal{Y}'_2 &=& \wp' \left( \frac{z_1-z_2}{2\tau} ; - \frac{1}{2\tau}\right)  \, ,
\end{eqnarray}
which are related to the original elliptic curve variables (\ref{ellipticso5variables}) by 
\begin{eqnarray}
 -e_1(\tau) - \mathcal{X}_1 - \mathcal{X}_2 + \frac{1}{4} \left( \frac{\mathcal{Y}_1-\mathcal{Y}_2}{\mathcal{X}_1-\mathcal{X}_2}\right)^2 &=& \frac{1}{16 \tau^2} \left( \frac{\mathcal{Y}'_1}{\mathcal{X}'_1-e_1(-\frac{1}{2\tau})}\right)^2 \\
 -e_1(\tau) - \mathcal{X}_1 - \mathcal{X}_2 + \frac{1}{4} \left( \frac{\mathcal{Y}_1+\mathcal{Y}_2}{\mathcal{X}_1-\mathcal{X}_2}\right)^2 &=& \frac{1}{16 \tau^2} \left( \frac{\mathcal{Y}'_2}{\mathcal{X}'_2-e_1(-\frac{1}{2\tau})}\right)^2 \, .
\end{eqnarray}
The sum of these relations is 
\begin{eqnarray}
 -2e_1(\tau) - 2\mathcal{X}_1 -2 \mathcal{X}_2 +  \frac{\mathcal{Y}_1^2+\mathcal{Y}_2^2}{2(\mathcal{X}_1-\mathcal{X}_2)^2}= \frac{1}{16 \tau^2} \left( \frac{\mathcal{Y}'_1}{\mathcal{X}'_1-e_1(-\frac{1}{2\tau})}\right)^2 +\frac{1}{16 \tau^2} \left( \frac{\mathcal{Y}'_2}{\mathcal{X}'_2-e_1(-\frac{1}{2\tau})}\right)^2 \, . \nonumber 
\end{eqnarray} 
After these preparations, we will now show that if we choose a point on the dual of the massless branch, namely a point satisfying the equation
$\mathcal{X}'_1+\mathcal{X}'_2=2e_1(-\frac{1}{2\tau})$, that this is consistent with the original variables lying
on the original massless branch. 
Indeed, this equality implies that the sum of the relations becomes
\begin{eqnarray}
 2(e_1(\tau) + \mathcal{X}_1 + \mathcal{X}_2) -  \frac{\mathcal{Y}_1^2+\mathcal{Y}_2^2}{2(\mathcal{X}_1-\mathcal{X}_2)^2} + \frac{(\mathcal{Y}'_1)^2 + (\mathcal{Y}'_2)^2 }{4 \tau^2 (\mathcal{X}'_1-\mathcal{X}'_2)^2} = 0  \, . 
\end{eqnarray}
Taking into account the elliptic curve equation, we can simplify this to
\begin{equation}
\frac{(\mathcal{Y}'_1)^2 + (\mathcal{Y}'_2)^2 }{4 \tau^2 (\mathcal{X}'_1-\mathcal{X}'_2)^2} = \frac{3}{4\tau^2} (\mathcal{X}'_1+\mathcal{X}'_2) = \frac{3}{2\tau^2} e_1 \left(-\frac{1}{2\tau} \right) = -3e_1(\tau) \, ,
\end{equation}
and we end up with 
\begin{equation}
\frac{(-2 e_1+\mathcal{X}_1+\mathcal{X}_2) \left(-2 e_1 e_2+e_1 \mathcal{X}_1+e_1 \mathcal{X}_2-2 e_2^2+2
   \mathcal{X}_1 \mathcal{X}_2\right)}{(\mathcal{X}_1-\mathcal{X}_2)^2} = 0 \, .
\end{equation}
Finally, we see that this equality is implied by the original point being on the original
branch $-2 e_1+\mathcal{X}_1+\mathcal{X}_2=0$, and we have therefore obtained a non-trivial check of the 
statement that the massless branch is invariant under $S_2$-duality.

\subsection{The Moduli Space of Vacua for the Different Gauge Theories}
\label{moduliSpaceso5}

In this subsection we obtain the global structure of the moduli space of massless vacua for the different theories with $so(5)$ gauge algebra, taking into account various discrete identifications. We also perform a consistency check on these global properties by showing how $S_2$-duality acts on these moduli spaces, thus completing the results in subsection \ref{globalso5}. 

We wish to characterize the branch by extracting the positions $z_i$ from the elliptic curve variables (\ref{ellipticso5variables}), and this should be done up to Weyl equivalence. The Weyl group is generated by two reflections: the reflection about $\alpha_1$ leads to the identification
\begin{equation}
(\mathcal{X}_1 , \mathcal{Y}_1 , \mathcal{X}_2 , \mathcal{Y}_2 ) \equiv (\mathcal{X}_2 , - \mathcal{Y}_2 , \mathcal{X}_1 , - \mathcal{Y}_1) \, , 
\end{equation}
while the reflection about $\alpha_2$ gives the identification
\begin{eqnarray}
(\mathcal{X}_1 , \mathcal{Y}_1 , \mathcal{X}_2 , \mathcal{Y}_2) \equiv (\mathcal{X}_1 , - \mathcal{Y}_1 , \mathcal{X}_2 , \mathcal{Y}_2) \, . 
\end{eqnarray}
This shows that the sign of the variables $\mathcal{Y}_i$ is irrelevant, and we no longer need to keep track of them. The Weyl symmetry therefore implies that we can study the manifold described by the variables $(\mathcal{X}_1 ,\mathcal{X}_2 )$ subject to the identification $(\mathcal{X}_1 ,  \mathcal{X}_2 ) \equiv (\mathcal{X}_2 ,  \mathcal{X}_1 )$. The branch of massless vacua of the $SO(5)_+$ theory, for which there is no additional identification, is described by  $\lambda \in \mathbb{C}^\ast/\mathbb{Z}_2$, where the $\mathbb{Z}_2$ action is $\lambda \leftrightarrow - \lambda$. This is a sphere with two excised points.

In the $SO(5)_-$ theory we have the additional identification $(z_1,z_2) \equiv (z_1 + \omega_1 + 2\omega_2 , z_2 + \omega_1)$. On the manifold parametrized by $\lambda$ it corresponds to $\lambda \equiv \lambda'=\pi^4 \theta_4^8 (2\tau) / \lambda$. This follows from the observation that if $\wp(z_1) = e_1 + \lambda$, then 
\begin{eqnarray}
\wp (z_1 + \omega_1) &=& - e_1 - (e_1 + \lambda) + \frac{\wp '(z_1)^2}{4\lambda^2} \\
 &=& e_1 + \frac{3e_1^2 - \frac{1}{4} g_2}{\lambda} + \frac{4e_1^3 - g_2 e_1 - g_3}{\lambda^2} \\
 &=& e_1 + \lambda ' \, ,
\end{eqnarray}
and similarly if $\wp(z_2) = e_1 - \lambda$ then $\wp (z_2 + \omega_1) = e_1 - \lambda '$. Note that the function $\theta_4(\tau)$ doesn't vanish on the upper-half plane,\footnote{The zeros of $\theta_4(z,\tau)$ are given by $z=n+(m+1/2) \tau $ with $n,m \in \mathbb{Z}$} so that $\lambda \mapsto \lambda'$ is a well-defined involution everywhere in the moduli space. 

For a given $\lambda \in \mathbb{C}^\ast$, the $SO(5)_-$ theory has two non-equivalent vacua $(z_1,z_2)$ and $(z_1 + 2 \omega_2 , z_2)$ which correspond to this $\lambda$. These two vacua are respectively equivalent to $(z_1+\omega_1 + 2\omega_2, z_2 + \omega_1)$ and $(z_1+\omega_1 , z_2 + \omega_1)$, which are associated to the same $\lambda '$. Therefore the branch of massless vacua is a double cover of $\mathbb{C}^\ast / \mathbb{Z}_2$. 

As for the $Spin(5)$ theory, we also need to take a double cover of the quotiented sphere $\mathbb{C}^\ast / \mathbb{Z}_2$. For a generic $\lambda \in \mathbb{C}^\ast$, the two vacua $\lambda$ and $\lambda '$ in the $SO(5)_+$ theory are inequivalent. They are mapped by $S_2$ to inequivalent vacua that share the same dual $\lambda_D$, or equivalently the same $\lambda '_D$. We see that $S_2$-duality cancels the cover and the quotient to recover the manifold for $SO(5)_+$ which is just $\mathbb{C}^\ast$.

\section{Conclusions}
\label{conclusions}
In this paper, we  further explored the vacuum structure and duality properties of ${\cal N}=1^\ast$ gauge theories.
We also found new phenomena in elliptic integrable systems. In the latter, we identified limits which exhibit a staircase
structure -- degrees of freedom are fixed at various powers of the modular parameter $q$ of the integrable system. This has a counterpart in the instanton effects responsible for fixing vacuum expectation values in the gauge theory. For the gauge theory compactified on the circle, we clarified multiple phenomena.  
We have exhibited  massive vacua of ${\cal N}=1^\ast$ gauge theories associated to discrete
component groups of nilpotent orbits. We also found vacua that become massive due to discrete Wilson lines.
Moreover, we started the study of branches of massless vacua of the ${\cal N}=1^\ast$ theory. For the $su(3)$ gauge algebra, we gave
the equation for the massless branch, and identified the (Argyres-Douglas) singularity. We moreover plunged into the  elliptic function theory that enters the exact description of the corresponding equilibrium
positions of the elliptic integrable systems, and  their duality properties. We thus provided a physical application (and extension) of the
Eichler-Zagier formulas. Moreover, we laid bare the massless branch of vacua for the theory with $so(5)$ gauge group. Our analysis invoked a combination of the rich semi-classical limits of  elliptic integrable systems, numerical data, modular forms and elliptic function theory.

We have walked into a field which is littered with interesting open problems. Let us enumerate
just a few.
\begin{itemize}
\item Count and characterize massive vacua and  massless branches (of differing dimension) of vacua of ${\cal N}=1^\ast$
gauge theories on $\mathbb{R}^{2,1} \times S^1$, or ${\cal N}=1$ theories in general.
\item Compute the duality diagram for all the vacua. 
\item Understand the (vector valued) modular objects with monodromies in the interior and on the boundary 
of the fundamental domain that naturally appear as equilibrium positions, as well as those that appear as extremal
values of the potential of elliptic integrable systems. 
\item Analyze the desingularization of the effective superpotential when it develops monodromies.
\item Classify complex extrema of integrable systems.
\item Compute all possible staircase limits of elliptic integrable systems,
as well as their extrema, and generalize these limits, for instance to integrable systems with spin.
\item Identify all massive and massless vacua on $\mathbb{R}^{2,1} \times S^1$ from 
(a generalization to the compactified theory of) the Seiberg-Witten curve of the ${\cal N}=2^\ast$ theory.
\item Investigate the relation to the geometric Langlands program and nilpotent orbit theory as
applied to defect theories.
\item Complete the analysis of global aspects of the theory, including the global choice
of gauge group and the spectrum of line operators.
\end{itemize}
We hope to revisit this field fruitfully in the future.

\section*{Acknowledgments}
We would like to thank Antonio Amariti, Costas Bachas, David Berenstein, Nick Dorey and Amihay Hanany for useful
discussions. We would like to acknowledge support from the grant ANR-13-BS05-0001, and from the \'Ecole Polytechnique and the \'Ecole Nationale Sup\'erieure des Mines de Paris.

\appendix

\section{The Lie Algebra and the Group $G_2$}

 \begin{figure}[H]
 \centering
 \includegraphics[width=0.2\textwidth]{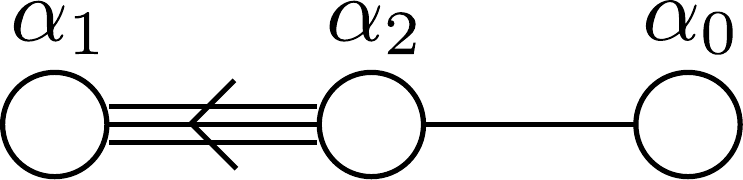}
 \caption{The Dynkin diagram of the affine algebra $G_2^{(1)}$. }
 \end{figure}

\label{g2app}
The  Cartan matrix of $G_2$ is
\begin{eqnarray}
A_{G_2} &=&
\left(
\begin{array}{cc}
2 & -1 
\\
-3 & 2
\end{array}
\right) \, .
\end{eqnarray}
The adjoint representation has dimension 14.
We can represent the root and weight system in terms of three linearly dependent vectors
$\epsilon_{1,2,3}$ satisfying $\epsilon_1+\epsilon_2+\epsilon_3=0$. 
The basis we use is $(\epsilon_1,\epsilon_2)$. It satisfies the relations $|\epsilon_1|^2=|\epsilon_2|^2=\frac{2}{3}$ and $\epsilon_1 \cdot \epsilon_2 = - \frac{1}{3}$.
(In fact, analogous relations hold for all $\epsilon_{i}$.) The roots are given
by $\epsilon_i \pm \epsilon_j$ ($ i \neq j$) and $\pm \epsilon_i$ for a total of $12$.
We have 6 positive roots.
There are three short positive roots, $e_1$, $\alpha_1 = -\epsilon_2$ and $\epsilon_1+\epsilon_2$, with length squared $\frac{2}{3}$ and three long roots, $\alpha_2 = \epsilon_1 + 2\epsilon_2$, $2\epsilon_1 + \epsilon_2$ and $\epsilon_1-\epsilon_2$ with length squared $2$. The ratio of lengths squared is equal to three.
The simple roots can be represented as $\alpha_1=-\epsilon_2$ and $\alpha_2 = \epsilon_2-\epsilon_3$.
The highest root is $\epsilon_1-\epsilon_3$, and it is also equal to the second fundamental weight.
The first fundamental weight is $\epsilon_1$.
The weight lattice is spanned by the $\epsilon_i$.

The co-roots $\alpha_1^{\vee}=-3\epsilon_2$ and $\alpha_2^{\vee}=\epsilon_1 + 2\epsilon_2$ have length
squared equal to $6$ and $2$ respectively. We deduce the fundamental co-weights 
$\pi_1^{\vee}=3\epsilon_1$ and $\pi_2^{\vee}=2\epsilon_1+\epsilon_2$ and the fundamental
weights $\pi_1=\epsilon_1$ and $\pi_2=2\epsilon_1+\epsilon_2$.
 Finally, the Weyl group has 12 elements, it is precisely 
 \begin{equation}
 \left\{ (r_1 r_2)^n (r_1)^{\epsilon} | 0 \leq n \leq 5 \quad \textrm{and} \quad \epsilon = 0,1 \right\} \, ,
 \end{equation}
 where $r_i$ are simple Weyl reflections.
 One of those elements, $(r_1 r_2)^2 r_1$, exchanges $\epsilon_1$ and $\epsilon_2$, meaning that in the bulk of the paper, extrema with $z_1$ and $z_2$ exchanged are  considered equivalent. A global sign flip is also allowed by $(r_1 r_2)^3=-1$. Finally $(r_1 r_2)^3 r_1$ acts as 
 \begin{equation}
 \begin{cases}
 \epsilon_1 \rightarrow \epsilon_1 + \epsilon_2  \\
 \epsilon_2 \rightarrow - \epsilon_2 \, .
 \end{cases}
 \end{equation}
 The final litany of useful facts includes that both the center of $G_2$ and its group of outer automorphisms  are trivial. 
The algebra $G_2$ is its own Langlands dual. The dual of the (non-twisted) affine algebra $G_2^{(1)}$ on the
other hand is $\left( G_2^{(1)} \right)^\vee = D_4^{(3)}$. This last algebra has two short simple roots 
and one long simple root whose length squared is three times larger.
The co-marks of $g_2^{(1)}$ are $(1,2,1)$.

 \section{Representations of the Vacua for $B_2$ Theories}
\begin{figure}[H]
\begin{minipage}{\linewidth}
\begin{tabular}{p{4.6cm}p{4.6cm}p{4.6cm}}
{
\setlength{\fboxsep}{8pt}
\setlength{\fboxrule}{0pt}
\fbox{\includegraphics[width=3.05cm]{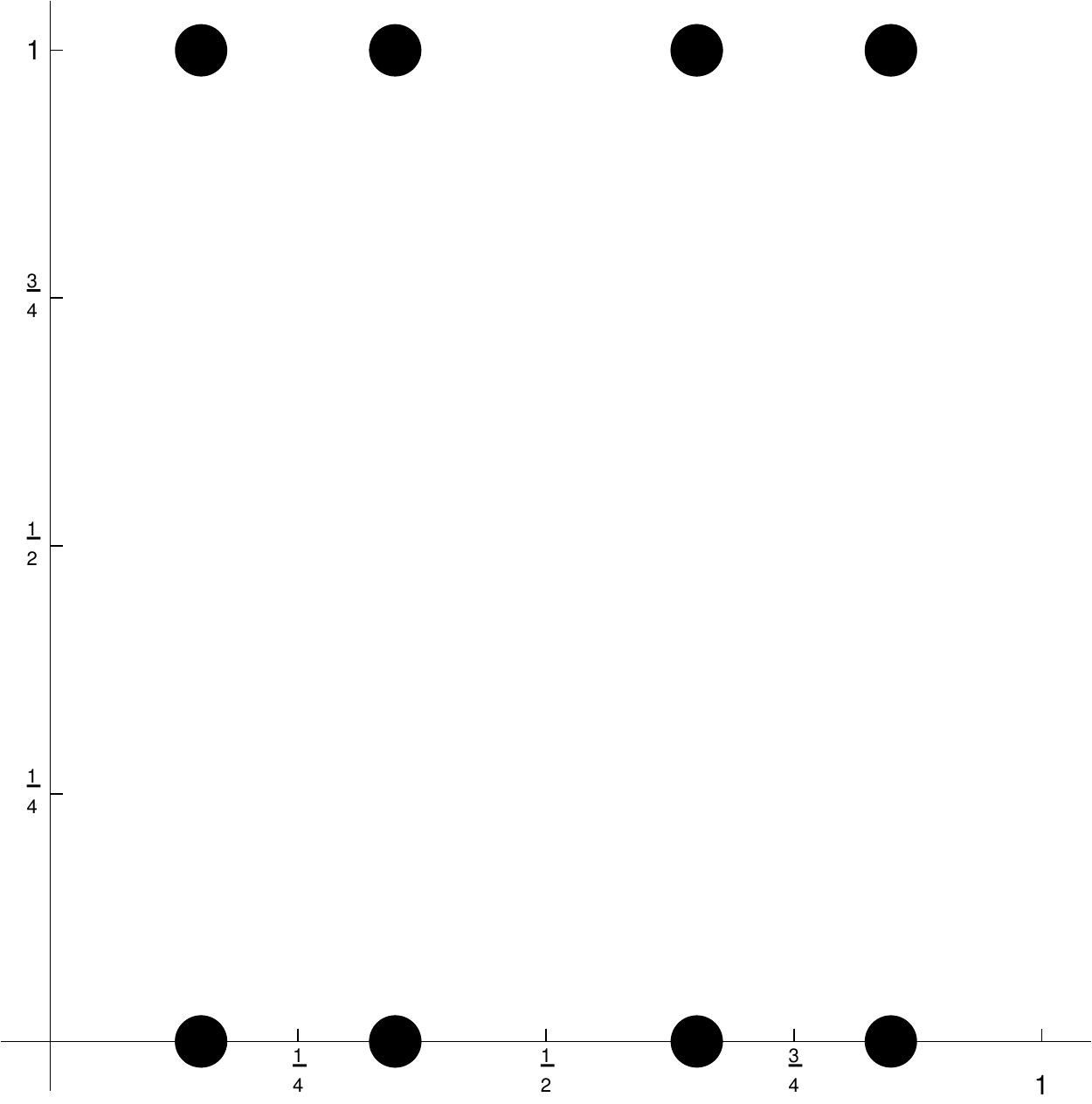}}
}
\begin{center}
Extremum 1
\end{center}
&
{
\setlength{\fboxsep}{8pt}
\setlength{\fboxrule}{0pt}
\fbox{\includegraphics[width=3.05cm]{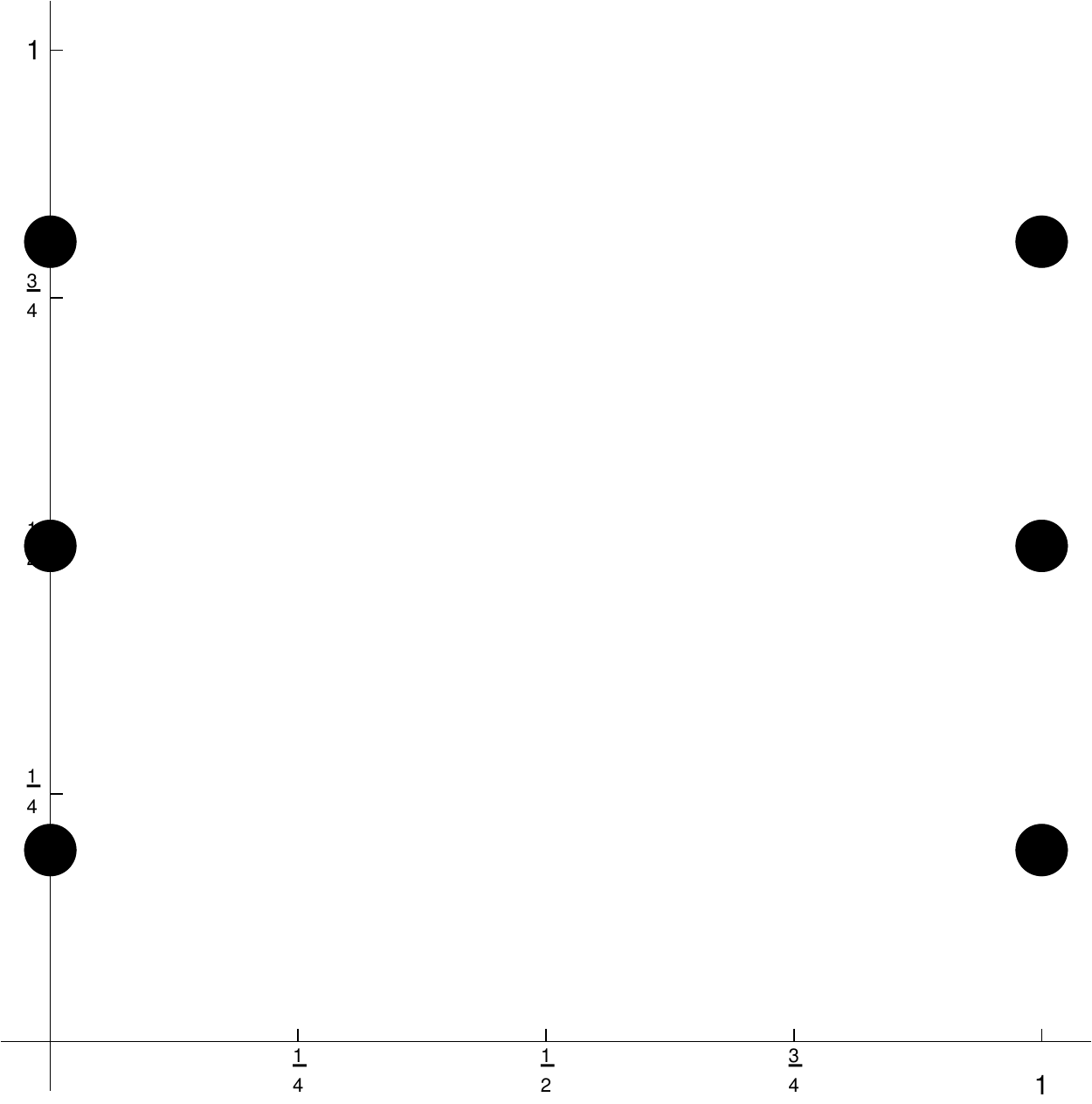}}
}
\begin{center}
Extremum 2
\end{center}
 &
 {
\setlength{\fboxsep}{8pt}
\setlength{\fboxrule}{0pt}
\fbox{\includegraphics[width=3.05cm]{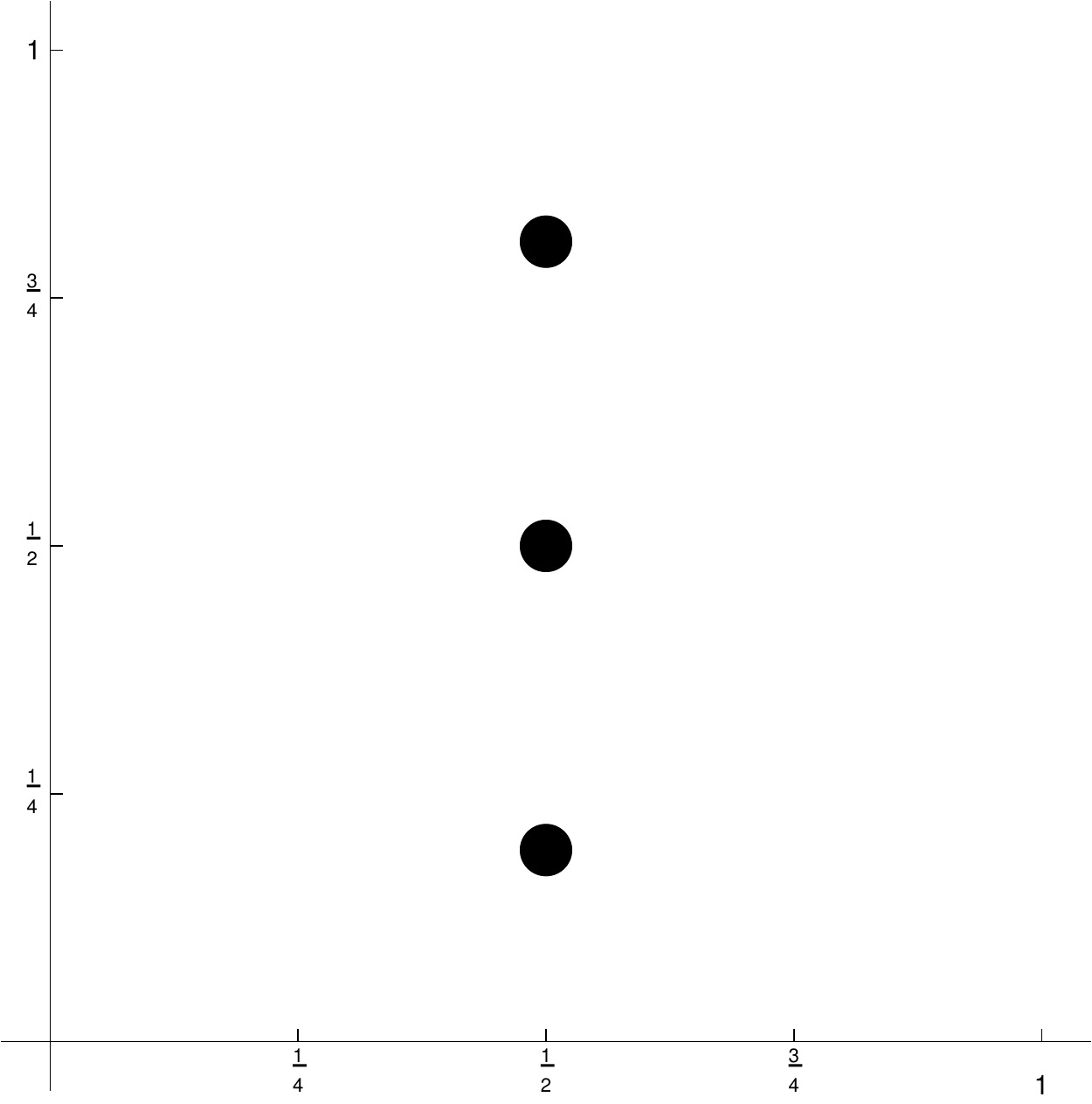}}
}
\begin{center}
Extremum 2b
\end{center}
\\
{
\setlength{\fboxsep}{8pt}
\setlength{\fboxrule}{0pt}
\fbox{\includegraphics[width=3.05cm]{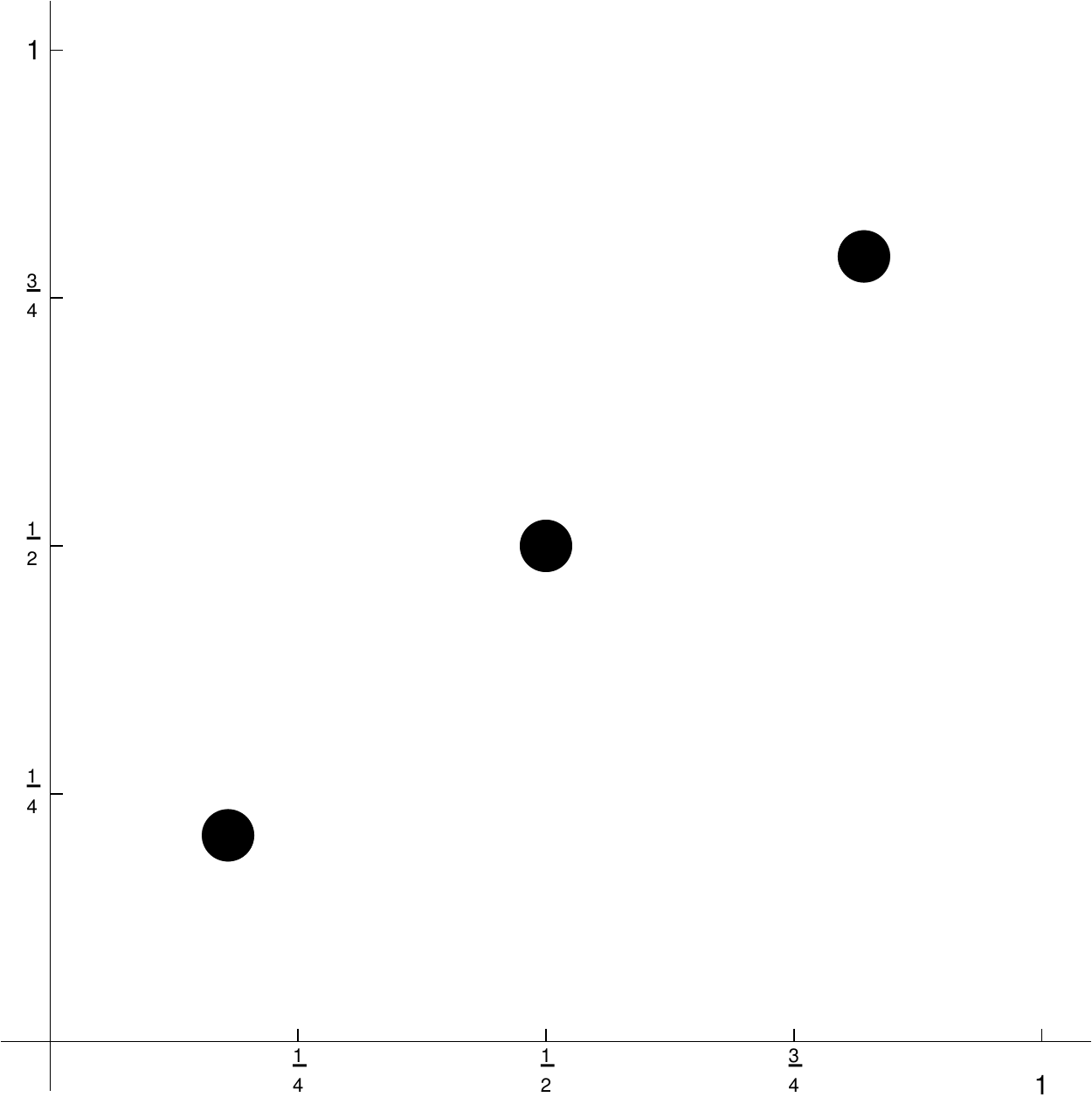}}
}
\begin{center}
Extremum 3
\end{center}
 &
 {
\setlength{\fboxsep}{8pt}
\setlength{\fboxrule}{0pt}
\fbox{\includegraphics[width=3.05cm]{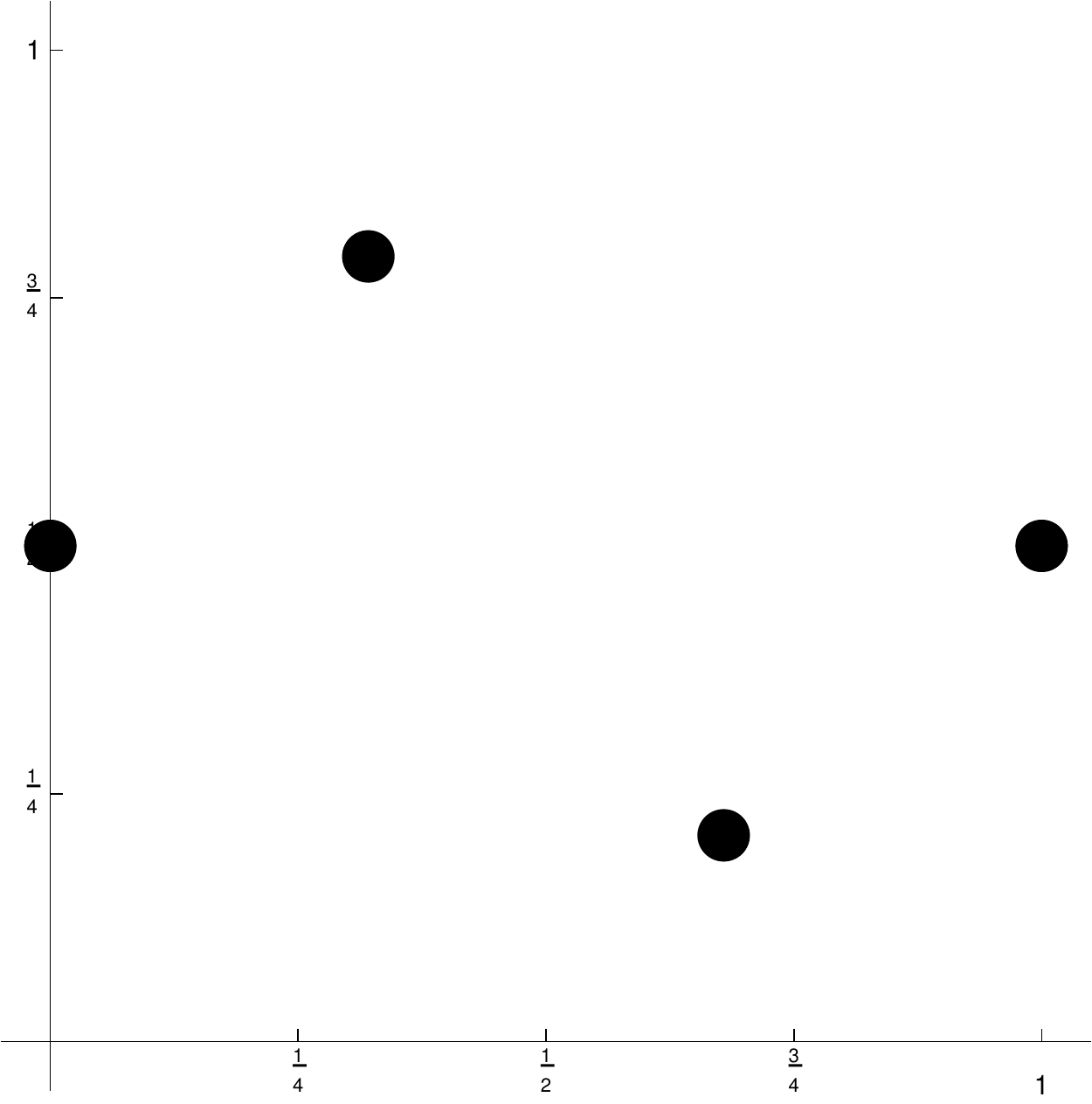}}
}
\begin{center}
Extremum 3b
\end{center}
&
{
\setlength{\fboxsep}{8pt}
\setlength{\fboxrule}{0pt}
\fbox{\includegraphics[width=3.05cm]{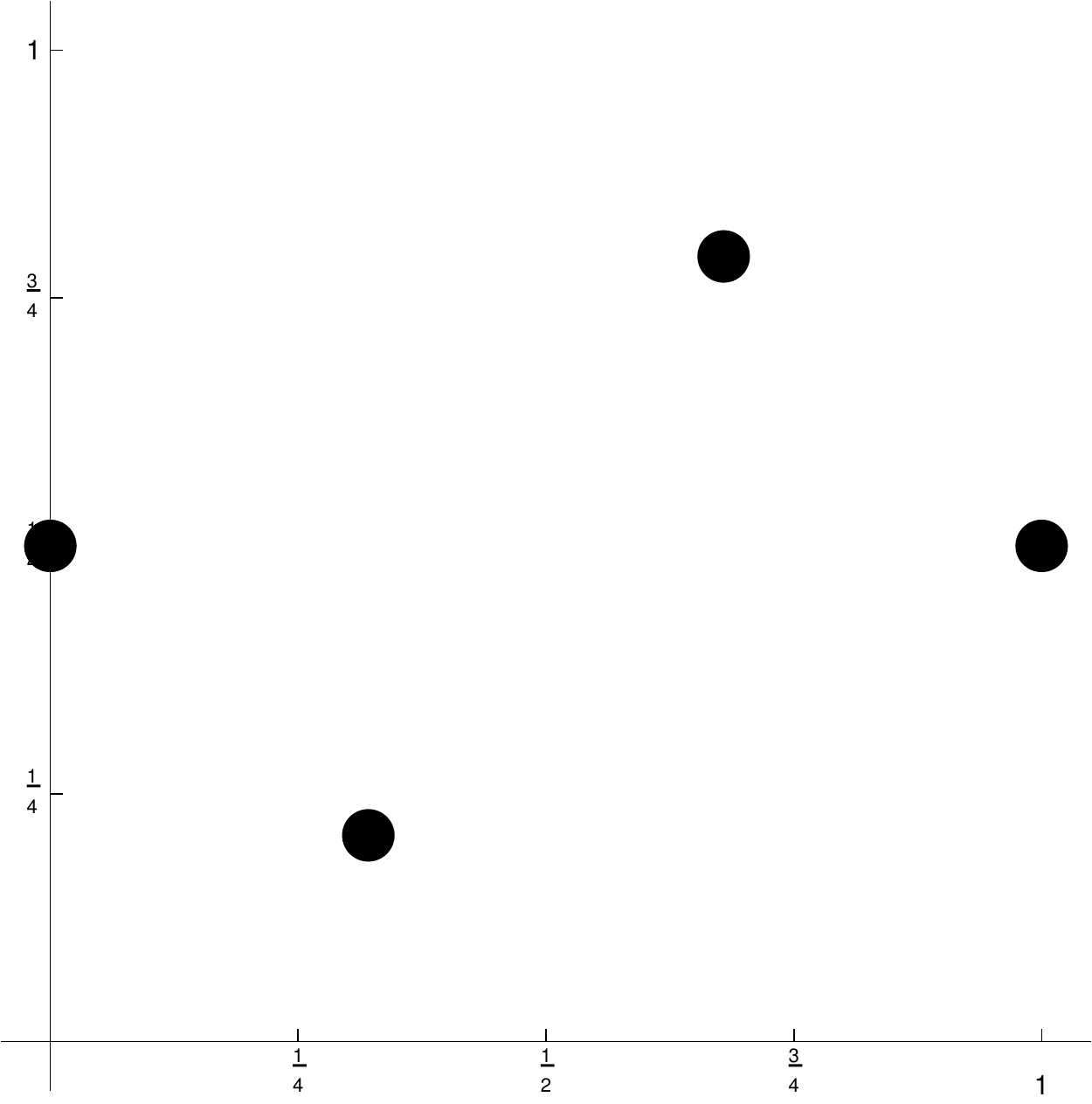}}
}
\begin{center}
Extremum 4
\end{center}
\\
 {
\setlength{\fboxsep}{8pt}
\setlength{\fboxrule}{0pt}
\fbox{\includegraphics[width=3.05cm]{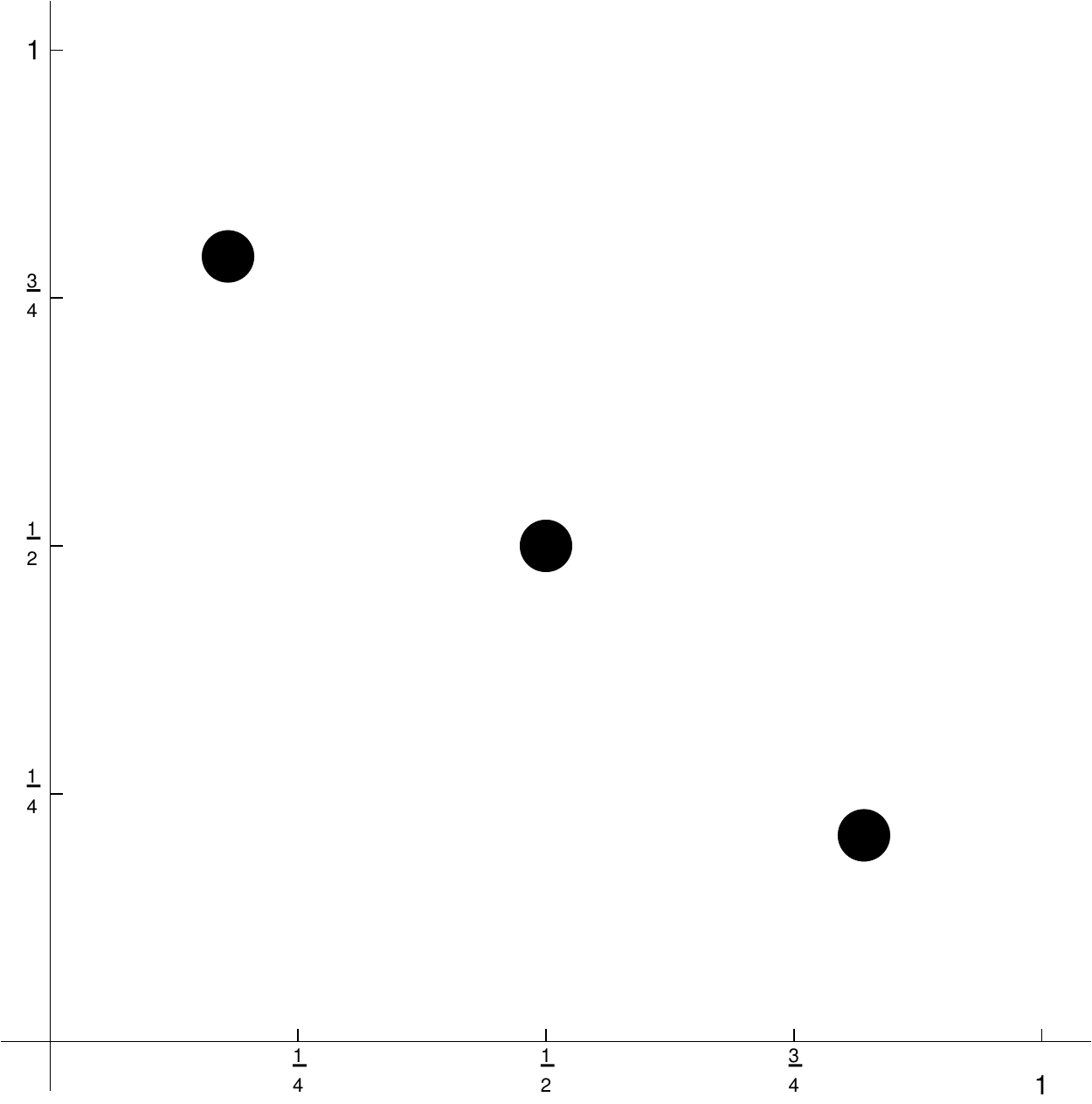}}
}
\begin{center}
Extremum 4b
\end{center}
&
{
\setlength{\fboxsep}{8pt}
\setlength{\fboxrule}{0pt}
\fbox{\includegraphics[width=3.05cm]{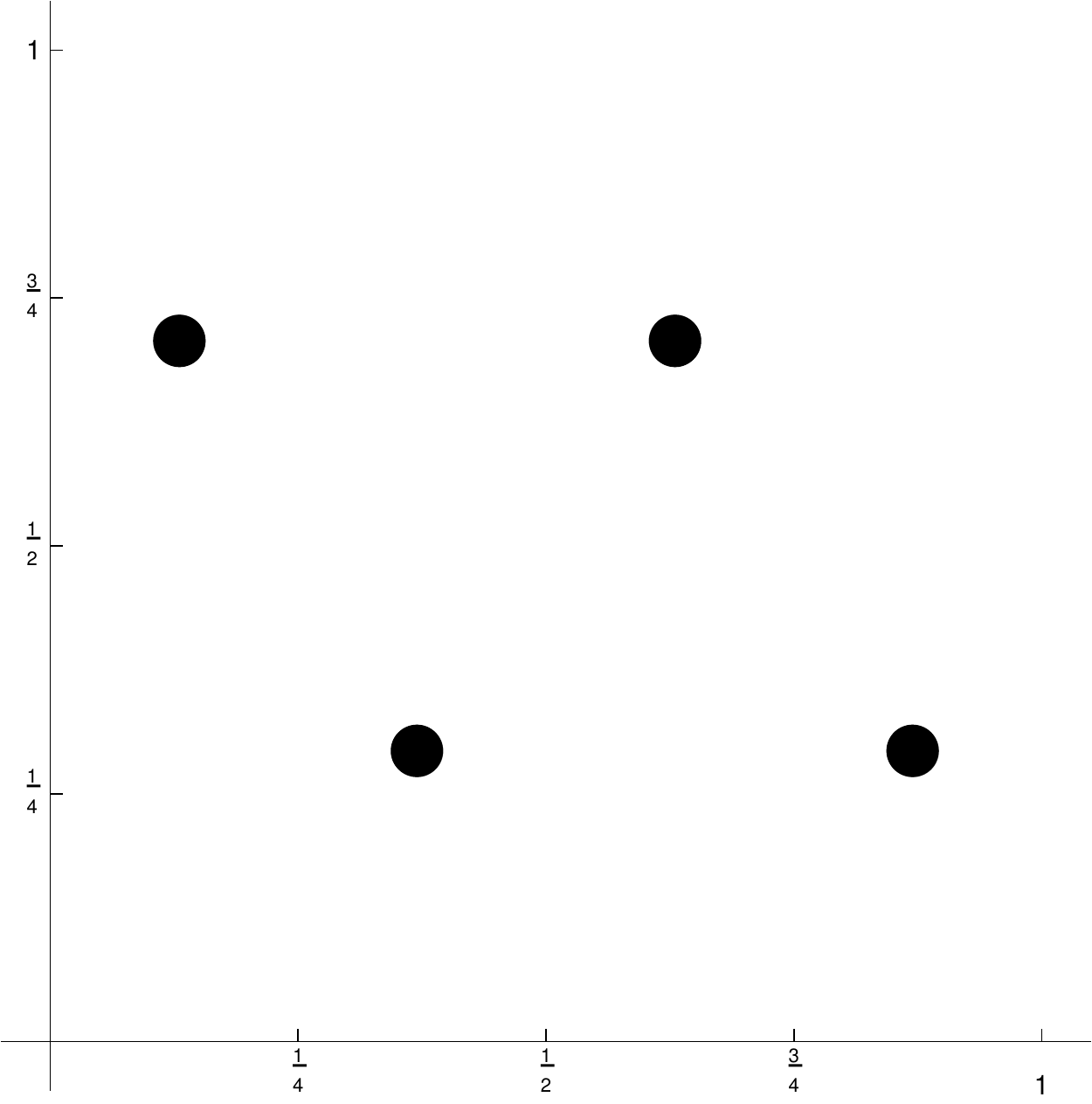}}
}
\begin{center}
Extremum 5
\end{center}
&
{
\setlength{\fboxsep}{8pt}
\setlength{\fboxrule}{0pt}
\fbox{\includegraphics[width=3.05cm]{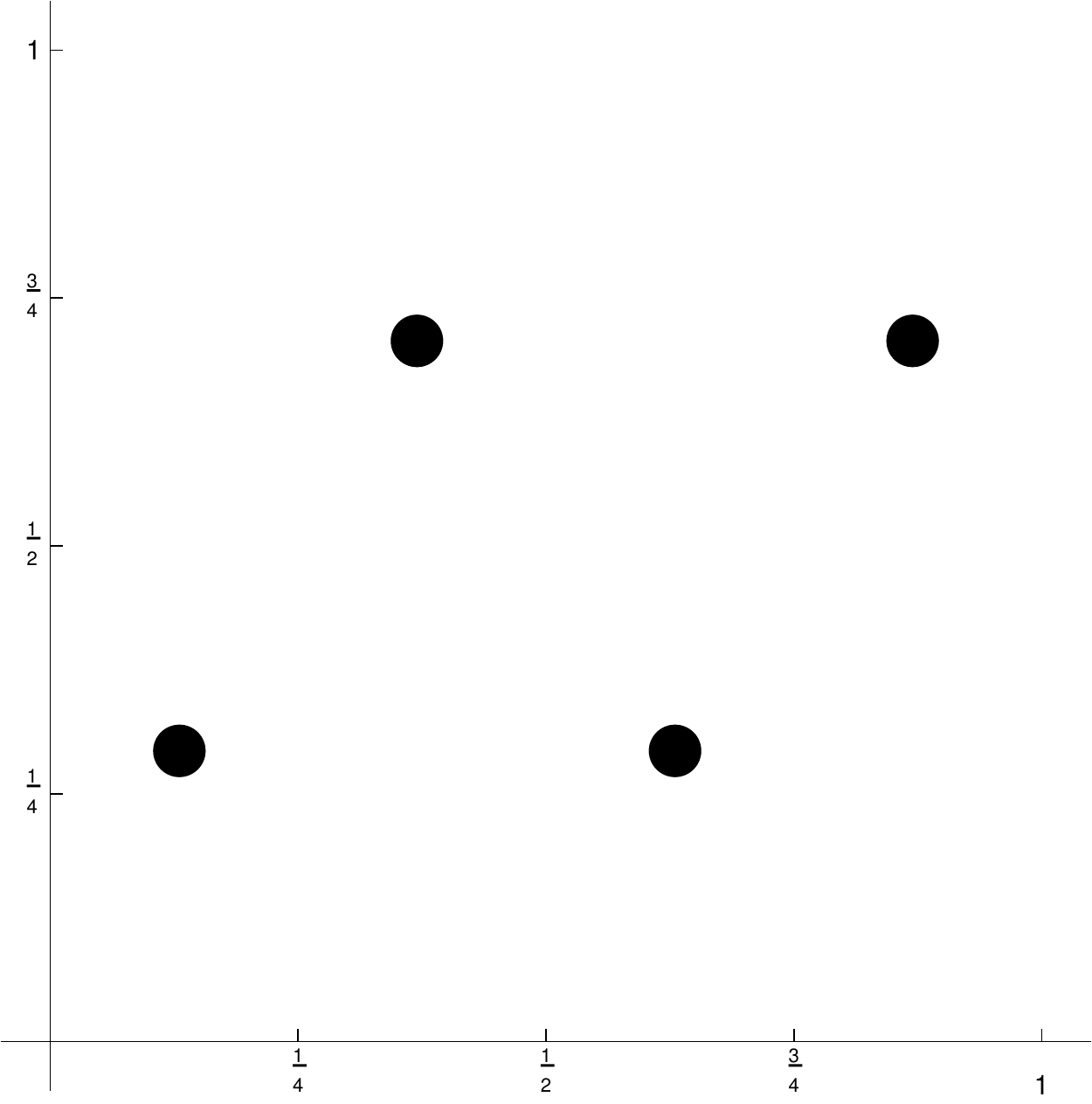}}
}
\begin{center}
Extremum 6
\end{center}
\\
{
\setlength{\fboxsep}{8pt}
\setlength{\fboxrule}{0pt}
\fbox{\includegraphics[width=3.05cm]{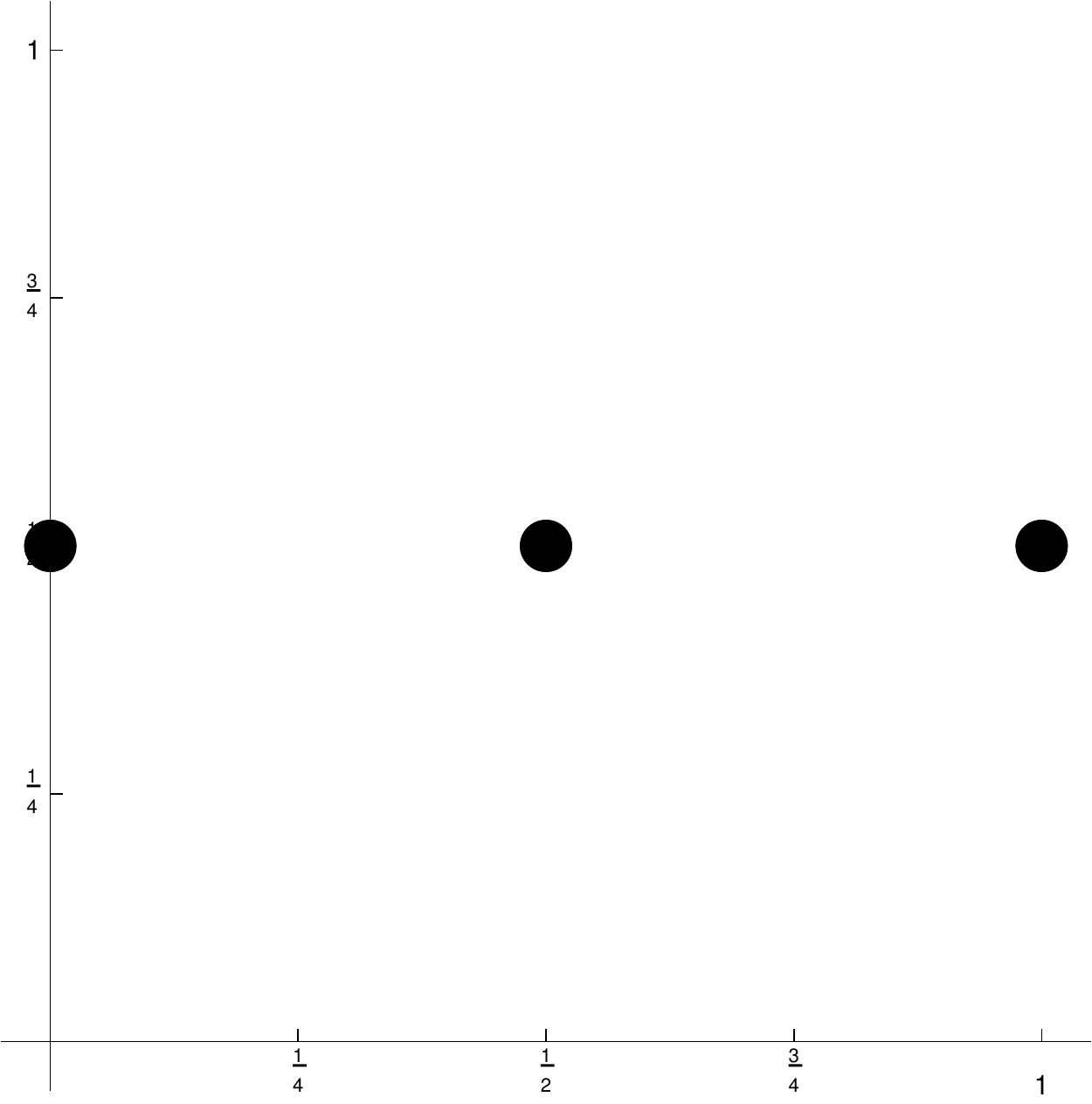}}
}
\begin{center}
Extremum 7
\end{center}
\end{tabular}
\end{minipage}
\caption{The extremal positions of the vacua for the $SO(5)_+$ theory.}
\label{extremaso5plus}
\end{figure}

\begin{figure}[H]
\begin{minipage}{\linewidth}
\begin{tabular}{p{4.6cm}p{4.6cm}p{4.6cm}}
{
\setlength{\fboxsep}{8pt}
\setlength{\fboxrule}{0pt}
\fbox{\includegraphics[width=3.05cm]{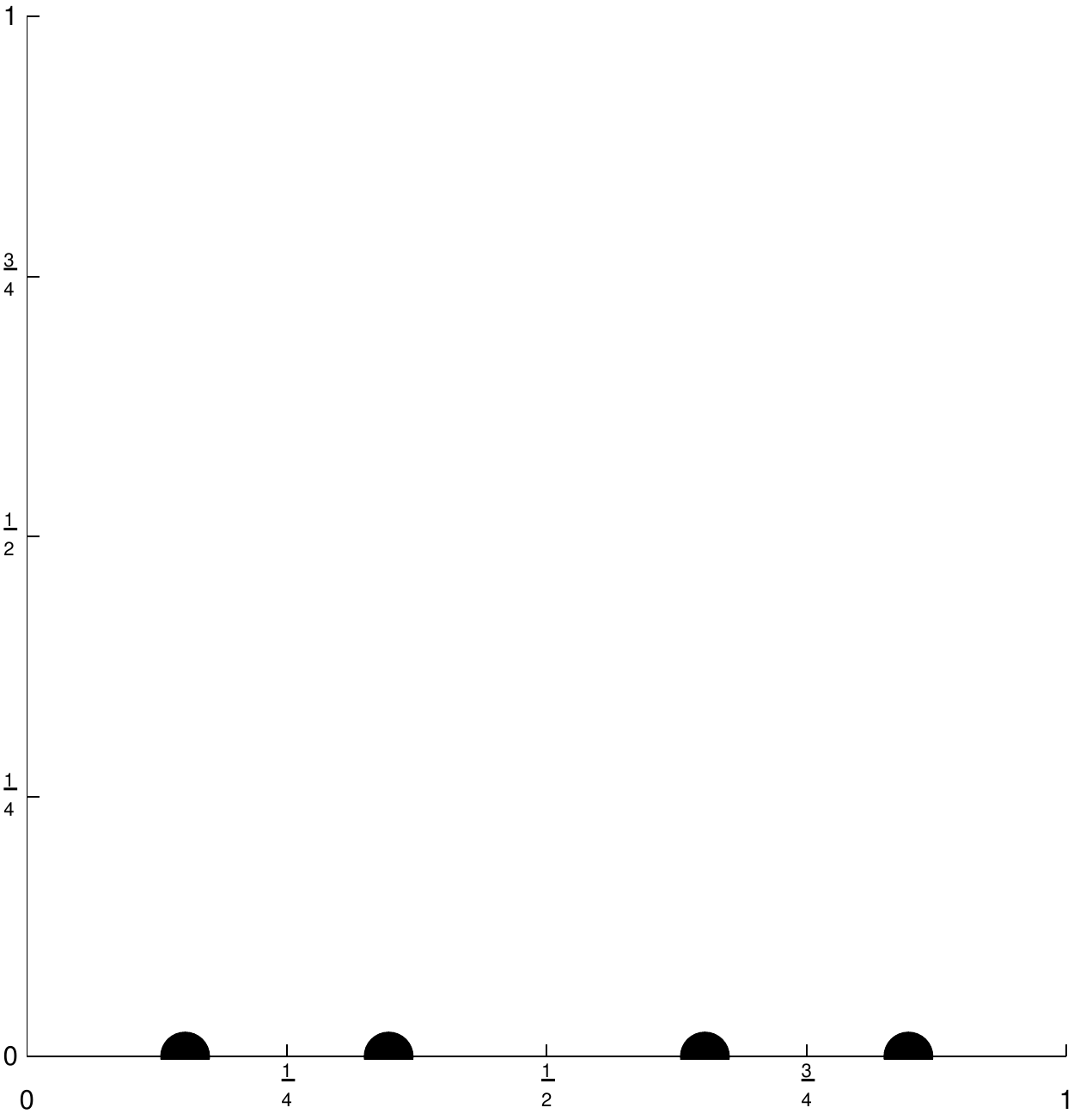}}
}
\begin{center}
Extremum 1
\end{center}
&
{
\setlength{\fboxsep}{8pt}
\setlength{\fboxrule}{0pt}
\fbox{\includegraphics[width=3.05cm]{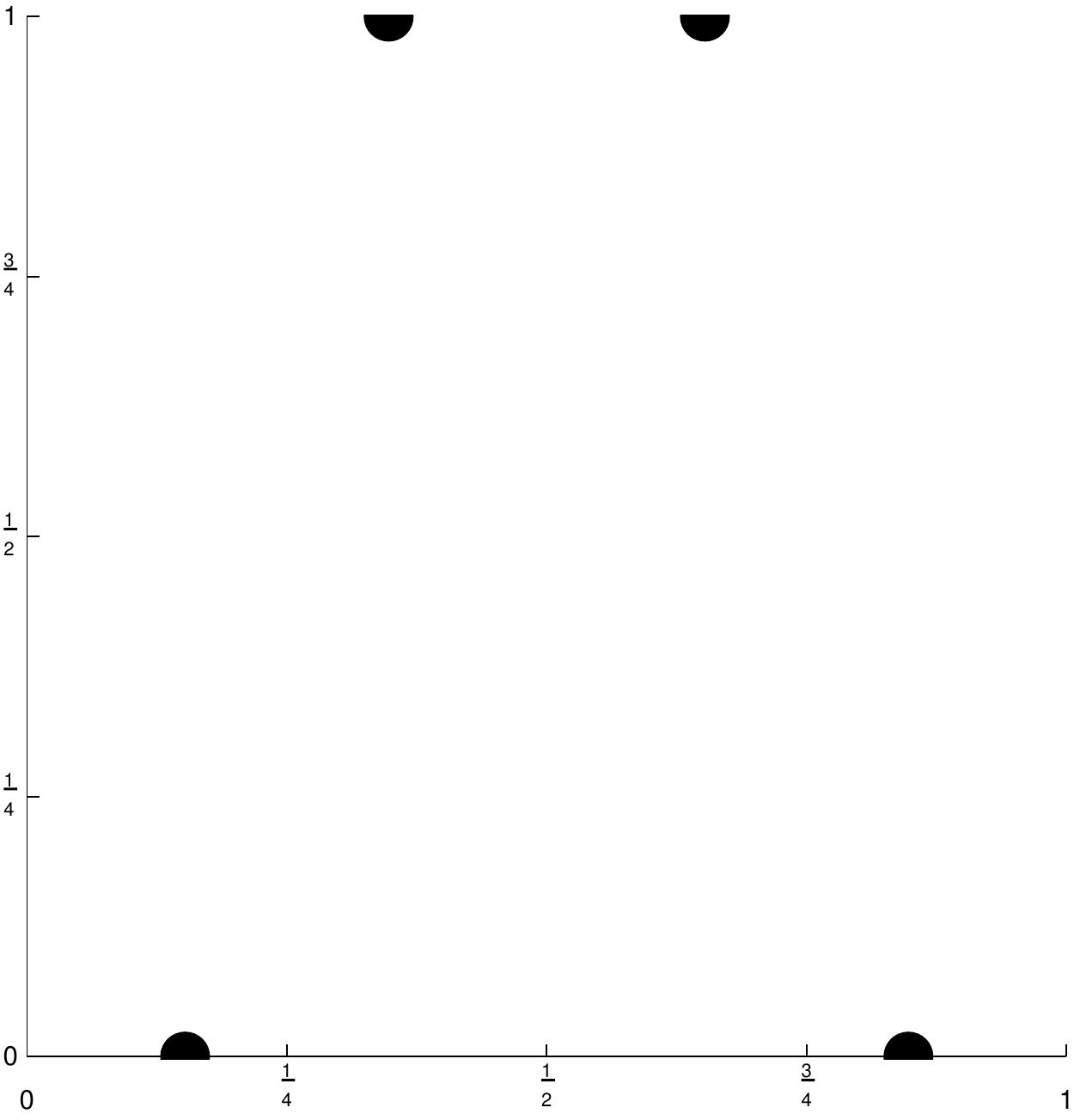}}
}
\begin{center}
Extremum 1b
\end{center}
 &
 {
\setlength{\fboxsep}{8pt}
\setlength{\fboxrule}{0pt}
\fbox{\includegraphics[width=3.05cm]{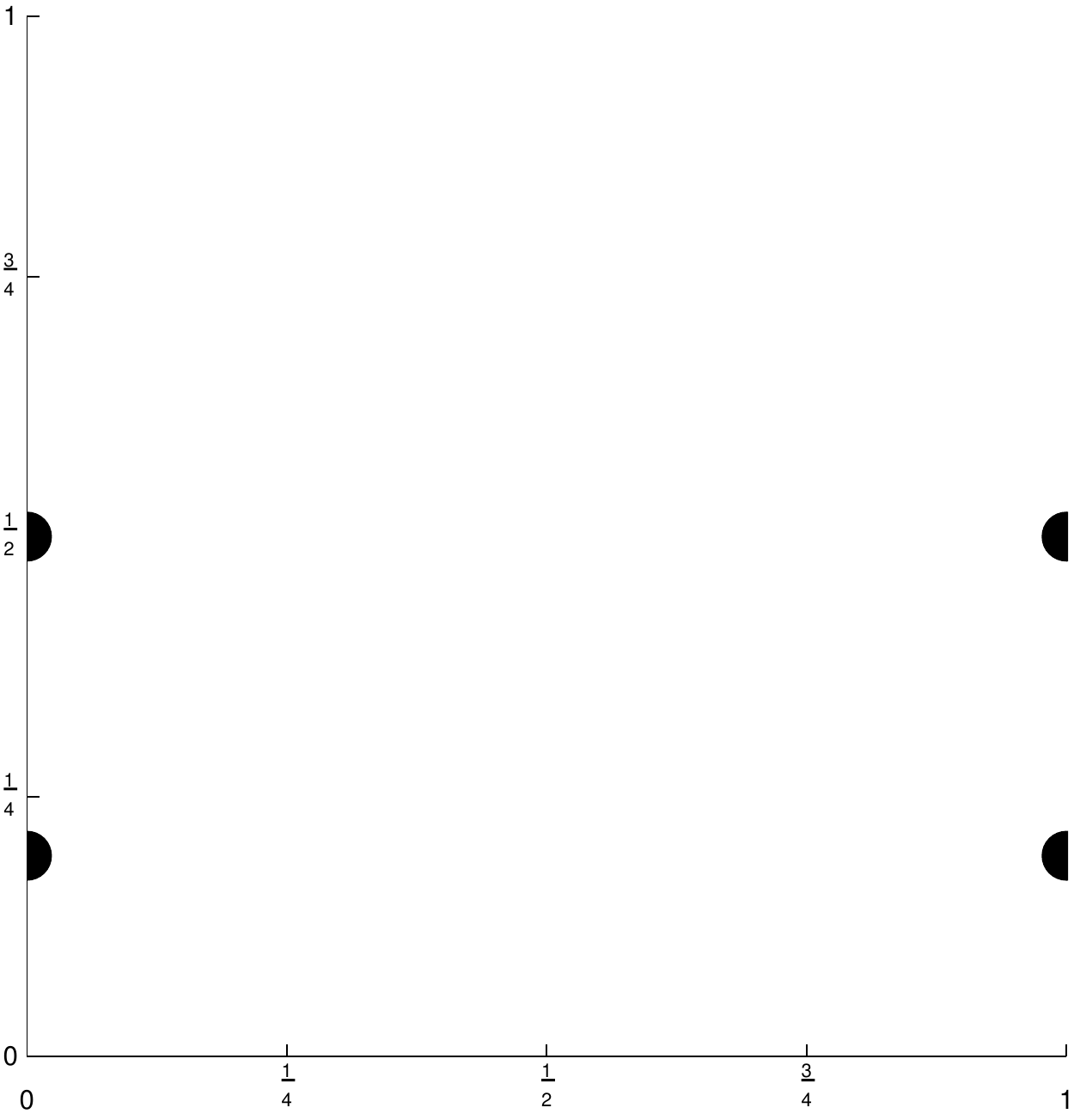}}
}
\begin{center}
Extremum 2
\end{center}
\\
{
\setlength{\fboxsep}{8pt}
\setlength{\fboxrule}{0pt}
\fbox{\includegraphics[width=3.05cm]{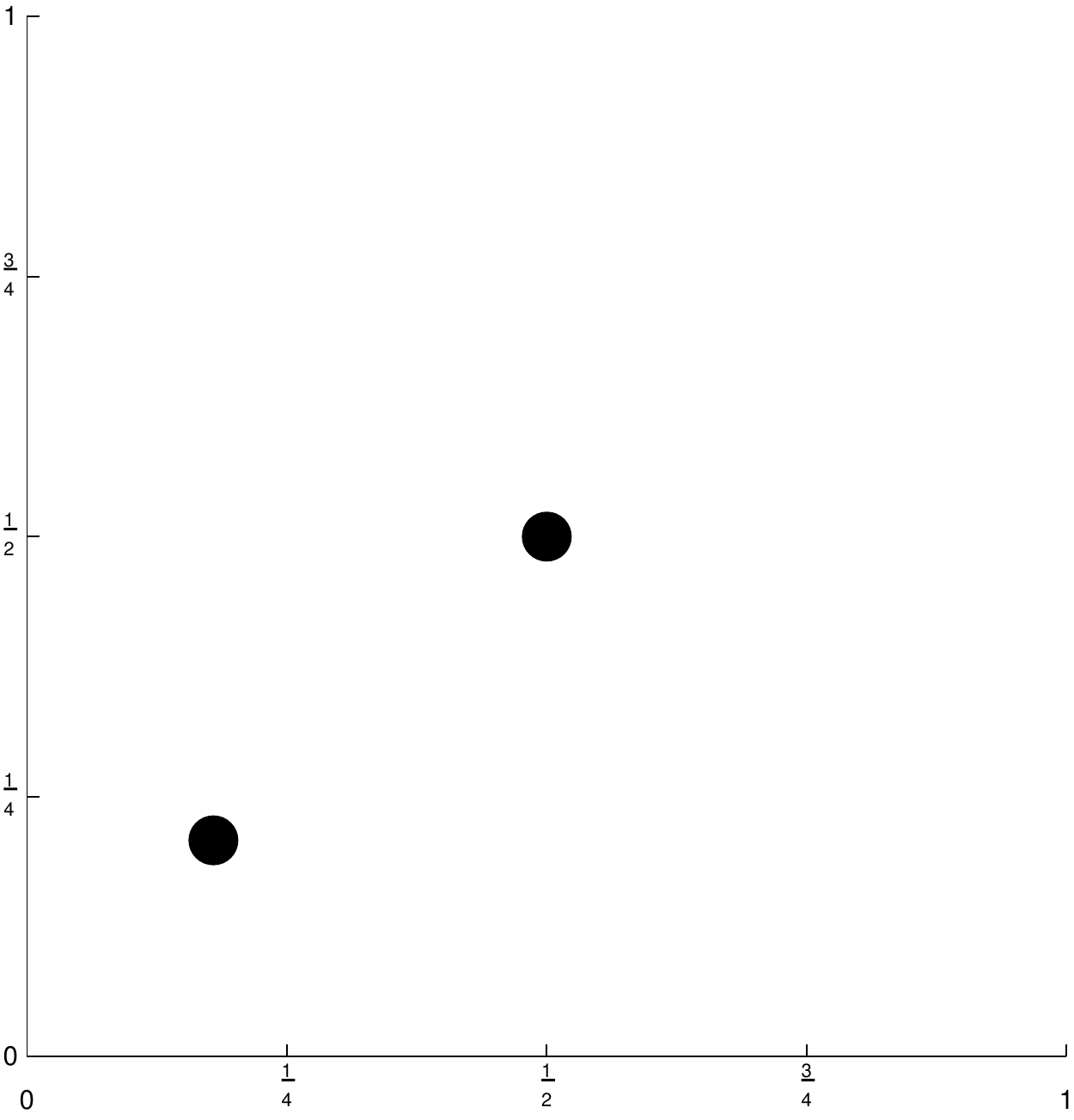}}
}
\begin{center}
Extremum 3
\end{center}
 &
 {
\setlength{\fboxsep}{8pt}
\setlength{\fboxrule}{0pt}
\fbox{\includegraphics[width=3.05cm]{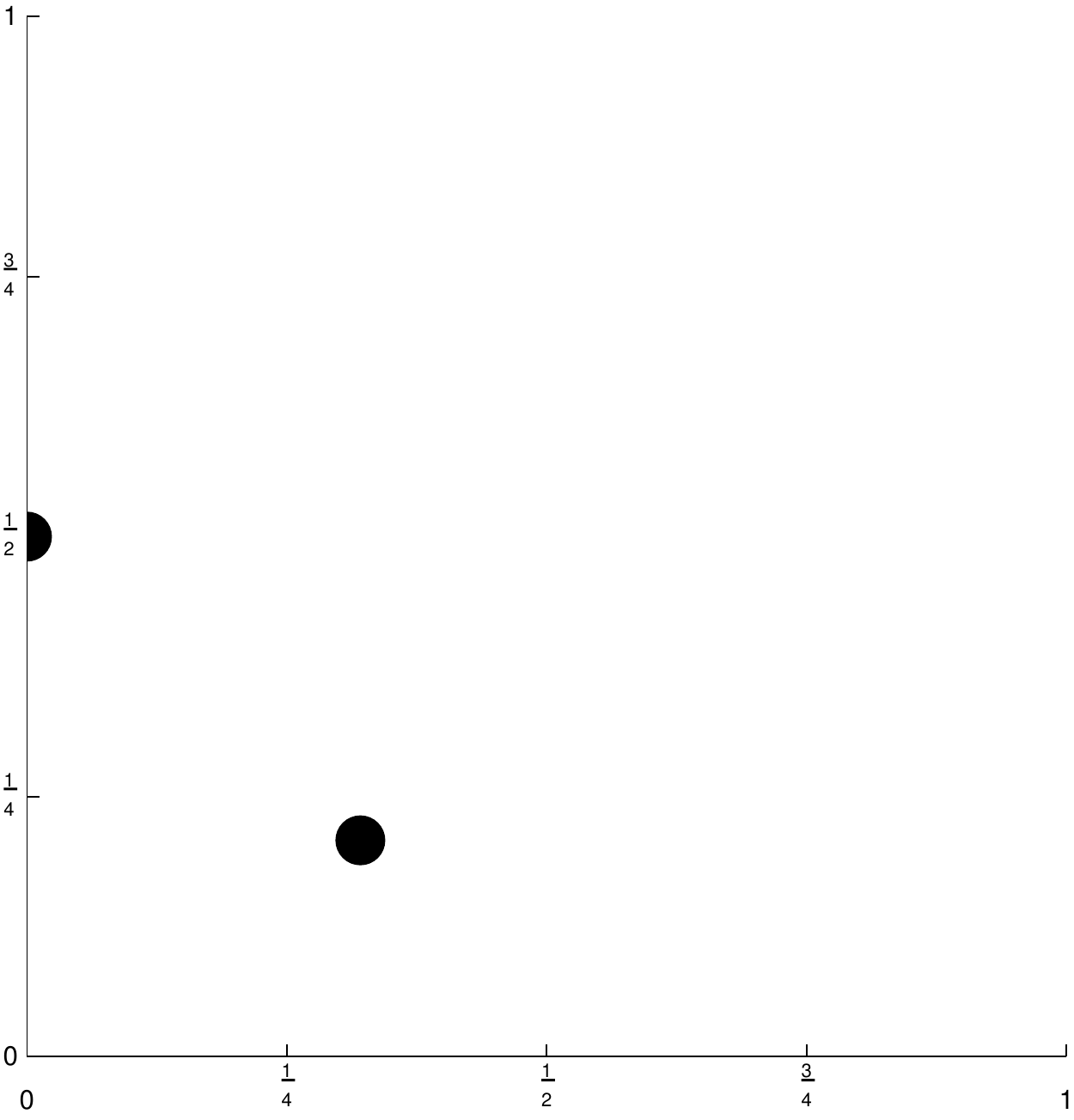}}
}
\begin{center}
Extremum 4
\end{center}
&
{
\setlength{\fboxsep}{8pt}
\setlength{\fboxrule}{0pt}
\fbox{\includegraphics[width=3.05cm]{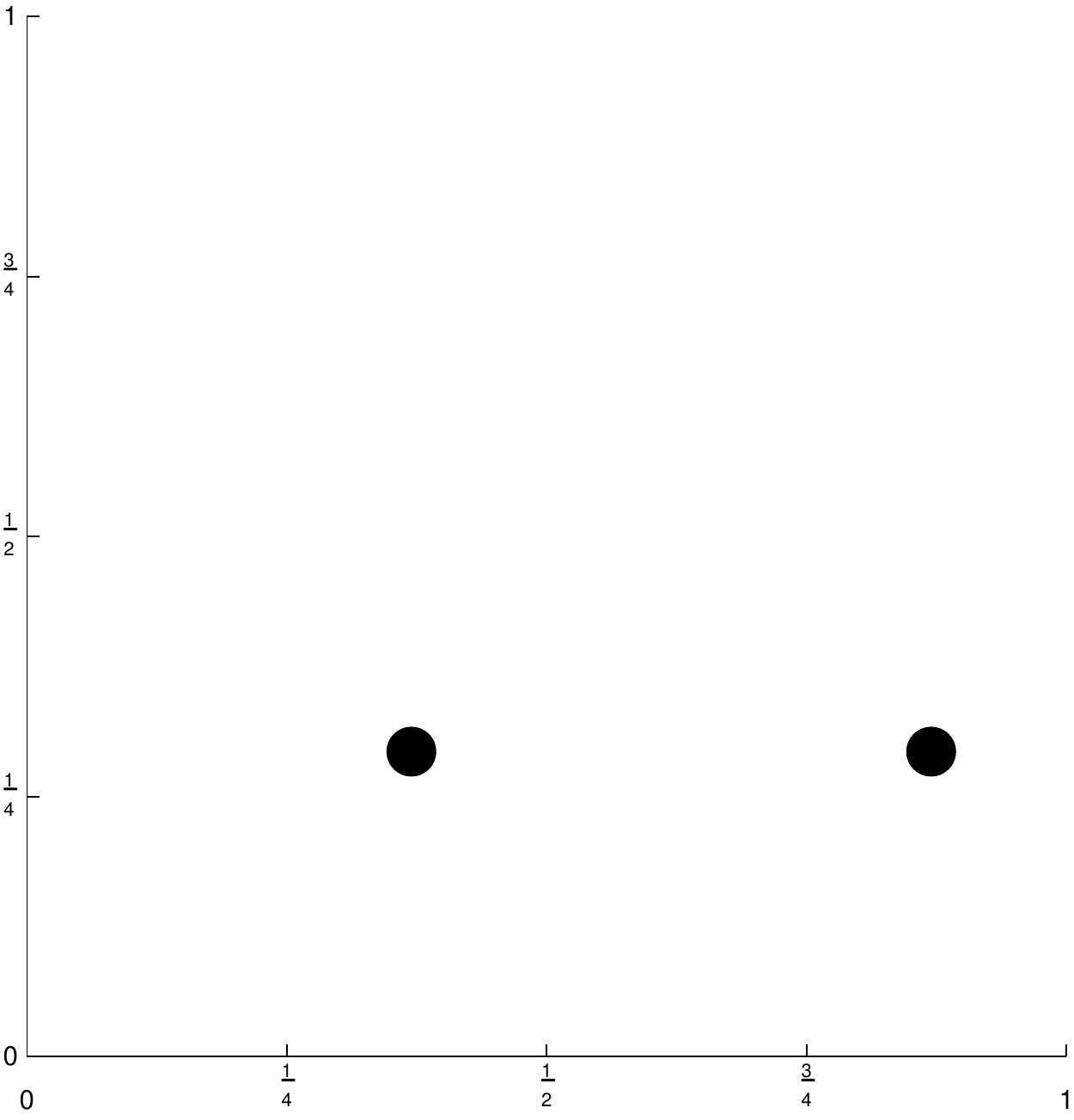}}
}
\begin{center}
Extremum 5
\end{center}
\\
 {
\setlength{\fboxsep}{8pt}
\setlength{\fboxrule}{0pt}
\fbox{\includegraphics[width=3.05cm]{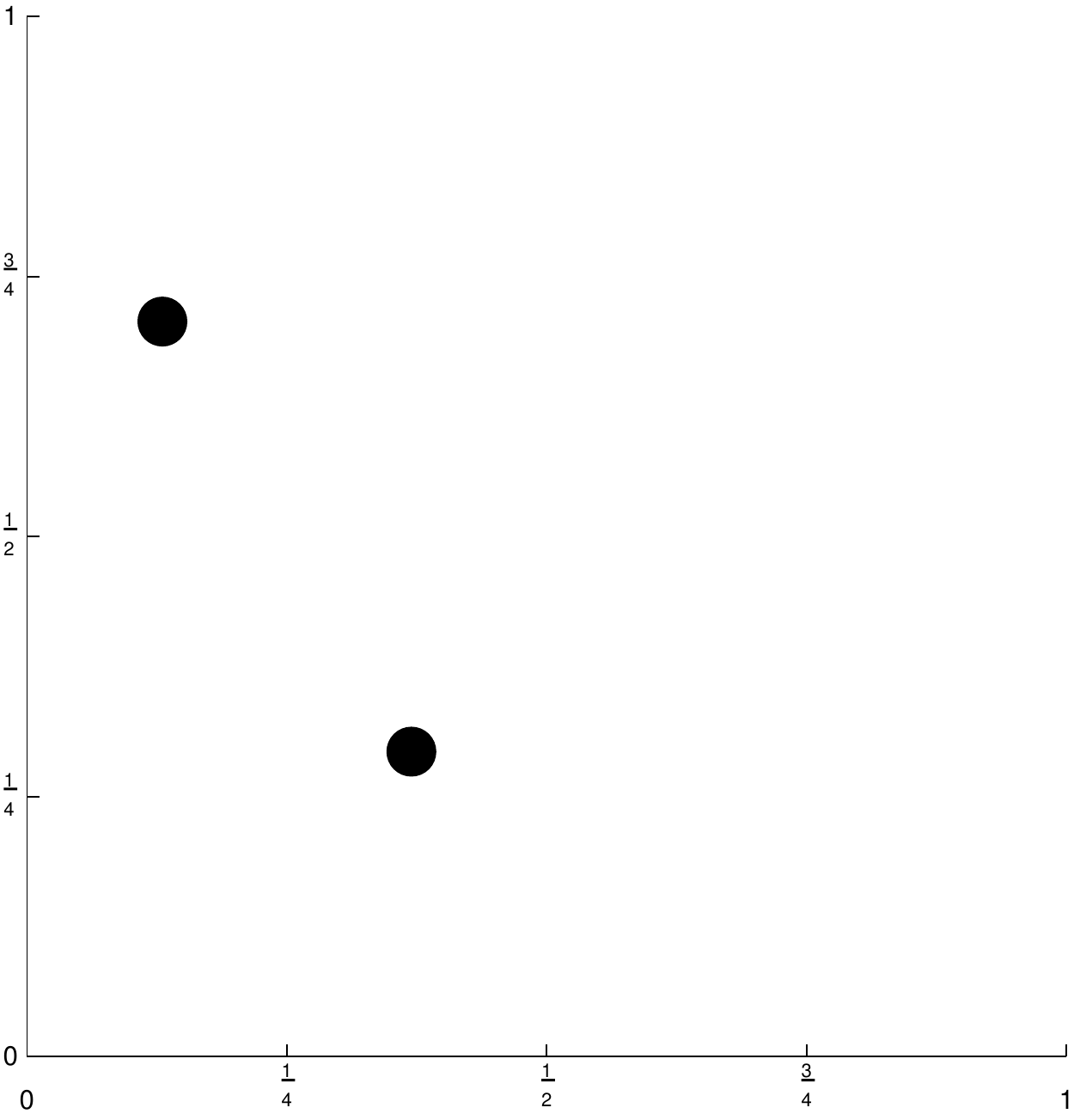}}
}
\begin{center}
Extremum 5b
\end{center}
&
{
\setlength{\fboxsep}{8pt}
\setlength{\fboxrule}{0pt}
\fbox{\includegraphics[width=3.05cm]{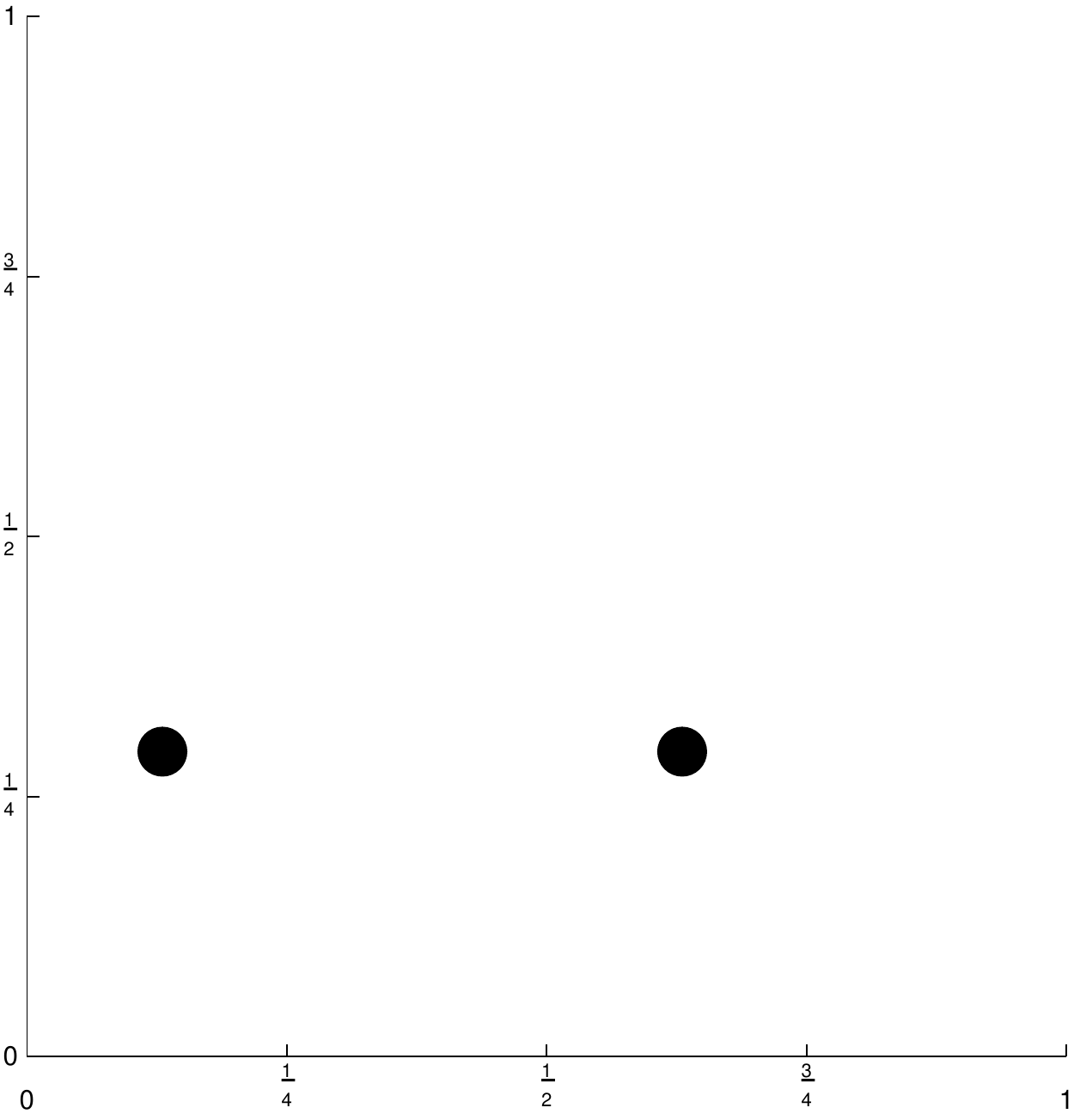}}
}
\begin{center}
Extremum 6
\end{center}
&
{
\setlength{\fboxsep}{8pt}
\setlength{\fboxrule}{0pt}
\fbox{\includegraphics[width=3.05cm]{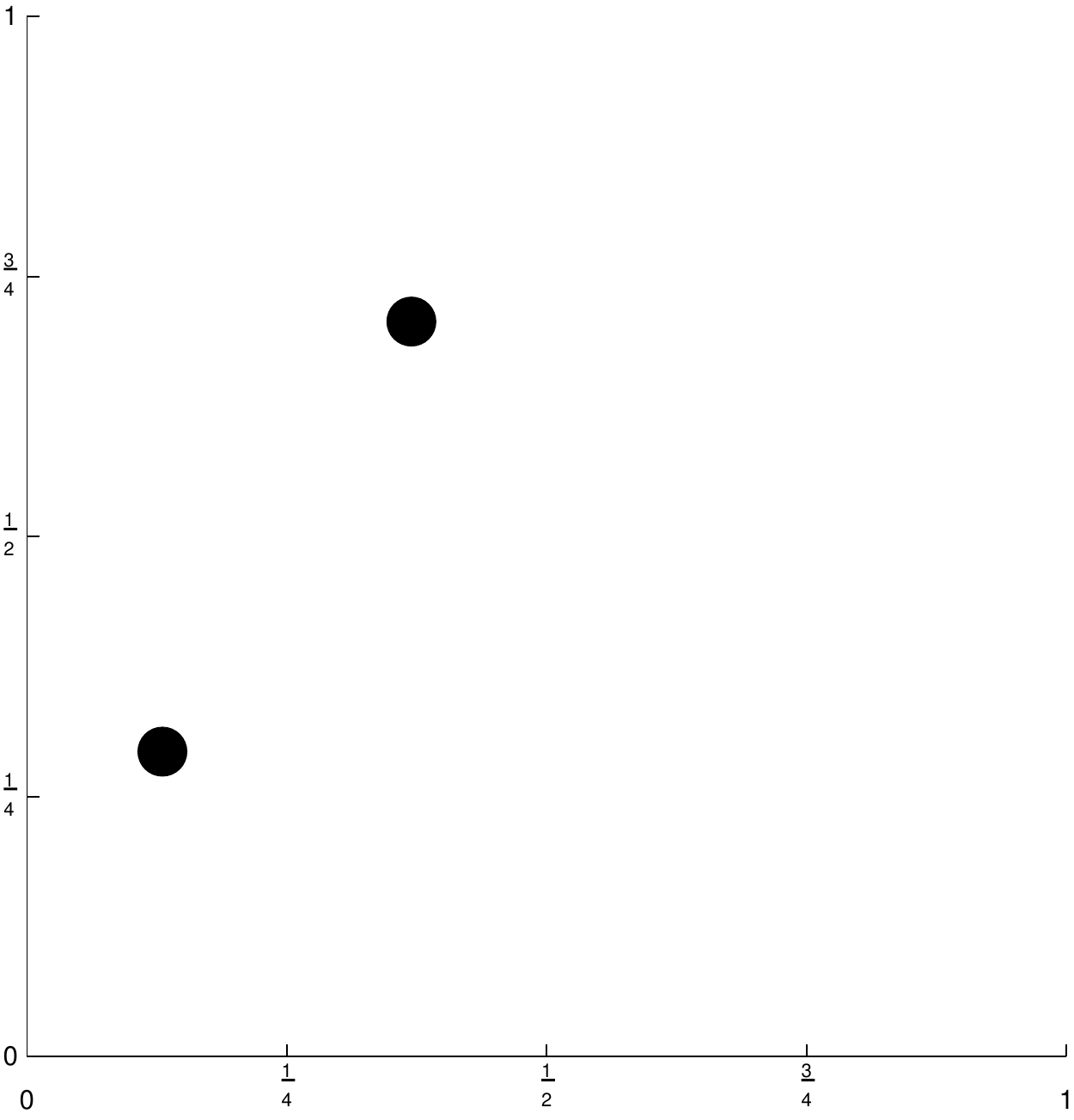}}
}
\begin{center}
Extremum 6b
\end{center}
\\
{
\setlength{\fboxsep}{8pt}
\setlength{\fboxrule}{0pt}
\fbox{\includegraphics[width=3.05cm]{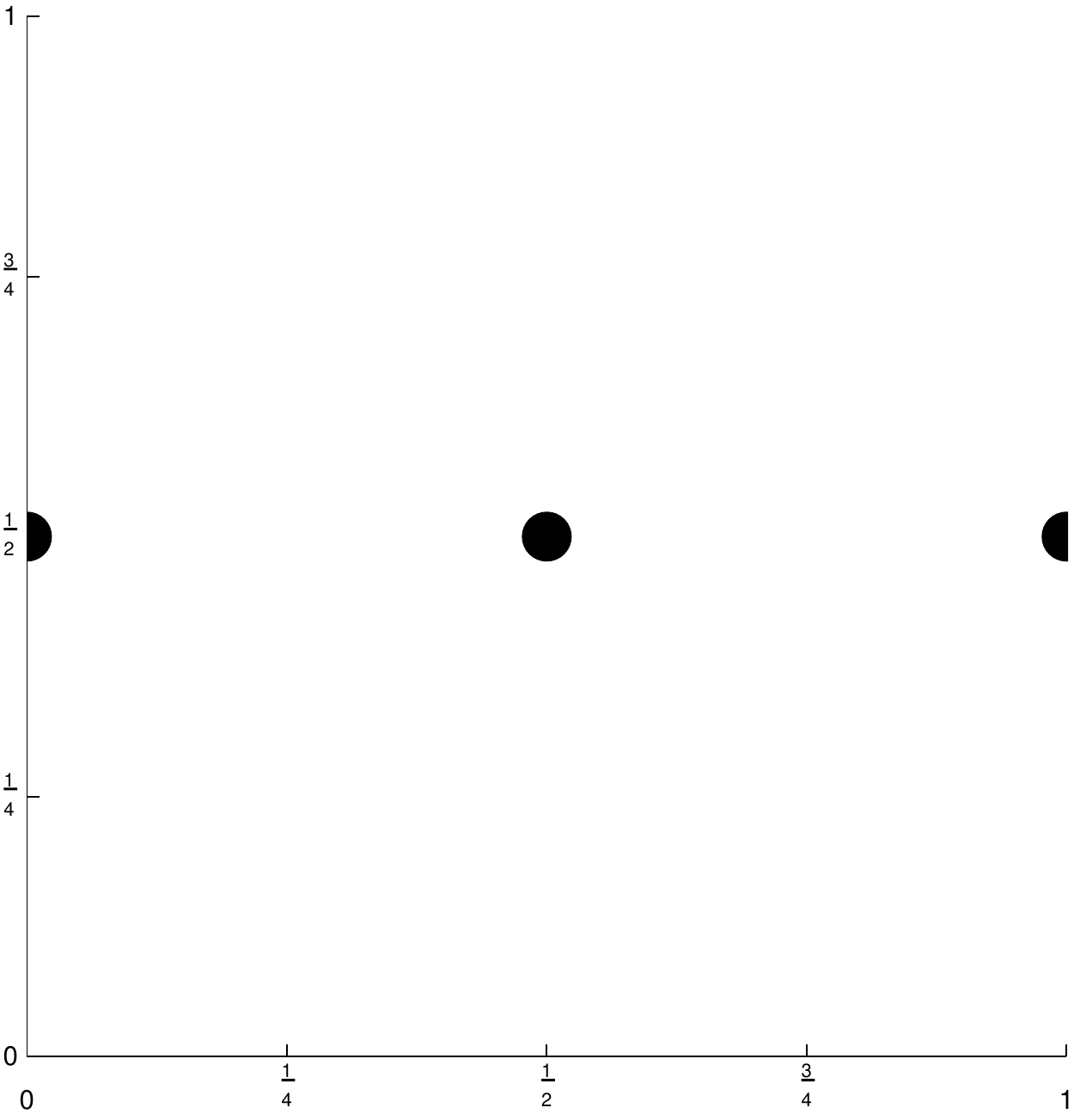}}
}
\begin{center}
Extremum 7
\end{center}
\end{tabular}
\end{minipage}
\caption{The extremal positions of the vacua for the $Spin(5)$ theory.}
\label{extremaspin5}
\end{figure}

 \end{document}